\documentclass[]{aa}
\usepackage{graphicx}
\usepackage{adjustbox}
\usepackage{txfonts}
\usepackage{booktabs} 
\usepackage{caption}
\usepackage{xcolor}

\makeatletter
\renewcommand*\aa@pageof{, page \thepage{} of \pageref*{LastPage}}
\makeatother
\usepackage{hyperref}
\hypersetup{
    colorlinks=true,
    linkcolor=blue,
    filecolor=magenta,      
    urlcolor=blue,
    citecolor=blue,
    }
\defcitealias{marinfranch2009}{MF09}
\defcitealias{dotter2010}{D10}
\defcitealias{vandenberg2013}{V13}
\defcitealias{tailo2020}{T20}

\begin{document} 

\title{Linking Photometry and spectroscopy: profiling multiple populations in globular clusters}
\subtitle{}

\author{E. Dondoglio\inst{1},
A. F. Marino\inst{1},
A. P. Milone\inst{1,2},
S. Jang\inst{3},
G. Cordoni\inst{4},
F. D'Antona\inst{5},
A. Renzini\inst{1},
M. Tailo\inst{1},
A. Bouras Moreno Sanchez\inst{2},
F. Muratore\inst{2},
T. Ziliotto\inst{2},
M. Barbieri\inst{2},
E. Bortolan\inst{2},
E. P. Lagioia\inst{6},
M. V. Legnardi\inst{2},
S. Lionetto\inst{2},
A. Mohandasan\inst{2}
}

\institute{
Istituto Nazionale di Astrofisica - Osservatorio Astronomico di Padova, Vicolo dell’Osservatorio 5, Padova, IT-35122 \\ \email{emanuele.dondoglio@inaf.it}
\and
Dipartimento di Fisica e Astronomia ``Galileo Galilei'', Univ. di Padova, Vicolo dell'Osservatorio 3, Padova, IT-35122
\and
Center for Galaxy Evolution Research and Department of Astronomy, Yonsei University, Seoul 03722, Republic of Korea
\and
INAF – Osservatorio Astronomico di Roma, Via Frascati 33, I-00040 Monte Porzio Catone, Roma, Italy
\and
Research School of Astronomy and Astrophysics, Australian National University, Canberra, ACT 2611, Australia
\and
South-Western Institute for Astronomy Research, Yunnan University, Kunming 650500, People's Republic of China
}

\titlerunning{Photometry and Spectroscopy of Multiple Populations} 
\authorrunning{Dondoglio et al.}

\date{Received XXX / Accepted XXX}

\abstract{
Our understanding of multiple populations in globular clusters (GCs) largely comes from photometry and spectroscopy: appropriate photometric diagrams are able to disentangle first and second populations (1P and 2P)—1P having chemical signatures similar to field stars, and 2P stars showing unique light-element variations—while spectroscopy enables detailed chemical abundances analyses of these populations. We combine multi-band photometry with extensive spectroscopic data to investigate the chemical composition of multiple populations across 38 GCs, yielding a chemical abundance dataset for stars with precise population tagging. This dataset provided the most extensive analysis to date on C, N, O, Na, Mg, and Al variations, revealing the largest sample yet of light-element spreads across GCs. We find that GC mass correlates with light-element variations, supporting earlier photometric studies.
We investigated iron differences among 1P stars, confirming their presence in 19 GCs, and finding a spread consistent with prediction based on photometry. Notably, in eight of them we detected a clear correlation between [Fe/H] values and the position in iron-sensitive photometric diagrams. More massive GCs display larger lithium depletion among 2P stars, which is instead consistent with zero at smaller masses. Notably, some 2P stars with the most extreme chemical differences with respect to 1P stars still show lithium comparable to 1P, suggesting that the 1P polluters have produced some amount of this element.
We analyzed the anomalous stars, a population characterized by enrichment in iron, s-process elements, and C$+$N$+$O, in ten GCs.
NGC\,1851, NGC\,5139 ($\omega$Cen), NGC\,6656, and NGC\,6715 display light-element inhomogeneities similar to 1P and 2P stars.
Iron and barium enrichment varies widely—negligible in some clusters and much larger than observational errors in others. Generally, these elemental spreads correlate with GC mass. 
In clusters with available data, anomalous stars show C$+$N$+$O enrichment compared to the non-anomalous stars.}

\keywords{Techniques: photometric - Stars: abundances - Stars: Population II - Globular Clusters: general.}

\maketitle
\section{Introduction}
\label{sec:intro}
The origin of multiple stellar populations in globular clusters (GCs) remains an unsolved astrophysical puzzle. Although the body of the observational properties associated with the diverse stellar populations has significantly increased in recent years, thanks above all to the contribution of spectroscopy and photometry, the formation mechanisms are still unclear \citep[see][for the most recent reviews on the phenomenon]{bastian2018, gratton2019, milone2022}.

The multiple-color plots, now nicknamed Chromosome Maps (ChMs), were originally introduced based on a combination of four {\textit{Hubble Space Telescope (HST)}} bands (F275W, F336W, F438W and F814W). This combination allows for the construction of a plane where the position of stars is sensitive to the chemical abundance of elements involved in the high-temperature H-burning, via the molecules that they form, as well as to the helium and iron abundance. 
The analysis of these maps has enabled us to define the main properties of the multiple population phenomenon as follows: 

\begin{enumerate}

    \item at least two groups of stars, dubbed first and second populations (1P and 2P), can be distinguished along the maps. 1P stars have chemical abundances similar to Halo field stars, 2P stars are O and C-depleted, and enhanced in He, N, Na, and Al \citep[][]{marino2008, carretta2009, milone2012, lee2021, lee2022}; 1P and 2P stars trace out the typical ChM sequence observed in most Milky Way GCs, named Type~I clusters;

    \item more massive clusters exhibit higher fractions of 2P stars \citep[e.g.,][]{milone2017, lagioia2024} and host He-enhanced 2P stars, which are generally more extreme in terms of all chemical differences;

    \item on the majority of the studied Galactic GCs, the fraction of 2P stars generally exceeds the 1P;

    \item the 1P itself is not-chemically homogeneous showing a sizeable spread in the $m_{\rm F275W}-m_{\rm F814W}$ color \citep[e.g.,][]{milone2017, dondoglio2021}, the $x$ axis of the ChMs. Both spectroscopic and photometric analysis suggests that relatively small variations in the overall metals are responsible for this spread \citep[][]{legnardi2022, lardo2022, marino2023};

    \item a significant fraction of GCs, almost 20\%, show peculiar ChMs. These clusters, named Type~II, display additional sequences running parallel to the main 1P/2P stream of the ChM -often referred to as anomalous stars- formed by stellar populations with enhanced metals and, in most cases, $s$-process elements abundances and total C$+$N$+$O content \citep[see][]{carretta2011, marino2015, yong2015, milone2017, mckenzie2022, dondoglio2023}.

\end{enumerate}

Exploiting spectroscopic elemental abundances from literature, \citet{marino2019} have explored the chemical content of the stellar populations as observed along the ChM of 29 GCs. This analysis shows that stars with different abundances of light elements populate distinct regions of the ChM. On the other hand, despite spanning a broad color range on the map, 1P stars have the same chemical abundances (relative to iron) for the same elements.
Because the ChMs were available only in relative small fields of view of the {\it{HST}} cameras, the chemical tagging on the ChMs has been confined, so far, to stars located in the GCs' most central regions.

Recently, \citet{jang2022} have constructed the ChM by using ground-based UBVI photometry, providing stellar population identifications based on maps for larger fields of view.
The availability of maps for larger regions across the clusters allows us the investigation of chemical and spatial properties of the multiple populations reaching the outermost cluster regions, those that could retain the fossil configuration of the system at the epoch of formation \citep[e.g.,][]{mastrobuono2016}. 

In this study, we take advantage of the two sets of ChMs published by \citet{milone2017} and \citet{jang2022} based on {\it{HST}} and ground-based photometry, respectively, along with new ChMs introduced in this work, and combine this information with literature spectroscopy abundances to explore the chemical pattern of multiple stellar populations. Our dataset includes 38 GCs, in which approximately 3,200 stars have chemical abundances in at least one of the 14 chemical species considered in this work, namely A(Li), [C/Fe], [N/Fe], [O/Fe], [Na/Fe], [Mg/Fe], [Al/Fe], [Si/Fe], [K/Fe], [Ca/Fe], [Ti/Fe], [Fe/H], [Ni/Fe], and [Ba/Fe].
The paper is organized as follows: Section~\ref{sec:2} presents the whole dataset exploited in this work, including both photometry and spectroscopy; Section~\ref{sec:3} illustrates the chemical composition of 1P and 2P stars among our sample of GCs; Section~\ref{sec:4} investigates the spread in different elements and its relation with clusters' mass and metallicity; Section~\ref{sec:5} focuses on the presence of iron variations among 1P stars; Section~\ref{sec:6} studies the lithium depletion among 2P stars; and Section~\ref{sec:7} analyzes in detail the anomalous population of Type II GCs. Finally, a summary and conclusions from our work are reported in Section~\ref{sec:9}.

\section{Dataset}
\label{sec:2}

\subsection{Photometric dataset}
\label{sec:2.1}

In this work, we exploit multi-facility photometry to identify the multiple populations harbored by GCs. In particular, we employ the following two collections of published ChMs:

\begin{enumerate}
    \item {\it{HST}} ChMs from \citet{milone2017}. In this work, Milone and collaborators combined  {\it{HST}} UV and optical photometry from the Ultraviolet and Visual Channel of the Wide Field Camera 3 (WFC3/UVIS) and Advanced Camera for Survey (WFC/ACS), in the F275W, F336W, and F438W and F814W, respectively, from multiple programs (see their Section 2). The published ChMs, allowed a clear-cut separation between 1P and 2P stars among red giant branch (RGB) stars of 57 Galactic GCs. The {\it{HST}} fields of view exploited in this work cover the innermost $\sim$2.7 arcmin, providing an effective separation in their central areas.
    Additionally, we take advance of the ChMs of NGC\,2419 and NGC\,6402 published by \citet{zennaro2019} and \citet{dantona2022}, respectively, obtained through the same procedure.

    \item Ground-based ChMs from \citet{jang2022}. This study exploited the photometric dataset published by \citet{stetson2019}, which produced UBVRI catalogs of 48 Milky Way GCs based on a large sample of archive ground-based observations taken with several facilities. Jang and collaborators exploited this dataset to build ChMs for 29 GCs from this sample, for which a separation between the distinct stellar groups is possible along the RGB.
    These ChMs are based on observations over a wider field of view than the {\it{HST}} one, thus providing a multiple population tagging up to around the outskirts of GCs.
\end{enumerate}

Moreover, we consider the U-band Wide-Field Imager (WFI) photometry of NGC\,5286 and NGC\,6656 from the SUrvey of Multiple pOpulations in GCs
(SUMO; programme 088.A-9012-A, PI. A. F. Marino), taken with the Max Planck 2.2m telescope at La Silla \citep[e.g.,][]{marino2015, marino2019}. This photometry, combined with archive observations in B, V, and I, served us to build ground-based ChMs for these two GCs.
Finally, we make use of the NGC\,5139 ($\omega$Cen) catalog recently published by \citet{haberle2024}. This dataset was built by reducing uniformly all the archive {\it{HST}} images in the innermost 10$\times$10 arcmin$^{\rm 2}$, thus allowing us to build a ChM to separate multiple populations at larger distances than what was possible by considering the ChM by \citet{milone2017} alone.
The ChMs that we introduce in this work are presented in Section~\ref{sec:3.1}.

\subsection{Spectroscopic dataset}
\label{sec:2.2}

\begin{figure*}
\includegraphics[width=18.4cm, clip, trim={0cm 0cm 0cm 0cm}]{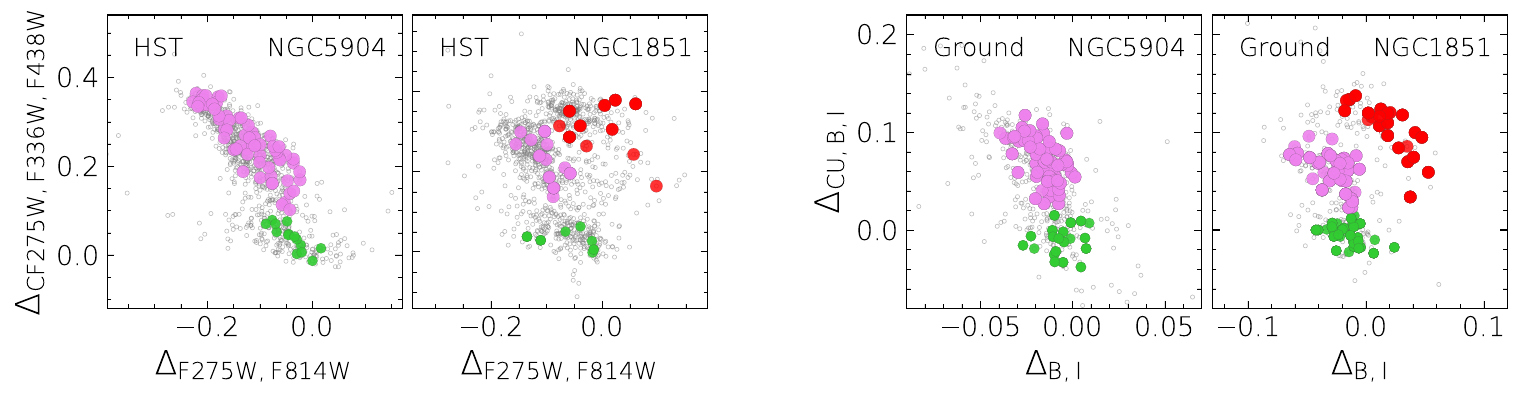}
    \caption{{\it{Left panels:}} from left to right, $\Delta_{\rm C F275W, F336W, F438W}$ vs. $\Delta_{\rm F275W, F814W}$ {\it{HST}} ChMs of NGC\,5904 and NGC\,1851. 1P, 2P, and anomalous stars with available spectroscopy measurements are highlighted with green, violet, and red dots, respectively.
    {\it{Right panels:}} same as left panels but with the ground-based $\Delta_{\rm C U, B, I}$ vs. $\Delta_{\rm B, I}$ ChM.
    }
    \label{fig:example}
\end{figure*}

To estimate the chemical properties of the multiple populations detectable from the ChM, stars with available chemical abundances were cross-matched with the diagrams described in Section~\ref{sec:2.1} to explore the chemical properties of each population identified through photometry. We consider the following elements in our study: Li, C, N, O, Na, Mg, Al, Si, K, Ca, Ti, Fe, Ni, and Ba. Our analysis incorporates data from three spectroscopic surveys:

\begin{enumerate}

    \item Apache Point Observatory Galactic Evolution Experiment (APOGEE). This survey, conducted utilizing the du Pont Telescope and the Sloan Foundation 2.5 m Telescope \citep{gunn2006} at Apache Point Observatory, observed over 700,000 stars with a resolution of R$\sim$22,500. The latest Data Release 17 \citep[DR17,][]{abdurro2022} offers chemical abundance insights into multiple populations of 21 GCs in our sample, making it the most numerous dataset of stars that overlap with our photometric sample. For our study, we select APOGEE stars with a signal-to-noise Ratio (SNR) $>$ 70 and exclude those flagged with {\tt{ASCAPFLAG}} set to {\tt{STAR\_BAD}}, indicative of spectrum-related issues such as a high number of bad pixels or non-stellar classification\footnote{See documentation {\url{https://www.sdss4.org/dr17/}} for details.}.
    The APOGEE dataset provides two carbon abundances, C and CI, measured trough molecules and neutral carbon lines, respectively. \citet{jonsson2020} proved that for metal-poor stars, CI is likely to be more accurate since molecular lines become particularly weak. For this reason, we consider the CI values as a tracer of carbon abundances in our sample.
    \citet{jonsson2020} also shown that APOGEE [Na/Fe] measurements exploit relatively weak lines, making this element one of the least reliable of this survey. Since sodium has been extensively studied in several other works in the literature, we do not consider [Na/Fe] estimates based on APOGEE spectra due to their relatively poor reliability.

    \item Gaia-ESO Survey (GES). This survey leverages spectroscopy data obtained from the FLAMES spectrograph \citep[][]{pasquini2002} on the Very Large Telescope (VLT), operating in multi-object spectroscopy mode with GIRAFFE and UVES, with spectral resolutions of R$\sim$18,000-35,500 (depending on the setup) and $\sim$47,000, respectively. The latest internal Data Release 6 \citep[iDR6; see][]{gilmore2022, randich2022, hourihane2023} encompasses data from over 114,000 stars, including members of 11 GCs featured in our sample. Abundance selection criteria involve the so-called simplified flags\footnote{{\url{https://www.eso.org/rm/api/v1/public/releaseDescriptions/191}}.}: {\tt{SNR}}, that exclude stars with inaccurate results caused by low signal-to-noise ratio, {\tt{SRP}}, to avoid stars that had spectral reduction problems and no abundances are available, and {\tt{NIA}}, that exclude those sourses with too few available lines for abundance determinations.

    \item Galactic Archaeology with HERMES (GALAH). This initiative derives chemical abundances using the High Efficiency and Resolution Multi-Element Spectrograph (HERMES) at the Anglo-Australian Telescope \citep[][]{sheinis2015}, achieving a resolution of R$\sim$28,000. Our analysis makes use of the most recent Data Release 4 \citep[GALAH+DR4, see][]{buder2024}, encompassing data more than 900,000 Milky Way stars. We identified RGB stars with ChM tagging in nine GCs. The selection process for the stellar sample adhered to criteria recommended in the GALAH+DR4 documentation\footnote{{{\url{https://www.galah-survey.org/dr4/overview/}}.}}, including SNR $>$ 30, {\tt{flag\_sp == 0}}, and {\tt{flag\_x\_fe == 0}}.

\end{enumerate}

As a general criterion, beyond the specific ones listed for each survey, we also exclude abundance measurements characterized by uncertainties much larger than their average values. 

Finally, we extensively utilize 56 publicly available spectroscopic catalogues from studies published within the last $\sim$20 years. These catalogues encompass abundance measurements of elements pertinent to our investigation, and are listed in Table~\ref{tab:data}.

In the analysis presented in the following Sections, we only consider abundance ratio measurements, without including upper limits, which among the datasets considered, where present for A(Li), [O/Fe], [Na/Fe], and [Al/Fe].

\subsection{Final dataset}
\label{sec:2.3}

In this work, we consider GCs' stars with both the population tagging ensured by the ChM and available spectroscopic abundances. Our final sample includes 38 GCs: 37 of them (all except NGC\,1904) have available {\it{HST}} ChM, 23 have ground-based ChM, with a total of 22 having both available. 

Detailed information regarding the selected GCs, the utilized public catalogs, and the number of common stars for each element is presented in Table~\ref{tab:data}. The numbers provided represent the combined count of 1P and 2P stars with spectroscopic abundance data. Notably, for Type II GCs (namely NGC\,0362, NGC\,1261, NGC\,1851, NGC\,5139, NGC\,5286, NGC\,6388, NGC\,6656, NGC\,6715, NGC\,6934 and NGC\,7089), this count also includes the anomalous population. Our final sample of stars with both ChM and chemical abundance information is made of 1,004 1P, 1,883 2P, and 303 anomalous, for a total of 3,190 members.

\section{Chemical abundances of multiple stellar populations}
\label{sec:3}

In this Section, we integrate population tagging from {\it{HST}}- and ground-based ChMs with literature spectroscopic measurements to assess the chemical abundances of various populations within GCs. Section~\ref{sec:3.1} involves identifying 1P and 2P stars (including the anomalous when present) in the ChM. This selection is then utilized in Section~\ref{sec:3.2}, where we merge it with our spectroscopic dataset to determine the median values and variability in the chemical abundances of the tagged populations.

\begin{figure*}
\includegraphics[height=5.0cm, clip, trim={-0.9cm 0cm 0cm 0cm}]{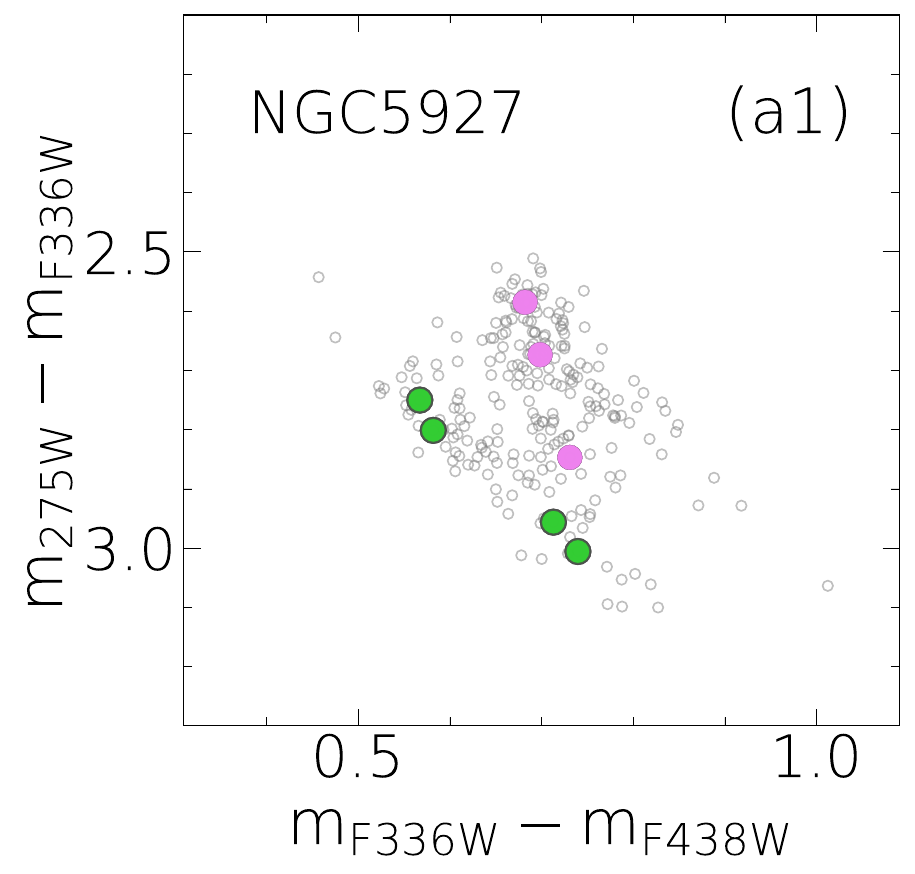}
\includegraphics[height=5.0cm, clip, trim={0cm 0cm 0cm 0cm}]{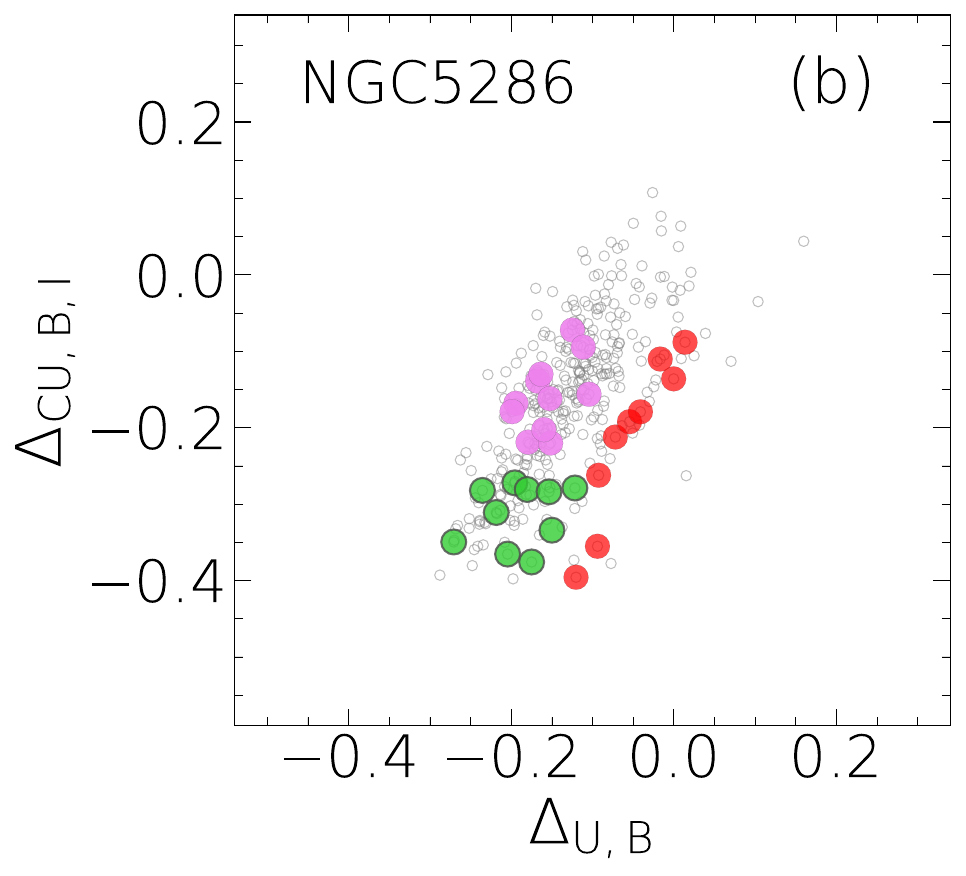}
\includegraphics[height=5.0cm, clip, trim={0cm 0cm 0cm 0cm}]{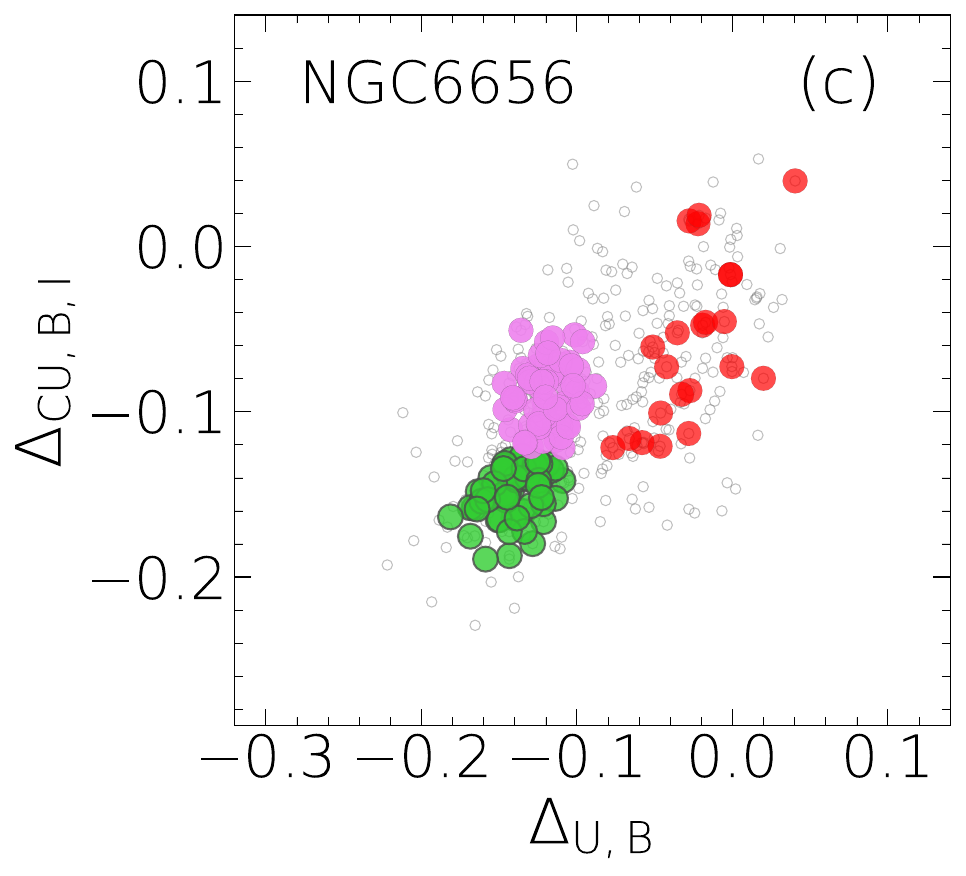}
\includegraphics[height=5.0cm, clip, trim={0cm 0cm 0cm 0cm}]{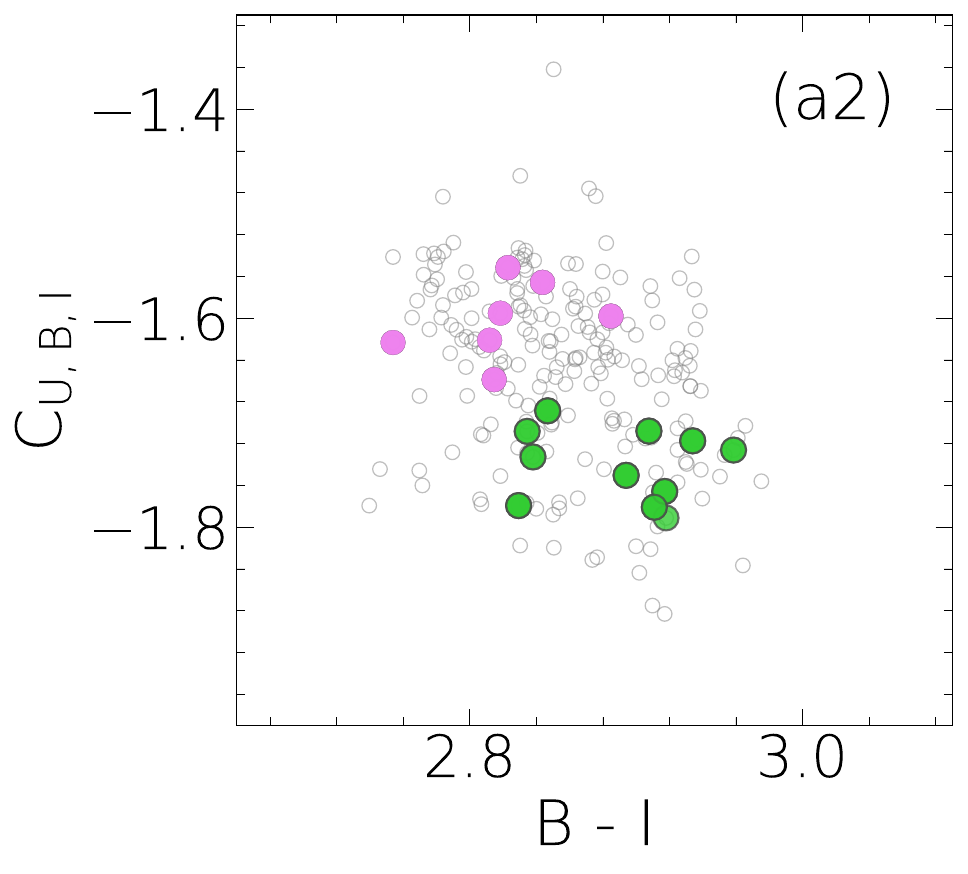}
\includegraphics[height=5.0cm, clip, trim={-2.4cm 0cm 0cm 0cm}]{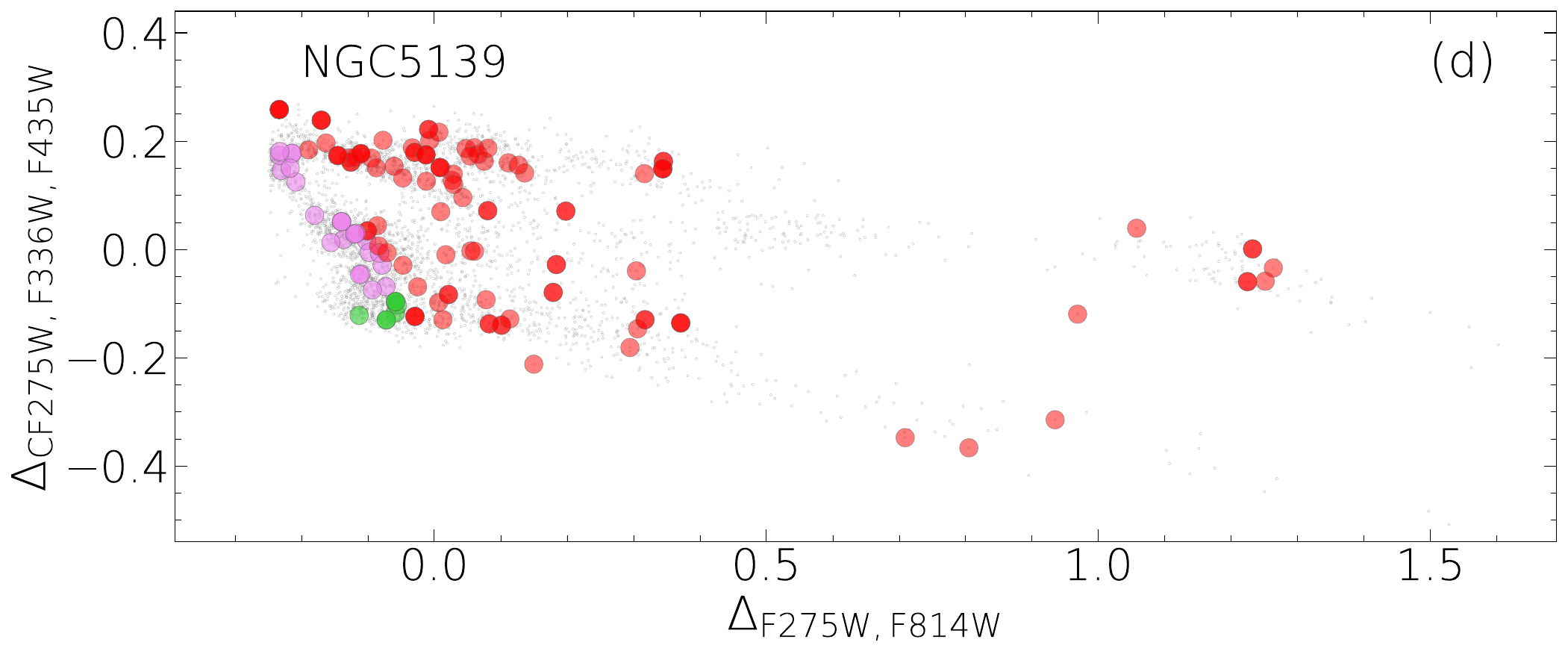}
    \caption{
    Summary of the photometric diagrams introduced in Section~\ref{sec:3.1}.
    {\it{Panel (a1) and (a2):}} $m_{\rm F275W}$-$m_{\rm F336W}$ vs. $m_{\rm F336W}$-$m_{\rm F438W}$ two-color diagram and $C_{\rm U,B,I}$ vs. $B-I$ pseudo two-color diagram for the red HB of NGC\,5927. 
    {\it{Panel (b):}} $\Delta_{\rm C U,B,I}$ vs. $\Delta_{\rm U,B}$ ChM of NGC\,5286. {\it{Panel (c):}} same but for NGC\,6656.
    {\it{Panel (d):}} $\Delta_{\rm C F275W, F336W, F435W}$ vs. $\Delta_{\rm F275W, F814W}$ ChM of $\omega$Cen derived from the catalog by \citet{haberle2024}.
    Stars highlighted with green, violet, and red dots represent 1P, 2P, and anomalous stars.}
    \label{fig:new_chm}
\end{figure*}

\subsection{Population tagging along the Chromosome Map}
\label{sec:3.1}

The initial step in acquiring the chemical characteristics of multiple stellar populations involves a precise selection process within the ChM. To accomplish this, we employ the methodology outlined by \citet{dondoglio2023} for distinguishing various populations within the ChM of NGC\,1851. This approach entails constructing ellipses that enclose the blobs of stars representing distinct groups, following these steps: (i) visually identifying genuine 1P and 2P members and using the median of the ChM coordinates as their center, (ii) determining the optimal orientation that minimizes their orthogonal dispersion to establish the major axis orientation of the ellipses, and (iii) deriving the lengths of the semi-major and semi-minor axes by considering 2.5 times the dispersion of the stars along their respective directions determined in the preceding step.
We consider as anomalous stars the ones selected by \citet{milone2017} and \citet{jang2022} based on their position on the $m_{\rm F336W}$ vs. $m_{\rm F336W}$-$m_{\rm F814W}$ and $I$ vs. $U$-$I$ color magnitude diagrams (CMDs), respectively\footnote{In our study of NGC\,1851, we choose to utilize the ChMs provided by \citet{milone2017} and \citet{jang2022} over the more recent developed by \citet{dondoglio2023} because of their wider magnitude coverage, which suits our goal of tagging a larger number of stars.}.
The anomalous GCs NGC\,6715 and NGC\,7089 host an additional population \citep[see][and references therein]{milone2017}. For the first, this population has been associated with the host Sagittarius dwarf galaxy \citep[e.g.,][]{siegel2007}, while for the latter these stars have likely a similar origin \citep[e.g.,][]{milone2015a}. 
We do not include them among the anomalous stars, and we consider them as a separate population from the other GC populations in the case of NGC\,7089, while for NGC\,6715 no spectroscopic measurements are available for the stars in the ChM.
For the ground-based catalogs of NGC\,5286 and NGC\,6656, we rely on the anomalous stars tagging performed by \citet{marino2019}.

The left panels of Figure~\ref{fig:example} display the {\it{HST}} ChMs of NGC\,5904 and NGC\,1851, which are Type I and II GCs, respectively, as an example of the results of our cross-matching. The two right panels represent the same tagging but by using ground-based ChMs.
Stars with available spectroscopic measurements in at least one of the datasets reported in Table~\ref{tab:data} are colored in green, violet, and red according to their belonging to the 1P, 2P, and anomalous populations, respectively. If a star has population tagging in both ChMs, we consider the photometric information from the {\it{HST}} data. 
In Appendix~\ref{sec:ap1}, we display all the ChMs published by \citet{milone2017} and \citet{jang2022} for the clusters in our sample, where the stars with available spectroscopy are color-coded following the prescriptions introduced in Figure~\ref{fig:example}. 

\begin{figure*}
\includegraphics[width=17.2cm, clip, trim={0cm 0.2cm 0cm 0.25cm}]{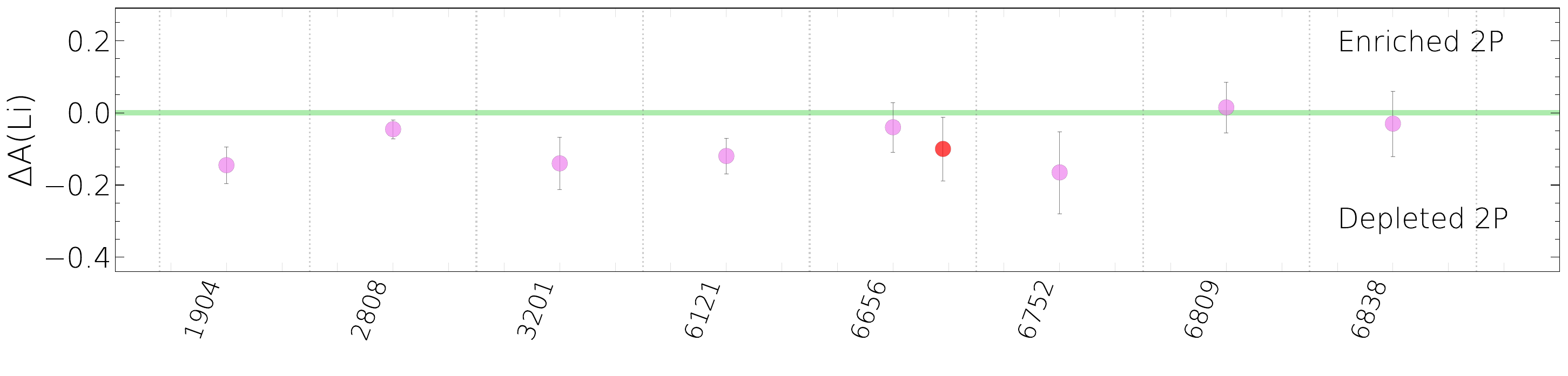}
\includegraphics[width=17.2cm, clip, trim={0cm 0.2cm 0cm 0.25cm}]{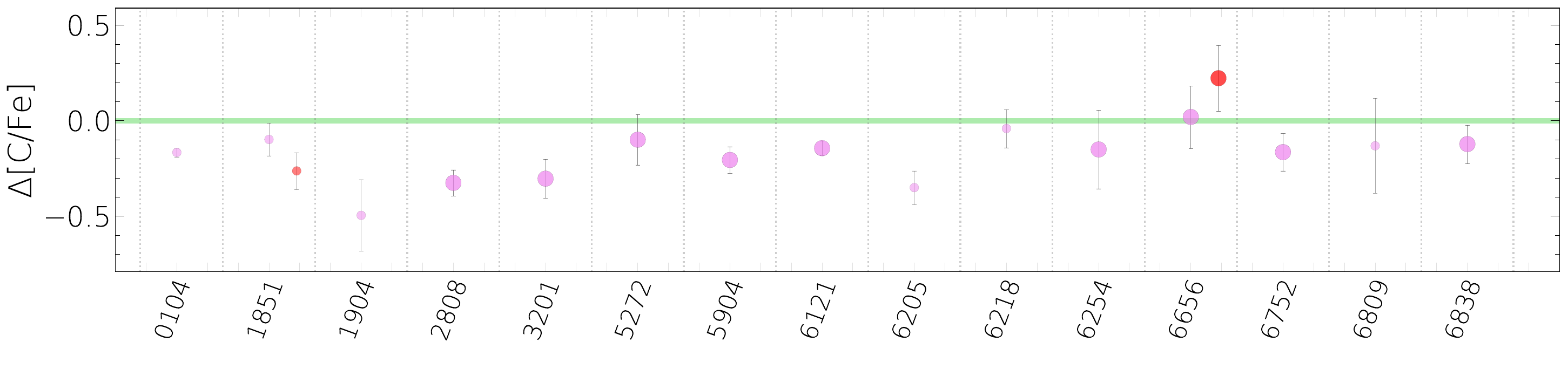}
\includegraphics[width=17.2cm, clip, trim={0cm 0.2cm 0cm 0.25cm}]{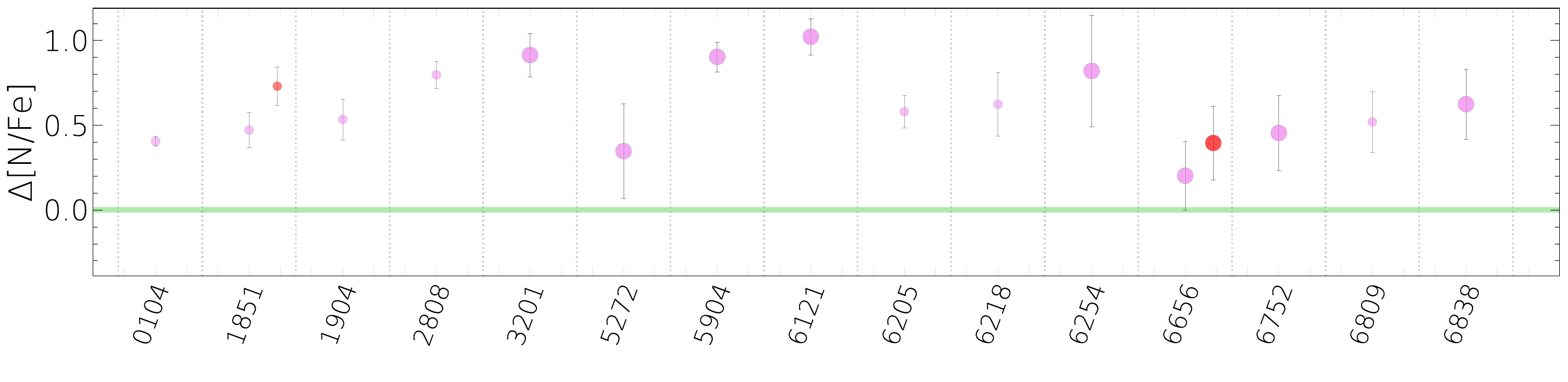}
\includegraphics[width=17.2cm, clip, trim={0cm 0.2cm 0cm 0.25cm}]{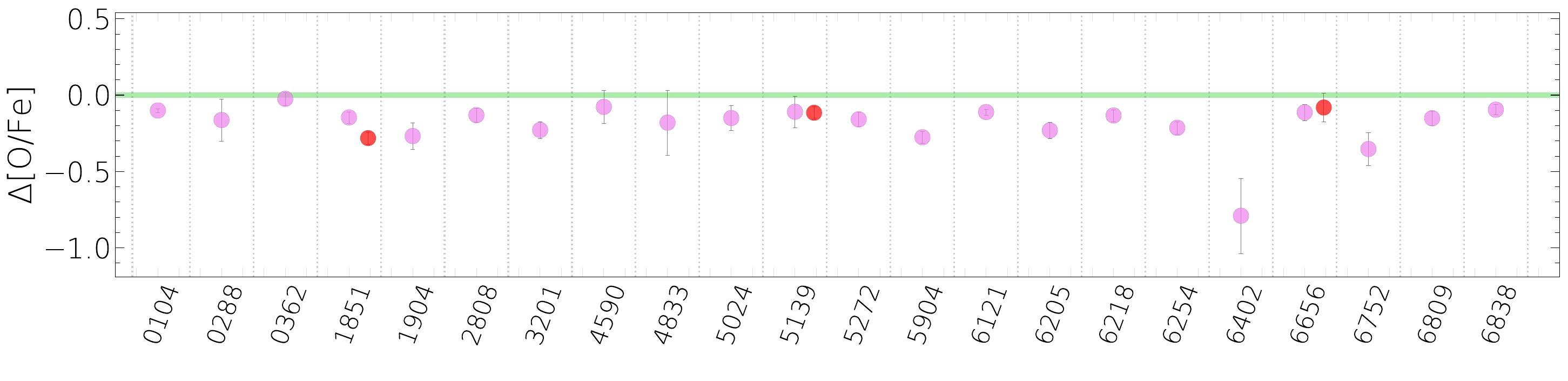}
\includegraphics[width=17.2cm, clip, trim={0cm 0.2cm 0cm 0.25cm}]{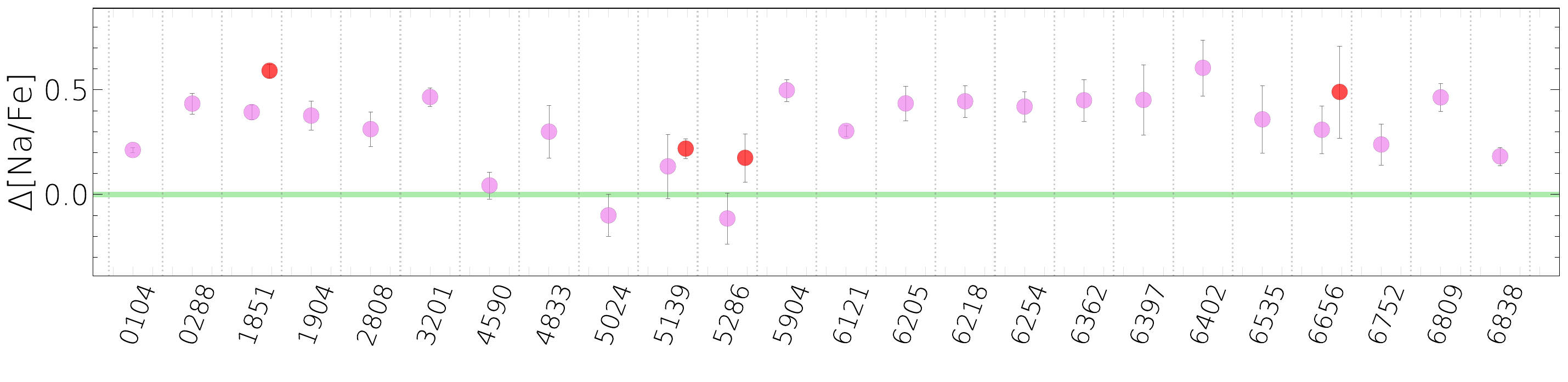}
    \caption{Differences between 1P and 2P (violet) and anomalous (red) median abundances of A(Li), [C/Fe], [N/Fe], [O/Fe], and [Na/Fe]. The green lines separate the regime where 2P and anomalous stars are enriched or depleted (see text for details). y-axis scale vary from plot to plot for clearer representation purposes. The smaller dots indicate GCs where the $\Delta$[C/Fe] and $\Delta$[N/Fe] measurements are based in RGB stars brighter than the bump.}
    \label{fig:ab1}
\end{figure*}

\begin{figure*}
\includegraphics[width=17.2cm, clip, trim={0cm 0.2cm 0cm 0.25cm}]{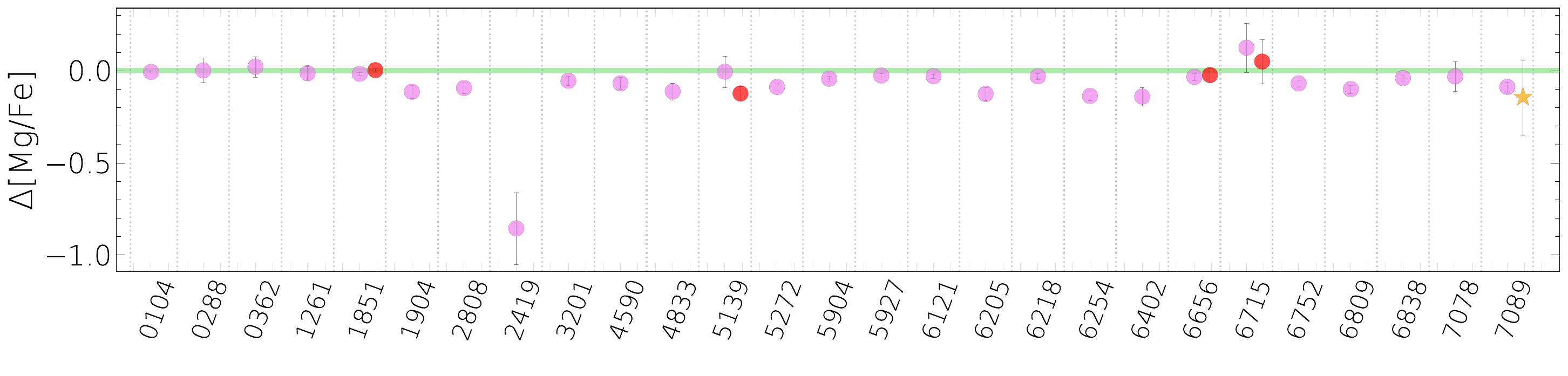}
\includegraphics[width=17.2cm, clip, trim={0cm 0.2cm 0cm 0.25cm}]{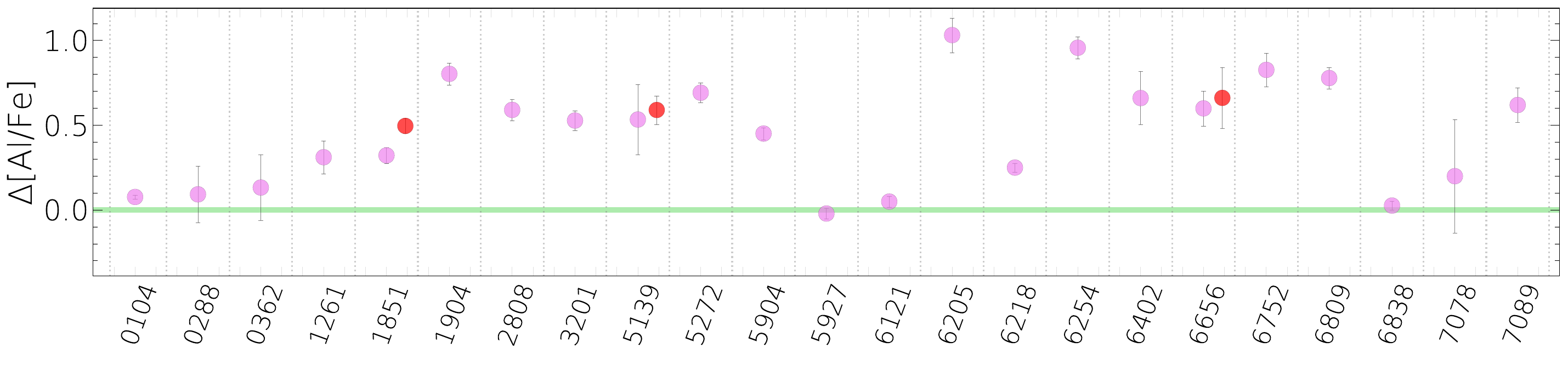}
\includegraphics[width=17.2cm, clip, trim={0cm 0.2cm 0cm 0.25cm}]{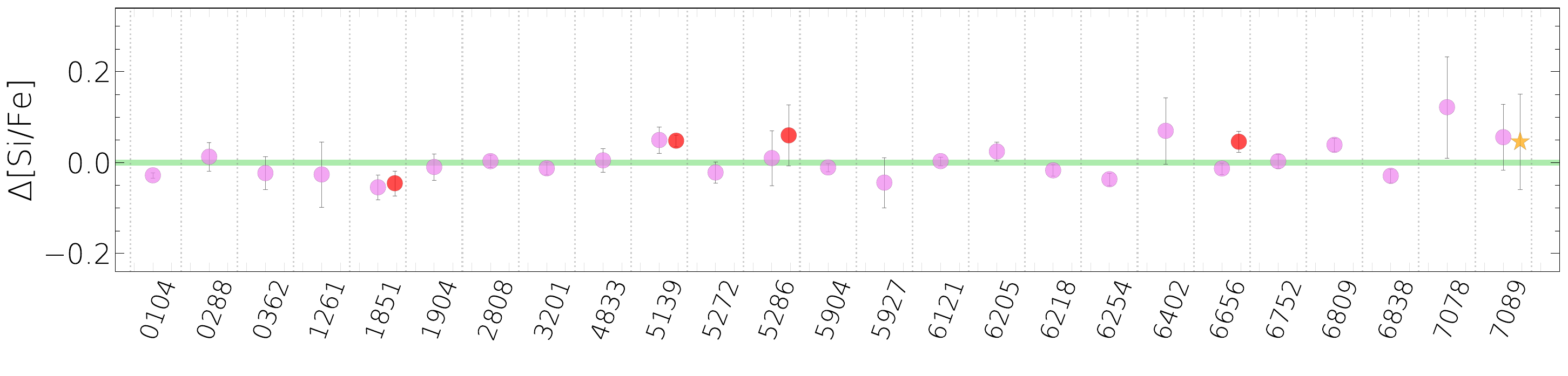}
\includegraphics[width=17.2cm, clip, trim={0cm 0.2cm 0cm 0.25cm}]{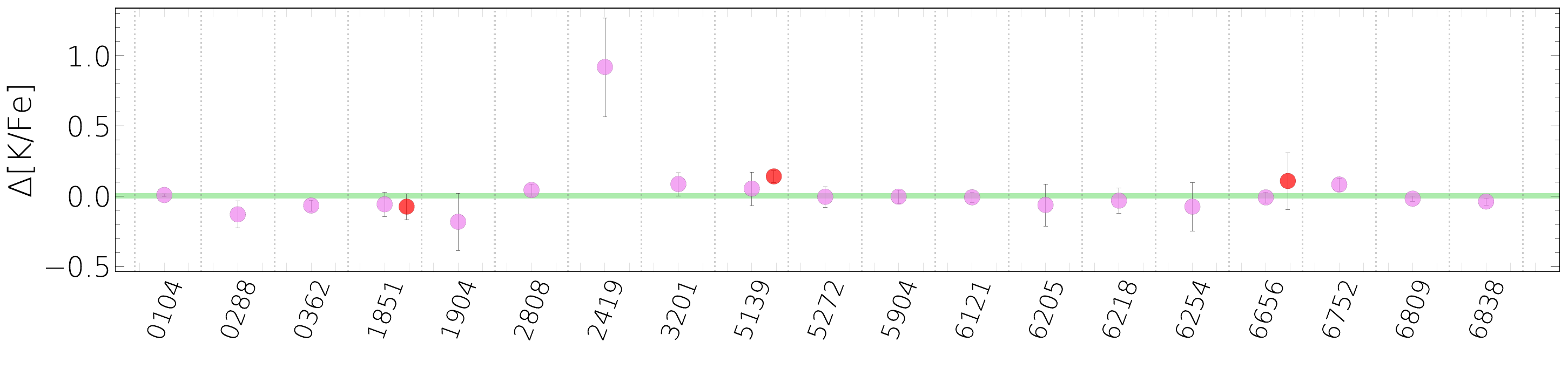}
\includegraphics[width=17.2cm, clip, trim={0cm 0.2cm 0cm 0.25cm}]{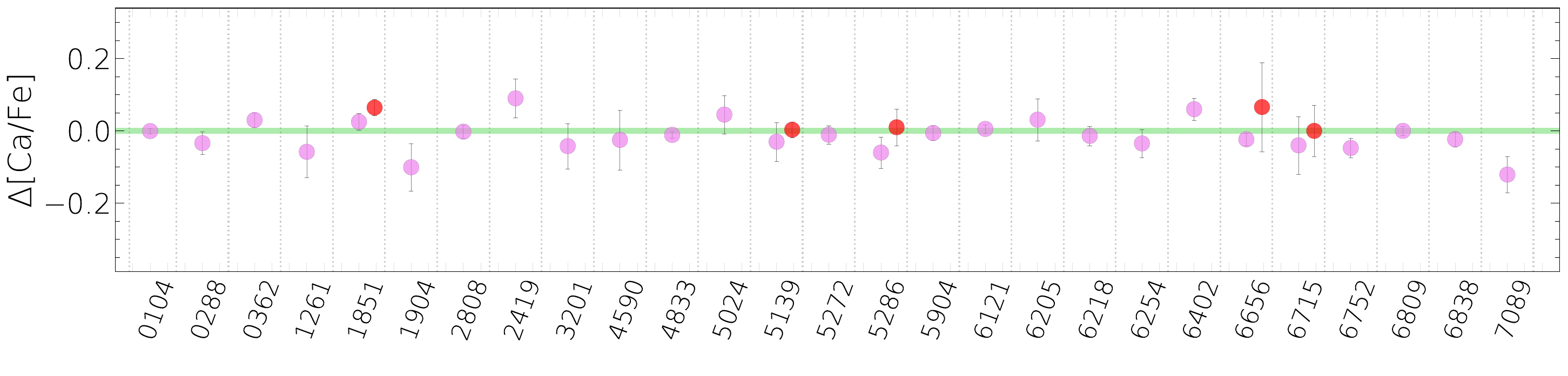}
    \caption{Same as Figure~\ref{fig:ab1} but for the [Mg/Fe], [Al/Fe], [Si/Fe], [K/Fe], and [Ca/Fe] abundances. The orange starred symbol refers to host galaxy remnant of NGC\,7089.}
    \label{fig:ab2}
\end{figure*}

As anticipated in Section~\ref{sec:2}, we also consider additional photometric diagrams, which are summarized in Figure~\ref{fig:new_chm}.
Panels (a1) and (a2) represent the diagrams exploited for NGC\,5927 to separate 1P and 2P stars in its red Horizontal Branch HB by using {\it{HST}} and UBI photometry, respectively. Since no spectroscopic measurements of RGB stars with ChM tagging are documented in the literature, we consider the red HB evolutionary phase, which has been examined in the context of chemical abundance analysis.
\citet{dondoglio2021} have demonstrated the efficacy of {\it{HST}} and ground-based multi-band photometry in distinguishing 1P and 2P stars among red HB stars, thanks to the $m_{\rm F275W}$-$m_{\rm F336W}$ vs. $m_{\rm F336W}$-$m_{\rm F438W}$ two-color diagram presented in panel (a1), where 1P stars occupy the lower-left sequence, and the pseudo-two-color diagram $C_{\rm U,B,I}$ vs. $B-I$ in panel (a2), in which 1P and 2P stars are distributed below and above $C_{\rm U,B,I}\sim -$1.7, respectively. 
We then build the ground-based ChMs of NGC\,5286 (panel (b)) and NGC\,6656 (panel (c)) by following the same approach of \citet{jang2022}, considering the $U$-$B$ color and the $C_{\rm U, B, I}$ pseudo-color from the datasets presented in Section~\ref{sec:2.1}.
Finally, we present the $\Delta_{\rm C F275W, F336W, F435W}$ vs. $\Delta_{\rm F275W, F814W}$ ChM of $\omega$Cen derived from the \citet{haberle2024} catalog in panel (d), obtained by applying the iterative procedure introduced in the Appendix A of \citet{milone2017}. This diagram will be exploited for the populations' tagging among stars outside $\sim$2 arcmin from the cluster center, while for the innermost area we consider the ChM by Milone and collaborators.
In all these diagrams, stars with available chemical abundances are depicted as in Figure~\ref{fig:example}.

\begin{figure*}
\includegraphics[width=17.2cm, clip, trim={0cm 0.2cm 0cm 0.25cm}]{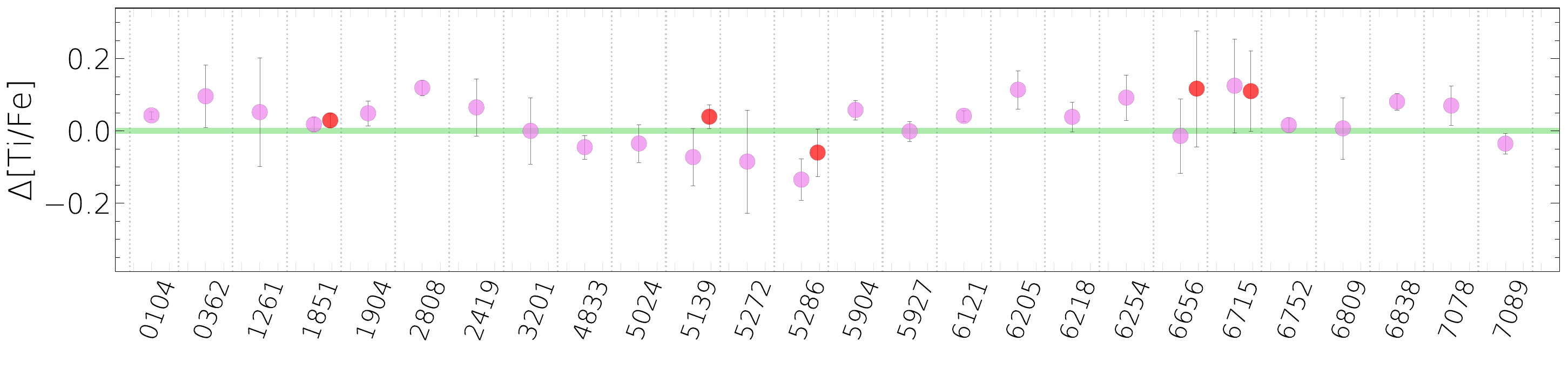}
\includegraphics[width=17.2cm, clip, trim={0cm 0.2cm 0cm 0.25cm}]{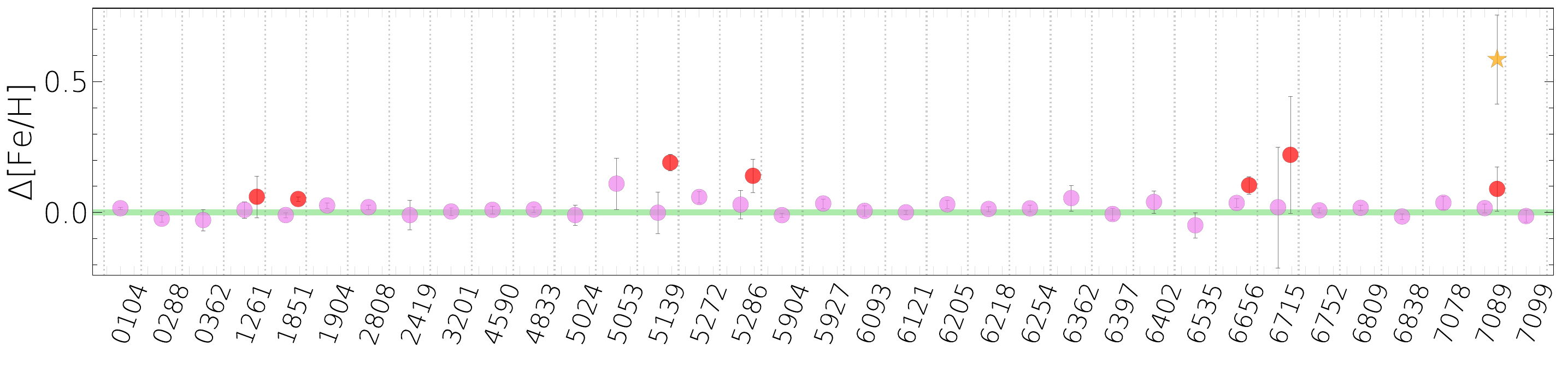}
\includegraphics[width=17.2cm, clip, trim={0cm 0.2cm 0cm 0.25cm}]{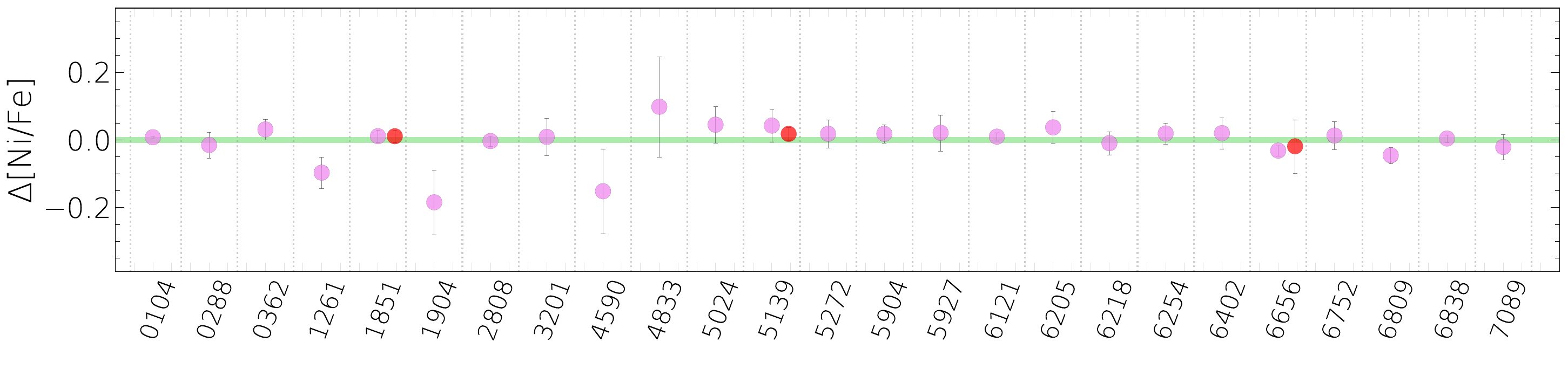}
\includegraphics[width=17.2cm, clip, trim={0cm 0.2cm 0cm 0.25cm}]{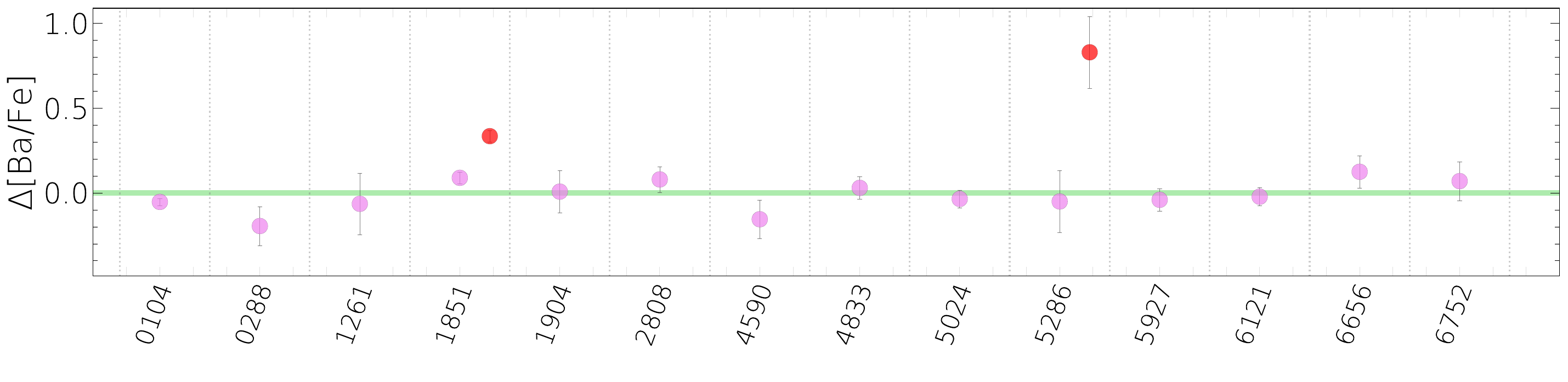}
    \caption{Same as Figure~\ref{fig:ab1} but for the [Ti/Fe], [Fe/H], [Ni/Fe], and [Ba/Fe] abundances.}
    \label{fig:ab3}
\end{figure*}

\subsection{Element abundances and distribution}
\label{sec:3.2}

With the tagging introduced in Section~\ref{sec:3.1}, we can now investigate the presence of chemical differences among the distinct populations in our sample of GCs. Results for A(Li) [C/Fe], [N/Fe], [O/Fe], and [Na/Fe] are displayed in Figure~\ref{fig:ab1}, while Figure~\ref{fig:ab2} illustrates [Mg/Fe], [Al/Fe], and [Si/Fe], [K/Fe], and [Ca/Fe], and Figure~\ref{fig:ab3} [Ti/Fe], [Fe/H], [Ni/Fe], and [Ba/Fe].
The spectral features exploited in the considered works cover two wavelength regimes: the optical range, covering roughly 3500-8800 \AA, and the IR regime exploited by APOGEE, which covers the H band (1.15-1.70 $\mu$m).
In each plot, we show in the y-axis the $\Delta X$ quantity -where X is any of the considered elemental ratio- measured as the difference between the median abundance of 2P stars (violet points) or anomalous stars (red points) and the 1P stars. The values relative to the population in NGC\,7089 associated with the remnant of its host galaxy are indicated with the orange starred symbols. Positive and negative $\Delta X$ values indicate that the 2P (or anomalous) stars are enriched and depleted in a given specie compared to the 1P population, respectively, as depicted in the upper panel of Figure~\ref{fig:ab1}. The horizontal green line delimits the two regimes and, by construction, lies at $\Delta X =$ 0.
$\Delta X$ has been measured only by considering the datasets with at least four 1P, four 2P, and (when present) four anomalous stars, in order to have enough statistics for inferring reliable median abundances of each population. NGC\,5986, NGC\,6388, and NGC\,6934 do not pass these criteria for any of the elements involved, thus they never appear in the analysis presented in the current Section. However, we still include them in our dataset, as they will be involved in the analysis of Section~\ref{sec:4} and~\ref{sec:7}.

In Table~\ref{tab:deltas}, we display all the $\Delta X$ and their uncertainty, derived by propagating the 1P, 2P and anomalous median abundance errors, estimated as $r.m.s. / \sqrt{N - 1}$, with $r.m.s.$ and $N$ being the root mean square and the number of stars of a certain population. For GCs with abundances from more than one dataset in a given element, we measured separately $\Delta X$ for each dataset, and then we combined them by considering their weighted average\footnote{When considering spectroscopic estimates from different works, it is unavoidable to deal with systematics between the various datasets. However, $\Delta X$ is measured by subtracting median abundances of stars from the same dataset (therefore affacted by the same systematics), thus erasing any systematics.}.

\begin{figure*}
\includegraphics[width=17.5cm, clip, trim={0cm 0cm 0cm 0cm}]{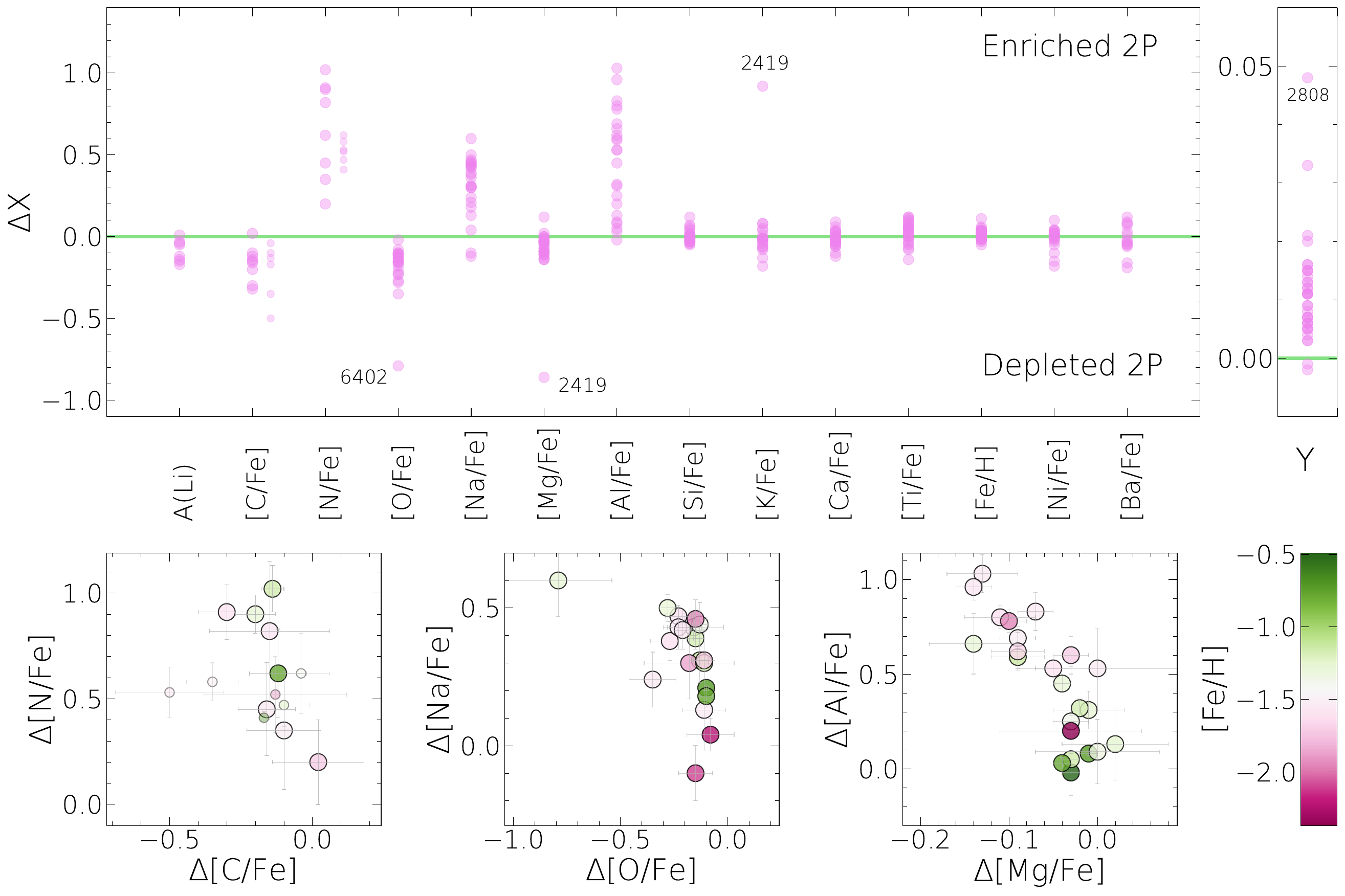}
    \caption{{\it{Upper panels:}} Difference between the median of the 2P and 1P abundance of each analyzed species. 
    Positive and negative values (above and below the green line), represent 2P stars enrichment and depletion, respectively.
    In the right panel, we represent the 2P-1P helium difference (in mass fraction) from \citet{milone2018}.
    {\it{Lower panels:}} $\Delta$[N/Fe] vs. $\Delta$[C/Fe] (left), $\Delta$[Na/Fe] vs. $\Delta$[O/Fe] (middle), and $\Delta$[Al/Fe] vs. $\Delta$[Mg/Fe] (right) diagrams. Each point is color-coded according to their [Fe/H], following the scheme represented in the colorbar.
    Smaller dots indicate $\Delta$[C/Fe] and $\Delta$[N/Fe] based on RGB stars above the bump.
    }
    \label{fig:deltas}
\end{figure*}

Based on the results presented in Figure~\ref{fig:ab1},~\ref{fig:ab2}, and~\ref{fig:ab3}, and in Table~\ref{tab:deltas}, we now discuss the general trend element by element:

{\it{Lithium.}} We consider A(Li) abundances only for stars below the RGB bump since brighter stars experience extra mixing that produces a decrease trend of lithium amount with the magnitude, which may introduce a bias in the relative behavior of the different populations\footnote{The lithium abundance in RGB stars below the bump is not primordial but has been depleted due to the first dredge-up in the RGB phase \citep[e.g.,][]{karakas2014}. Different populations may experience varying degrees of lithium depletion depending on their stellar parameters. However, \citet{schiappacasse-ulloa2022} observed the same A(Li) vs. [Al/Fe] anticorrelation in MS turn-off and RGB stars, indicating that evolutionary lithium depletion has minimal or undetectable impact on the 1P-2P Li trend in the lower RGB phase.}. 
All the lithium measurements in the works considered exploited lines in the Li I doublet at $\sim$6708 \AA.
From a visual inspection of Figure~\ref{fig:ab1}, we notice a slight decrease among 2P stars, with the exceptions of NGC\,6656, NGC\,6809 and NGC\,6838. Overall, $\Delta$A(Li) spans a range from $\sim$0.0 to $\sim$-0.2 dex. In NGC\,6656, the anomalous population has an A(Li) distribution similar to 2P stars. We will dedicate Section~\ref{sec:6} to a more detailed investigation of the lithium distribution among multiple populations.

{\it{Carbon.}} The features involved in the [C/Fe] measurements exploited in this work are the CH and CH G bands in the 3800-4200 \AA range, the CI line at $\sim$6588 \AA, and the CI lines in the H-band.
The extra mixing after the first dredge-up affects, beyond lithium, also carbon and nitrogen amount \citep[e.g.,][]{charbonnel1998, shetrone2019, lee2023},
introducing a decrease and an increase, respectively, dependent on the magnitude. For this reason, results based on [C/Fe] and [N/Fe] abundances of stars below the RGB bump are generally more reliable. $\Delta$[C/Fe] measurements based only below the bump were possible for nine GCs, which tend to exhibit negative values or being consistent with zero (like in NGC\,5272, NGC\,6254, and NGC\,6656), meaning that, as expected, 2P stars have generally lower carbon abundance than 1P stars. On the other hand, the anomalous population of NGC\,6656 is  significantly carbon-richer than its 1P stars.
In this plot, we also display with smaller dots the $\Delta$[C/Fe] based on stars brighter than the RGB bump. Even if these estimates are generally less reliable due to their magnitude dependence, we still point out how their value are qualitatively in agreement with a carbon depletion among 2P stars. Moreover, anomalous stars in NGC\,1851 appear to be, on average, more carbon-poor than 2P stars.

{\it{Nitrogen.}}
[N/Fe] where obtained by exploiting the CN absorption bands (once known C) present in both in the optical and in the H-band ranges.
We represent $\Delta$[N/Fe] following the same approach adopted for carbon, as also this element is affected by extra mixing, indicating with bigger and smaller dots the measurements based on stars below and above the RGB bump. 
Here, all the clusters have positive values (i.e., are N-enhanced), ranging from about 0.3 up to around 0.9 dex. Nitrogen displays one of the largest elemental variations between 1P and 2P stars. Anomalous stars in both NGC\,1851 and NGC\,6656 exhibit larger median [N/Fe] values than their 2P stars.

{\it{Oxygen.}} The considered studies measured the oxygen abundance by exploiting, in the optical region, the forbidden [O I] lines (6300 and 6364 \AA) and the oxygen triplet around 7770 \AA, and the NIR OH absorption bands for APOGEE data.
$\Delta$[O/Fe] is consistent with a depletion among 2P stars in almost all studied clusters, except NGC\,0362, NGC\,4590, and NGC\,6535 (for which is nil). It ranges from $\sim$-0.05 to $\sim$-0.35 dex, with the remarkable exception of NGC\,6402, that exhibit a very large depletion of $\sim$-0.8 dex.
Anomalous stars in NGC\,1851 display smaller [O/Fe] compared to 2P stars, while in NGC\,5139 and NGC\,6656 they span similar ranges.
    
{\it{Sodium.}} All the [Na/Fe] estimates were obtained through the sodium doublets at 5682–5688 \AA and 6154–6160 \AA. Similar to Nitrogen, 2P stars exhibit an overall enhancement in this element, with only NGC\,4590, NGC\,5024, and NGC\,5286 being consistent with a nil $\Delta$[Na/Fe] within uncertainties. NGC\,6402 presents the largest enhancement, valuing $\sim$0.6 dex.

\begin{figure*}
\includegraphics[width=18.7cm, clip]{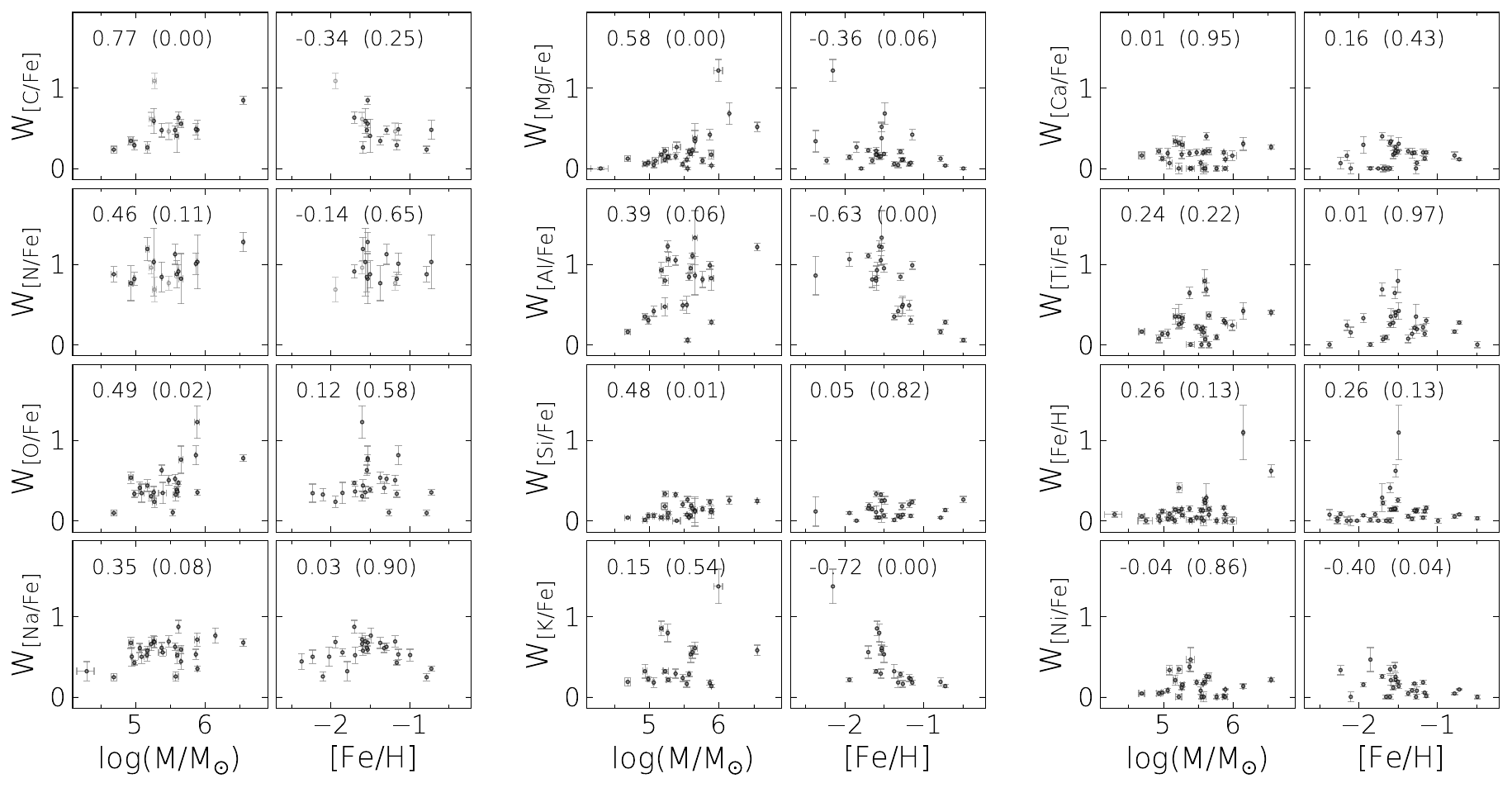}
    \caption{Total spread of [C/Fe], [N/Fe], [O/Fe], [Na/Fe], [Mg/Fe], [Al/Fe], [Si/Fe], [K/Fe], [Ca/Fe], [Ti/Fe], [Fe/H], and [Ni/Fe] with respect to the clusters' mass and average metallicity. In each plot, we report the Spearman correlation coefficient and its significance (p-value) between the round parenthesis.
    Gray points are relative to W$_{\rm [C/Fe]}$ and W$_{\rm [C/Fe]}$ measurements based on stars below the RGB bump.
    }
    \label{fig:width}
\end{figure*}

{\it{Magnesium.}} Magnesium abundances were determined by exploiting the several Mg I lines in the 5300-8800 \AA range (mostly the 5711 \AA line) and in the H band.
As shown in Figure~\ref{fig:ab2}, magnesium differences are very small and in 8 out of 27 studied GCs are consistent with zero within the uncertainty. Only in one case (NGC\,6715) $\Delta$[Mg/Fe] lies above the green line, while in the rest of our sample we detect a modest but significant depletion ($\lesssim$ 0.2 dex) among 2P stars, as previously suggested in the literature \citep[e.g.,][]{carretta2015b, pancino2017}. NGC\,2419 stands out as a notable outlier, with a much larger depletion of $\sim$-0.9 dex.
Anomalous stars generally similar behavior than 2P stars, with the exception of $\omega$Cen, which displays a smaller average magnesium.
Stars in NGC\,7089 host galaxy have average [Mg/Fe] closer to 2P stars, even though the large uncertainty (having only four stars) prevents from excluding a 1P-like abundances.

{\it{Aluminum.}} Aluminum was mostly measured through the doublets in the 6700-8800 \AA range and by exploiting the Al features in H band. [Al/Fe] spans the largest abundance interval, ranging from a small-to-nil difference (like in NGC\,5927 and NGC\,6121), up to a $\sim$1 dex enhancement, as in NGC\,6205 and NGC\,6254. Anomalous stars in NGC\,1851 and NGC\,5139 are the Al-richest stars in their host GCs, while in NGC\,6656 they cover a similar range than the 2P stars.

{\it{Silicon.}} No evident [Si/Fe] trend between populations arise from inspecting the distributions portrayed in Figure~\ref{fig:ab2}. Indeed, $\Delta$[Si/Fe] ranges values around zero, from 0.12 to -0.06 dex.
Notably, a small but significant silicon 2P stars depletion is present in NGC\,0104, NGC\,1851, $\omega$Cen, NGC\,6254, and NGC\,6838, while NGC\,6809 and NGC\,7078 exhibit a modest enhancement.
Anomalous stars in NGC\,1851 and $\omega$Cen show a similar depletion than 2P stars, while in NGC\,6656 they are slightly Si-richer than 1P. NGC\,7089's host galaxy star show average silicon consistent with the cluster stars.
    
{\it{Potassium.}} [K/Fe] measurements come from the K resonance line at $\sim$7699 \AA in the optical regime, and from different K I lines (such as the ones at 15163 and 15168 \AA) in the H band.
No clear overall potassium variation trends arise from Figure~\ref{fig:ab2}. Even though $\Delta$[K/Fe] spans a range from $\sim$-0.15 to $\sim$0.10, its large errors (due to a relatively small sample of stars with these measurements) prevent us from detecting any significant difference. Cases with significant K-enhancement are NGC\,0104, NGC\,6752, and the remarkable NGC\,2419, which exhibits a very high ($\sim$0.9 dex) enhancement among 2P stars. On the other hand, 2P stars of NGC6838 are modestly depleted in potassium,
Finally, anomalous in $\omega$Cen are significantly K-richer than the 1P stars.

{\it{Calcium.}} Overall, no particular trend is observed by the values of $\Delta$[Ca/Fe]. From the values reported in Table~\ref{tab:deltas}, we point out a small Ca-depletion among the 2P stars of NGC\,1904, NGC\,5286, NGC\,6752, NGC\,6809, NGC\,7078, and NGC\,7089, and Ca-enhancement in NGC\,0362, NGC\,1851, NGC\,2419, and NGC\,6402, all significant at a 1-$\sigma$ level.
Anomalous stars in NGC\,1851 are slightly Ca-richer than the 1P, while in other Type II GCs $\Delta$[Ca/Fe] is consistent with zero.

{\it{Titanium.}} The [Ti/Fe] are very similar between the different populations in the studied GCs. We report modest but significant 2P stars enhancement in NGC\,0104, NGC\,0362, NGC\,2808, NGC\,6121, NGC\,6205, NGC\,6254, NGC\,6838, and NGC\,7078, and depletion in NGC\,4833, NGC\,5286, NGC\,6752, and NGC\,7089.

{\it{Iron.}} [Fe/H] is the most measured element, with at least four measurements for each populations in 35 out of 38 GCs of our sample.
2P stars show smaller-to-nil differences compared to 1P stars. As expected, anomalous stars are the most iron-rich stars in several Type II GCs, namely NGC\,1851, NGC\,5139, NGC\,5286, NGC\,6656, NGC\,6715, and NGC\,7089. Stars belonging to the NGC\,7089's host galaxy display a large iron enrichment (about 0.5 dex, in agreement with the estimate from \citealt{yong2014}).
    
{\it{Nickel.}} No significant variation or evident specific trend is detectable for [Ni/Fe] between the different populations in our sample of GCs, which on average share a similar behavior. Notable exceptions are 2P stars of NGC\,1261, NGC\,1904, and NGC\,4590, which are slightly Ni-depleted compared to 1P stars.

{\it{Barium.}} Ba II lines at $\sim$5854 and 6497 \AA have been used for all the measurements of [Ba/Fe] here presented.
The only populations with large variations compared to 1P stars are, as expected, the anomalous stars. Indeed, NGC\,1851 and NGC\,5286 display large enhancement in [Ba/Fe]. Several Type II GCs are not present in Figure~\ref{fig:ab3} because not all their populations pass the criteria adopted in this Section. In particular, NGC\,5139 has less than four 1P stars, even though a large number of anomalous stars has barium measurements. A more detailed analysis of anomalous stars is left for Section~\ref{sec:7}.

The upper panels of Figure~\ref{fig:deltas} summarizes our findings on relative differences between 1P and 2P stellar abundances. For each element, we plot the $\Delta X$ of all the GCs available with violet bullets.
$\Delta$[C/Fe] and $\Delta$[N/Fe] measurements based on stars brighter than the RGB bump are portrayed with smaller dots, arbitrarily shifted from the other measurements for clarity.
For completeness, we include in the upper-right panel the helium difference (in mass fraction) between 2P and 1P stars inferred by \citet{milone2018}. The names of the GCs that are outliers are also reported in the Figure.
In the lower panels, we show the $\Delta$[N/Fe] vs. $\Delta$[C/Fe] (left), $\Delta$[Na/Fe] vs. $\Delta$[O/Fe] (middle), and $\Delta$[Al/Fe] vs. $\Delta$[Mg/Fe] (right) diagrams, with each point color coded according to the cluster's [Fe/H].
The represented trends demonstrate that the more 2P stars are enhanced in nitrogen, sodium, and aluminum, the more the are depleted in carbon, oxygen, and magnesium. This behavior is particularly evident when comparing $\Delta$[Mg/Fe] and $\Delta$[Al/Fe].
In the magnesium-aluminum plot, it is clear the influence of the [Fe/H] on these two quantities, as the metal rich GCs of the sample generally have smaller values -sometimes consistent with zero- in both elements, while the most metal poor tend to display larger 1P-2P differences. This is likely due to the high temperature required to ignite the Mg-Al cycle (above $\sim$70 MK, thus more efficiently reached in the interiors of metal-poor stars), that is prime suspect of producing the Mg-depleted and Al-enriched material that contributed forming 2P stars.
A notable exception is NGC\,7078 which, despite being the most metal poor of the sample, has value consistent with zeros in both quantities.

\section{Spread of elements among globular clusters}
\label{sec:4}

\begin{figure*}
\includegraphics[width=18.0cm]{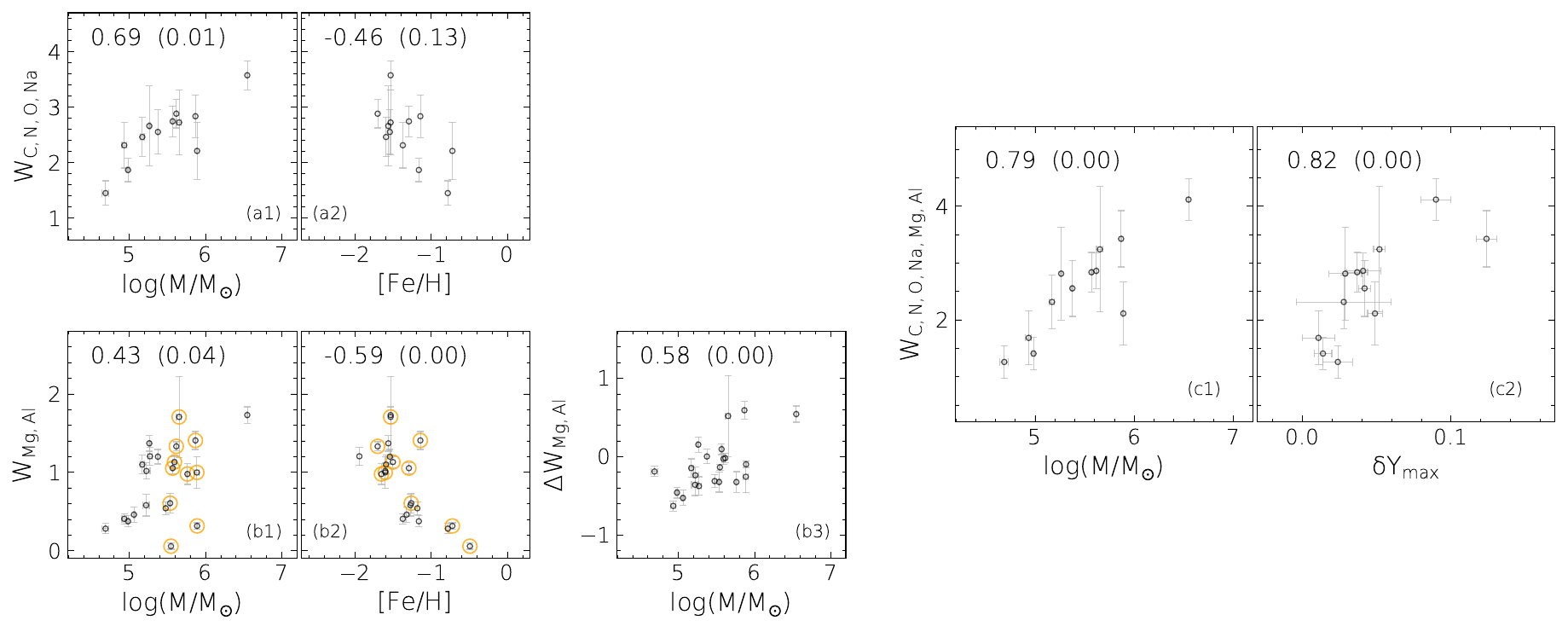}
    \caption{
    {\it{Panels (a1) and (a2):}} $W_{\rm C,N,O,Na}$ vs. $log(M/M_{\rm \odot})$ and [Fe/H], respectively.
    {\it{Panels (b1) and (b2):}} same as panels (a1) and (a2) but for $W_{\rm Mg,Al}$. GCs within the 5.5 $< log(M/M_{\rm \odot}) <$ 5.9 mass interval used to remove the metallicity dependence (see text for details) are encircled in yellow.
    {\it{Panel (b3):}} resulting width after removing [Fe/H] dependence ($\Delta W_{\rm Mg,Al}$) vs. mass.
    {\it{Panels (c1) and (c2):}} total width of light elements involved in the multiple population phenomenon $W_{\rm C,N,O,Na,Mg,Al}$ vs. mass and maximum helium spread measured by \citet{milone2018}, respectively. 
    The Spearman correlation coefficient and its p-value are reported in each plot as in Figure~\ref{fig:width}.
    }
    \label{fig:total}
\end{figure*}

As shown in Figure~\ref{fig:deltas}, the difference in 1P and 2P abundances for a given element can vary significantly from one cluster to another. For instance,  $\Delta$[Al/Fe] ranges from almost zero in NGC\,6838 to nearly 1 dex in NGC\,6254, while oxygen depletion is negligible in NGC\,4590 but around 0.8 dex in NGC\,6402. This indicates that the abundance ratios can exhibit vastly different spreads across various GCs. In this Section, we explore this variability, with particular focus on the elements that show clear patterns of distinct average abundances, specifically carbon, nitrogen, oxygen, sodium, magnesium, and aluminum.

However, $\Delta X$ is not the most effective metric in this context, as it may underestimate the total abundance spread. This is particularly true in cases where 2P stars span a wide range of light-element abundances, as observed in clusters like NGC\,2808, NGC\,6218, and NGC\,6752 \citep[e.g., ][]{carretta2007b, johnson2012, carlos2023}, or display the most extended ChMs \citep{milone2017}. To address this, we introduce a more appropriate measure, the total width $W_{\rm X}$, where X represents any given abundance ratio. $W_{\rm X}$ is calculated as the difference between the 90th and 10th percentiles of the elemental distribution of all the stars in the photometric catalogs with spectroscopically measured chemical abundances, even those without the ChM tagging (therefore not considering their belonging to a given population). This approach allows us to use a larger sample of stars in each cluster compared to the analysis in Section~\ref{sec:3}. Notably, this allows us to also include NGC\,5986, NGC\,6388, and NGC\,6934, which did not pass the strict selection applied in Section~\ref{sec:3}.
We account for observational errors by subtracting in quadrature the contribution of the average uncertainty, measured by applying our procedure to a simulated distribution of points scattered by the average uncertainty provided by the source datasets.
For abundances with multiple determinations, we calculate the final $W_{\rm X}$ by averaging the individual widths, weighted by their errors.
Similarly to what we have done in Section~\ref{sec:3}, to calculate $W_{\rm [C/Fe]}$ and $W_{\rm [N/Fe]}$ we only consider RGB stars fainter than the bump. Measures that were possible only with stars above the bump are indicated, for completeness, in gray, and are not included in the analysis performed in the rest of this Section.

In Figure~\ref{fig:width}, we present the width distributions relative to cluster mass and metallicity\footnote{The mass of GCs is taken from \citet{baumgardt2018}, while the overall [Fe/H] come from \citet[][2010 edition]{harris1996}. We assume [Fe/H] as a tracer of the star overall metals' abundances.}. Lithium and barium are excluded, as their distributions will be addressed separately in Sections~\ref{sec:6} and~\ref{sec:7}, respectively. For each plot, we provide the Spearman rank correlation coefficient along with the p-value (in parentheses), which represents the probability that two intrinsically independent variables could produce the observed distribution.
The spreads of carbon, nitrogen, oxygen, and magnesium  show Spearman coefficients and p-values consistent with a correlation with cluster mass. Additionally, silicon also shows a mild correlation with $log(M/M_{\rm \odot})$.
Our results are in qualitative agreement with the analysis recently published by \citet{lee2025}, who also detected that the dispersion in [C/Fe] and [N/Fe] abundances increase with the cluster mass, while our dataset does not suggest any correlation between the nitrogen spread and [Fe/H], contrary to what derived by Lee and collaborators.

The two most massive GCs, NGC\,5139 and NGC\,6715, are both Type II clusters with significant iron enhancement, clearly reflected in their large $W_{\rm [Fe/H]}$ values. 
Moreover, iron content plays a significant role, as the overall [Fe/H] is found to anticorrelate with $W_{\rm [Al/Fe]}$, $W_{\rm [K/Fe]}$, and (mildly) $W_{\rm [Ni/Fe]}$.
Potassium represents an intriguing case: the $W_{\rm [K/Fe]}$ distribution remains almost flat for [Fe/H]$>-$1.5 dex but increases sharply at lower metallicities. 
This distinct behavior may be due to the activation of the Ar-K chain, which could lead to K-enhanced 2P stars, as suggested by previous studies \citep[e.g., ][]{mucciarelli2015, meszaros2020}.
However, in Figure~\ref{fig:ab2}, we clearly observed this phenomenon only in NGC\,2419, and to a less extent in NGC\,3201, NGC\,2808, and NGC\,6752, likely due to the small sample size of stars with both [K/Fe] measurements and ChM tagging in other GCs, which limits statistical significance. NGC\,7078, one of the most metal-poor clusters in the sample, stands out as an exception with a very small $W_{\rm [K/Fe]}$.

For some abundance ratios, upper limits were reported in certain datasets beyond the observational measurements. Specifically, [O/Fe] obtained through the [O I] forbidden lines at 6300.3 and 6363.8 Å frequently present such cases. As noted in Section~\ref{sec:2}, we did not include upper limits in our width estimates. While this exclusion could, in principle, lead to an underestimation of the elemental spread—particularly if some GCs host stars with extremely low oxygen abundances—we evaluated this potential bias by re-measuring elemental spread with upper limits included. Our findings indicate that, neglecting the upper limits, the differences in the spreads are negligible in most cases, but sometimes they can reach up to $\sim$0.10 mag.
We overall consider this bias negligible, and comparable to the typical observational errors (0.03–0.20 mag for W$_{\rm [O/Fe]}$). 
Nevertheless, although we have verified that all the stellar populations, including the most-extreme ones in the ChMs, have been sampled with  O-abundances measurements, we cannot exclude the presence of stars with very-low (undetectable) O in clusters like NGC\,2808. In such few cases our measured spreads would be overestimated. 

\begin{figure*}
\includegraphics[trim={0cm 0cm 0cm 0cm},clip,width=16cm]{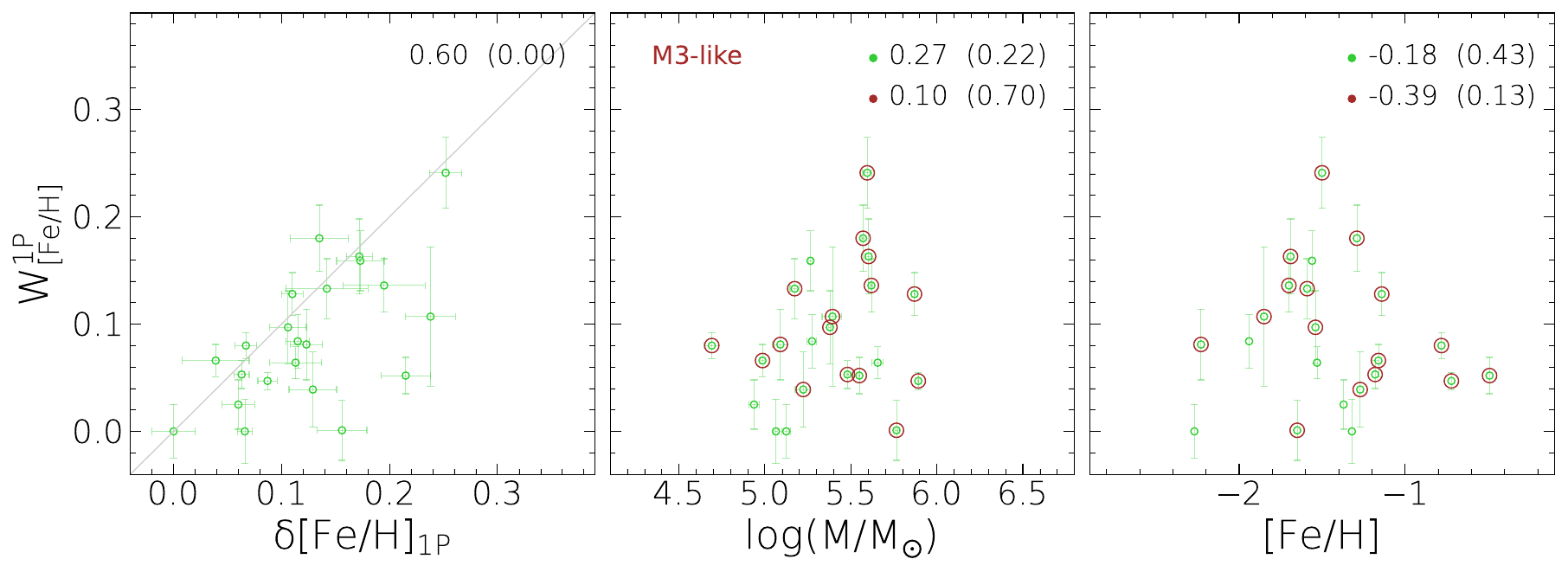}
    \caption{
    {\it{Left panel:}} 1P stars [Fe/H] spread from spectroscopy ($W_{\rm [Fe/H]}^{\rm 1P}$) vs. the predicted iron variations by \citet{legnardi2022} ($\delta [Fe/H]_{\rm 1P}$).
    {\it{Central and right panel:}} $W_{\rm [Fe/H]}^{\rm 1P}$ vs. the logarithm of cluster mass and 
    clusters' overall [Fe/H], respectively. Brown circles highlight M3-like GCs.
    In each panel, we report the Spearman's correlation coefficient for all the clusters and for M3-like only.}
    \label{fig:feh_1g}
\end{figure*}

In Figure~\ref{fig:total}, we focus on the elements that, according to our findings in Section~\ref{sec:3}, display evident and widespread variation patterns between different populations: carbon, nitrogen, oxygen, sodium, magnesium, and aluminum.
We separately analyze the total spread of these elements against cluster mass and metallicity for the two chains of reactions responsible for the observed multiple-population patterns: CNONa ($W_{\rm C,N,O,Na}$, panels (a1) and (a2)) and MgAl ($W_{\rm Mg,Al}$, panels (b1) and (b2)). The CNONa spread shows a clear correlation with cluster mass, whereas the MgAl spread demonstrates a weaker correlation. 
$W_{\rm Mg,Al}$ displays a statistically significant anticorrelation with [Fe/H], strongly supported by the metallicity dependence observed in the $\Delta$[Al/Fe] vs. $\Delta$[Mg/Fe] diagram in Figure~\ref{fig:deltas}. 
To account for metallicity effects on the $W_{\rm Mg,Al}$ vs. mass relation, we applied a method similar to that of \citet{milone2017}. Specifically, we selected a narrow mass range where GCs exhibit significant variation in $W_{\rm Mg,Al}$ values. Within this range, the influence of mass on $W_{\rm Mg,Al}$ is minimal, suggesting that the variability is primarily driven by metallicity differences. We chose the range 5.5 $ < log(M/M_{\rm \odot}) < $ 5.9, where $W_{\rm Mg,Al}$ spans values from approximately 0.3 mag to 1.7 mag. As expected, the selected GCs (highlighted in yellow in panels (b1) and (b2)) cover a wide [Fe/H] range of about 1 dex. 
To remove the metallicity effect, we fit these clusters in the $W_{\rm Mg,Al}$ vs. [Fe/H] plane with a straight line. The difference between $W_{\rm Mg,Al}$ and the best-fit line, denoted as $\Delta W_{\rm Mg,Al}$, was then calculated. As shown in panel (b3), this adjustment strengthens the correlation between $W_{\rm Mg,Al}$ and cluster mass.
Finally, we calculate the overall spread of the six light elements, $W_{\rm C,N,O,Na,Mg,Al}$, by summing the CNONa spread and the metallicity-corrected MgAl spread. As illustrated in panel (c1), this combined quantity exhibits a strong correlation with GC mass.

Helium is another element involved in the multiple population phenomenon. \citet{milone2018} calculated its total spread ($\delta Y_{\rm max}$) based on the ChM positions of 1P and 2P stars, showing that more massive GCs tend to have larger helium variations. In panel (c2) , we connect, for the first time, the spread of light elements derived from spectroscopy with the $\delta Y_{\rm max}$ for a large sample of GCs. We find a strong correlation between these two quantities, demonstrating the link between helium and other light elements in the multiple populations phenomenon.

\section{Metallicity variations among first-population stars}
\label{sec:5}

One of the most intriguing recent discoveries about GCs is the presence of chemical inhomogeneities among 1P stars. Indeed, their distribution along the x-axis of the ChM is significantly broader than what is expected by observational errors only \citep[e.g., ][]{milone2017, dondoglio2021, jang2022}.
Moreover, metallicity inhomogeneities where also detected in NGC\,0104, NGC\,5272, NGC\,6341, and NGC\,6656 \citep{lee2021, lee2022, lee2023, lee2024} through the JWL indices build ground-based photometry \citep[e.g.,][]{lee2017}.
Spectroscopic observations revealed that the most likely driver of this feature is the presence of star-to-star iron variation among them \citep[][]{marino2019a, legnardi2022, lardo2023, marino2023}. 
\citet{legnardi2022}, by assuming that the ChM spread is due to iron variations only, predicted the necessary 1P [Fe/H] spread, $\delta [Fe/H]_{\rm 1P}$, to justify their ChM dispersion.

Our dataset allows us to investigate the presence of iron variations among 1P stars and its relation with the elongation along the ChM's x-axis.
To do that, we calculate the spread of [Fe/H], $W_{\rm [Fe/H]}^{\rm 1P}$, among 1P stars by following the same recipe used to estimate the elemental spread in Section~\ref{sec:4}.
We considered only the clusters with spectroscopic datasets with at least ten 1P stars except NGC\,5024, which, despite fulfilling this criterion, has iron measurements only for stars that span a $\Delta_{\rm F275W, F814W}$ much smaller than the spread of 1P stars over the ChM, thus possibly leading to underestimating its [Fe/H] spread.

The left panel of Figure~\ref{fig:feh_1g} represents $W_{\rm [Fe/H]}^{\rm 1P}$ vs. $\delta [Fe/H]_{\rm 1P}$, proving that overall, our direct measurements are in agreement with the prediction by \citet{legnardi2022}, distributing around the one-to-one relation shown with the gray line. Moreover, the two quantities correlate with each other, as supported by the Spearman's coefficient of 0.60 and a p-value $<$0.01.
Our measurement of metallicity spread spans an interval from $\sim$ 0.00 to 0.25 dex, with NGC\,5272 (M\,3) exhibiting the largest value. 
We find that only three GCs are consistent within 1-$\sigma$ with a nil spread in [Fe/H] (NGC\,0288, NGC\,7089, and NGC\,7099), while for the other 19 we confirm the presence of non-negligible iron inhomogeneities.
The central and right panels represent $W_{\rm [Fe/H]}^{\rm 1P}$ vs. the logarithm of the cluster mass and average metallicity, respectively. The criterion provided by the Spearman's coefficient does not suggest that $W_{\rm [Fe/H]}^{\rm 1P}$ correlates with any of the two quantities. However, we point out that the GCs with the largest spread are mostly present at $log (M/M_{\rm \odot}) >$ 5.5, and that when considering the M\,3-like (classified as in \citealt{milone2014, tailo2020}) GCs only (encircled in brown), $W_{\rm [Fe/H]}^{\rm 1P}$ displays a mild anticorrelation with [Fe/H].
Overall, our results suggest that both cluster mass and average metallicity are in some way linked to the phenomenon (especially among M3-like GCs), qualitatively in agreement with the prediction by \citet{legnardi2022}.

\begin{figure*}
\includegraphics[width=18cm]{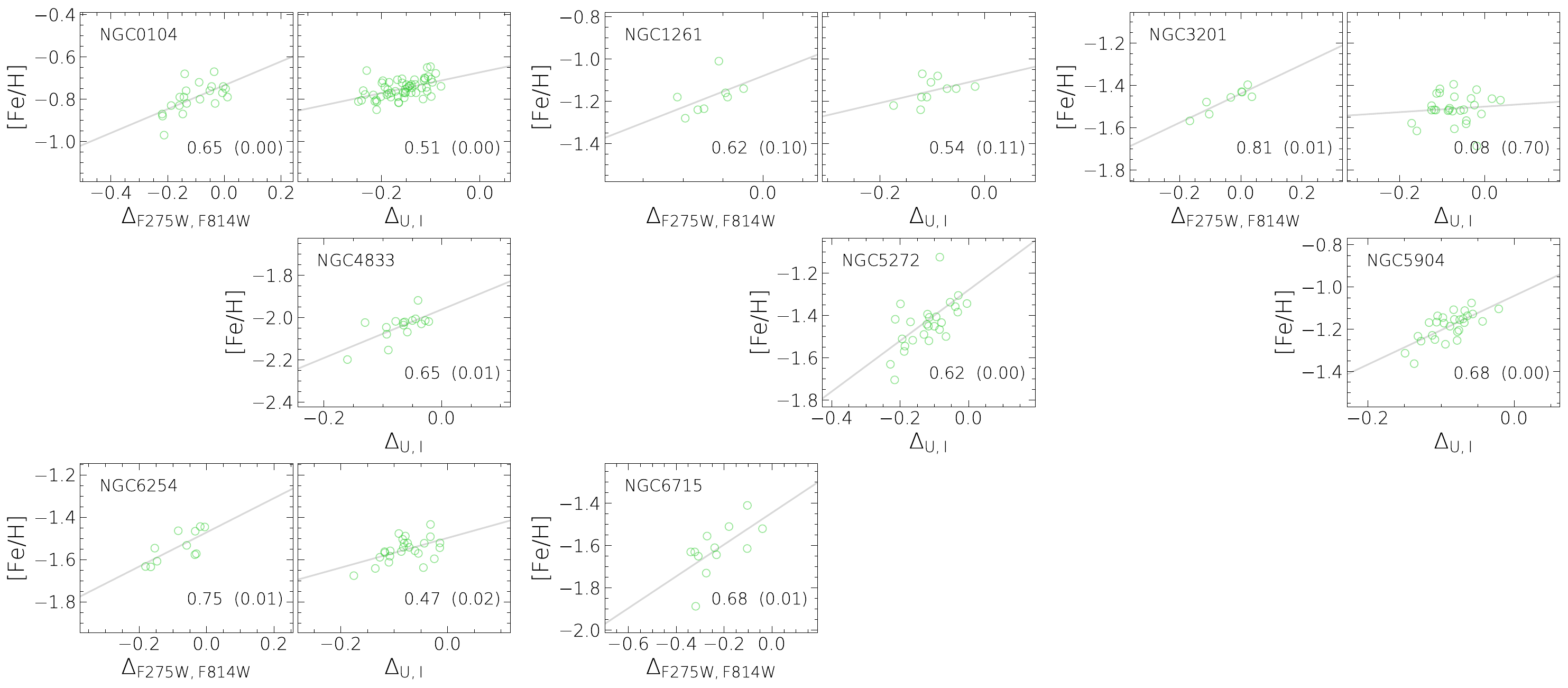}
   \caption{[Fe/H] abundance for 1P stars vs. $\Delta_{\rm F275W, F814W}$ and/or $\Delta_{\rm U,I}$ (depending on the availability of iron spectroscopic measurements) of NGC\,0104, NGC\,1261, NGC\,3201, NGC\,4833, NGC\,5272, NGC\,5904, NGC\,6254, and NGC\,6715.
   Grey lines indicate the best-fit straight line. Correlation coefficients are reported in each plot.
   Comparison of iron abundances with {\it{HST}} and ground-based x axis of the ChMs are aligned along the even and odd columns on the image, respectively.
   }
    \label{fig:1G_spread}
\end{figure*}

We now investigate the relation between the [Fe/H] distribution of 1P stars and their extension along the x-axis of the ChMs. A correlation between the two quantities has been found in NGC\,0104, NGC\,3201, NGC\,6254 \citep{marino2019, marino2019a, marino2023} by comparing spectroscopy abundances to {\it{HST}}-based ChMs. Here, we extend this analysis to a larger sample of GCs, also considering the extension of 1P stars along the ground-based ChM.
For the latter, we do not use the ChM built by \citet{jang2022}, but a ChM that exploits the $U-I$ instead of the $B-I$ color on its x-axis. The former filter combination ensures a wider color baseline than the latter, thus being more sensitive to possible metallicity variations.
To maximize the sample of stars with [Fe/H] measurement and ChM tagging, we combine, for each cluster, the iron measurements from different works. 
To account for eventual systematics between different studies, we consider the average [Fe/H] difference between the common stars of the dataset with the larger amount of stars and every other dataset and shifted each dataset by this quantity. 
Typically, the average shift between different datasets ranges in the $\sim$0.00-0.15 dex interval.
For stars present in more than one dataset, we consider the combined [Fe/H] derived by averaging the different measurements (after the shifting for eventual systematics).

\begin{figure*}
\includegraphics[trim={0cm 0cm 33cm 0cm},clip,height=6cm]{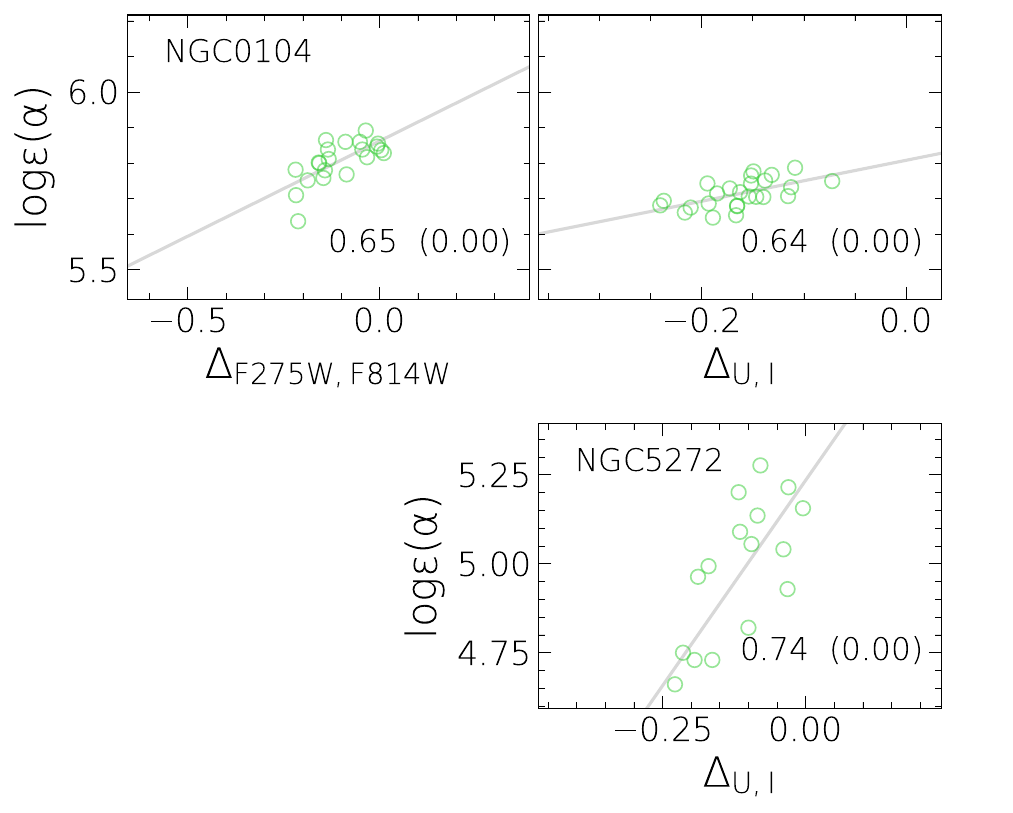}
\includegraphics[trim={23cm 0cm 17cm 0cm},clip,height=6cm]{FIGURES/5.Metallicity/11_EXT_alpha.pdf}
\includegraphics[trim={39cm 0cm 0cm 0cm},clip,height=6cm]{FIGURES/5.Metallicity/11_EXT_alpha.pdf}
   \caption{Same as Figure~\ref{fig:1G_spread} but for $log \epsilon (\alpha)$ vs. $\Delta_{\rm F275W, F814W}$ (only for NGC\,0104) and $\Delta_{\rm U,I}$ for NGC\,0104, NGC\,1261, NGC\,3201, NGC\,5272, NGC\,5904, and NGC\,6254. }
    \label{fig:alpha}
\end{figure*}

Figure~\ref{fig:1G_spread} represents our results for eight GCs in our sample, namely NGC\,0104, NGC\,1261, NGC\,3201, NGC\,4833, NGC\,5272, NGC\,5904, NGC\,6254, and NGC\,6715. We show the [Fe/H] vs. $\Delta_{\rm F275W, F814W}$ and $\Delta_{\rm U, I}$ for each cluster, when available. Comparison with {\it{HST}} and ground-based ChMs are aligned in the odd and even columns, respectively. NGC\,0104, NGC\,1261, and NGC\,6254 show that [Fe/H] increases by moving through redder color in both ChMs, with Spearman coefficients consistent with mild-to-strong correlations. NGC\,3201, the other GC with a significant number of 1P stars with iron abundances in both ChMs, share this behavior only in the innermost area covered by the {\it{HST}} observations while it does not display any significant trend with $\Delta_{\rm U, I}$. NGC\,4833, NGC\,5272, and NGC\,5904 have spectroscopic information only for the stars in the ground-based ChM, while for NGC\,6715, this analysis is possible only by exploiting {\it{HST}} photometry. These latter four GCs are consistent as well to have their $\Delta_{\rm F275W, F814W}$ and $\Delta_{\rm U, I}$ 1P extensions linked to their iron abundance.

\begin{figure*}
\includegraphics[trim={0cm 0cm 0cm 0cm},clip,width=18cm]{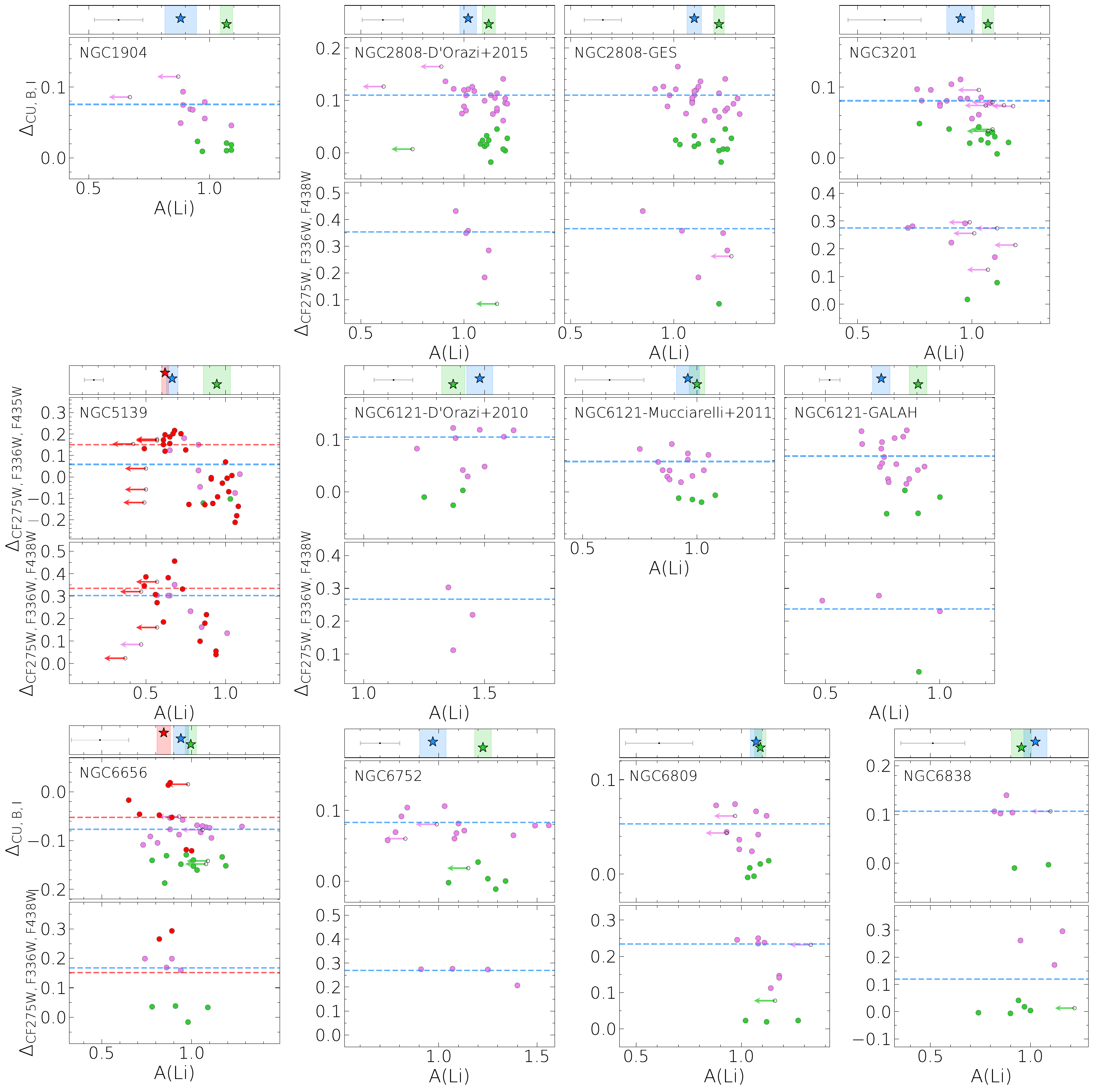}
    \caption{$y$-axis of the ChM vs. A(Li). Measurements and upper limit (arrows) are color-coded according to their population. For each GC, the bottom panel refers to stars in the innermost areas covered by HST observations, while the upper ChM regards stars outside the central field (covered by ground-based photometry in all GCs except $\omega$Cen). Horizontal blue and red dashed lines indicate the 70-th percentile level of the 1P+2P and 1P+anomalous stars distribution, respectively.
    Above the ChMs, we display with starred green and blue symbols the median 1P and 2P$_{\rm ext}$ A(Li), shifted arbitrarily on the y axis for showing purposes.}
    \label{fig:lithium_chm}
\end{figure*}

Moreover, \citet{marino2023} found that in NGC\,0104, the absolute abundances of other elements that they analyzed follow the same trend of [Fe/H]. In particular, they show that the average $\alpha$ elements abundance of increases with $\Delta_{\rm F275W, F814W}$ among 1P stars (see their Figure~6). Driven by this result, we investigate the same quantity in six of the GCs displayed in Figure~\ref{fig:1G_spread}, which are the ones with available Si, Ca, and Ti estimates, that we use as a proxy of $\alpha -$elements value. Results are illustrated in Figure~\ref{fig:alpha}, where we represent the average absolute $\alpha$ elements abundance $log \epsilon (\alpha)$\footnote{For a given X element, we remind that $log \epsilon (X) = [X/Fe] - [Fe/H] + log_{\rm \odot} \epsilon (X)$, with $log_{\rm \odot} \epsilon (X)$ being its absolute solar abundance.}.
NGC\,0104 is the only one with measurements of 1P stars in both {\it{HST}} and ground-based ChM, for which we exploited the \citet{marino2023} and APOGEE datasets, respectively. The APOGEE dataset also allows us to measure this quantity in the case of NGC\,3201, NGC\,5272, NGC\,5904, and NGC\,6254, while for NGC\,1261 we used the GES abundances.
In NGC\,0104, our data are consistent with a correlation between the $\alpha$ elements and both the $\Delta_{\rm F275W, F814W}$ and the $\Delta_{\rm U,I}$. NGC\,5272 and NGC\,6254 also display a correlation with the x-axis of the ChM, while for NGC\,1261 and NGC\,5904 (the other two clusters with a sign of [Fe/H] correlation detected in Figure~\ref{fig:1G_spread}), no clear correlation is suggested by the Spearman's coefficient. Finally, NGC\,3201 does not display an evident trend between these two quantities, in agreement with the lack of iron correlation with  $\Delta_{\rm U,I}$.

The results presented in this Section all indicate that the presence of 1P iron variations is widespread among GCs, and that these overall metallicity differences are the principal cause of the color spread observed in the ChM.

\section{Lithium among multiple populations}
\label{sec:6}

Lithium plays a critical role in constraining the origins of multiple stellar populations, particularly in identifying the nature of the polluters responsible for the processed material from which 2P stars formed. As noted by \citet{dantona2019}, polluters that generate gas processed through hydrogen burning in their convective envelopes are thought to completely destroy lithium. In that case, any lithium observed in 2P stars must result from dilution between pristine gas and the lithium-free material that formed these stars. In contrast, if intermediate-mass asymptotic giant branch (AGB) stars are the source of the processed gas, they could account for lithium in 2P stars without dilution, as they produce it through the Cameron-Fowler mechanism \citep{cameron1971}.

As discussed in Section~\ref{sec:3}, 2P stars generally show slightly lower lithium levels than 1P stars. Here, we focus on 2P stars with more extreme chemical compositions (2Pext), which are believed to have undergone minimal, if any, dilution \citep[e.g.,][]{dantona2016}. For this analysis, we use eight GCs from the first row of Figure~\ref{fig:ab1}, as well as $\omega$Cen, which has only two 1P stars with A(Li) measurement and therefore does not meet the criteria for detailed analysis in Section~\ref{sec:3}.

The first step is to identify a sample of 2Pext stars, which typically display the highest y-axis coordinates in the ChM. Numerous studies have shown that stars positioned higher on the y-axis of the ChM have chemical compositions that deviate more from 1P stars \citep[e.g.,][]{marino2019, carlos2023, carretta2024}. Figure~\ref{fig:lithium_chm} illustrates the relationship between the y-axis position in the ChM and A(Li) for each GC. We present both the {\it{HST}} and ground-based ChMs, as well as the two {\it{HST}} datasets for $\omega$Cen, for all GCs except NGC\,1904 (which only has ground-based data) and NGC\,6121 for the \citet{mucciarelli2011} catalog (which lacks lithium measurements in the {\it{HST}}-based diagram).
We define 2P${\rm ext}$ stars as those above the 70th percentile of the y-axis distribution, indicated by the blue dashed horizontal line. This calculation includes upper limits (excluded in Section~\ref{sec:3}), represented by arrows colored according to their population. For NGC\,2808 and NGC\,6121, where lithium measurements are available from multiple datasets, both are shown separately in the Figure. For each GC, we display the median A(Li) values for 1P and 2P${\rm ext}$ stars using green and blue starred symbols, respectively, with bands around the points representing uncertainties. 2P${\rm ext}$ stars exhibit significantly lower A(Li) than 1P stars in all GCs except NGC 6121, NGC 6809, and NGC 6838.
In the cases of $\omega$Cen and NGC\,6656, we apply a similar procedure to compare 1P stars with anomalous stars. Here, the 70th percentile line is shown in red, and the average A(Li) for extreme anomalous stars (Aext) is represented by a red starred symbol. In both clusters, anomalous stars are more lithium-depleted than the 2P${\rm ext}$ stars.

\begin{figure}
\includegraphics[trim={0cm 0cm 0cm 0cm},clip,height=3.8cm]{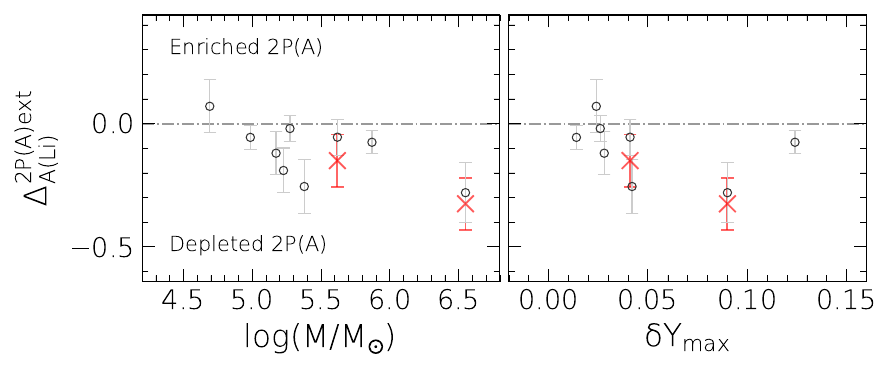}
    \caption{
    Difference between the most extreme 2P and the 1P stars lithium abundance (open black dots) vs. the cluster mass and maximum helium spread (right an left panels, respectively). Red crosses indicate the same quantity but measured for anomalous stars (see text for details).
    }
    \label{fig:lithium_delta}
\end{figure}

As in Section~\ref{sec:3.2}, we define $\Delta^{\rm 2Pext}$A(Li) (or $\Delta^{\rm Aext}$A(Li)) as the difference between the median A(Li) values of 2P${\rm ext}$ (or Aext) and 1P stars. For NGC\,2808 and NGC\,6121, we average the results from the separate datasets to obtain the final $\Delta^{\rm 2Pext}$A(Li). Figure~\ref{fig:lithium_delta} shows the trend of $\Delta^{\rm 2Pext}$A(Li) and $\Delta^{\rm Aext}$A(Li), represented by open black points and red crosses, respectively, with GC mass and the $\delta$Y$_{\rm max}$ measured by \citealt{milone2018}, in the left and right panels.

Overall, $\Delta^{\rm 2Pext}$A(Li) is wider (i.e., more negative) both at larger GC mass and helium spread. This trend may seem qualitatively consistent with the dilution hypothesis, as less massive and less He-enriched GCs are expected to host more diluted 2P stars, which would be more lithium-rich. However, we highlight two key points:
(i) in clusters where $\Delta^{\rm 2Pext}$A(Li) is close to zero, 2P stars would almost entirely consist of pristine material. This is difficult to reconcile with the large sodium and nitrogen differences observed in these same clusters.
(ii) Even in GCs with the largest $\Delta_{\rm A(Li)}^{\rm 2Pext}$, their 2Pext stars exhibit a wide lithium spread partially overlapping the 1P A(Li) distribution. NGC 2808, the most helium-enriched cluster in our sample, is a notable example of this, indicating that even the least diluted 2P stars formed from gas with a non-negligible amount of lithium.

These observational constraints qualitatively show that dilution alone cannot justify the amount of lithium observed in 2P stars, but an additional production mechanism in the polluter is necessary. AGB stars have been traditionally seen as the most likely candidates. However, it remains to be explored whether other candidates (e.g., massive interacting binaries) may also produce lithium at some point in their evolution).

\section{Type II globular clusters}
\label{sec:7}

In this Section, we examine the chemical composition of anomalous stars in detail, investigating their internal abundance variations and their distinctions from canonical stars (e.g., the bulk of 1P and 2P). Since our focus is on anomalous stars, we use data from the dataset with the highest incidence of these stars (i.e., the ratio of anomalous stars to the total spectroscopically analyzed stars) rather than averaging multiple datasets if others are available in the literature.

\begin{figure*}
\includegraphics[trim={0cm 0cm 0cm 0cm},clip,height=13.4cm]{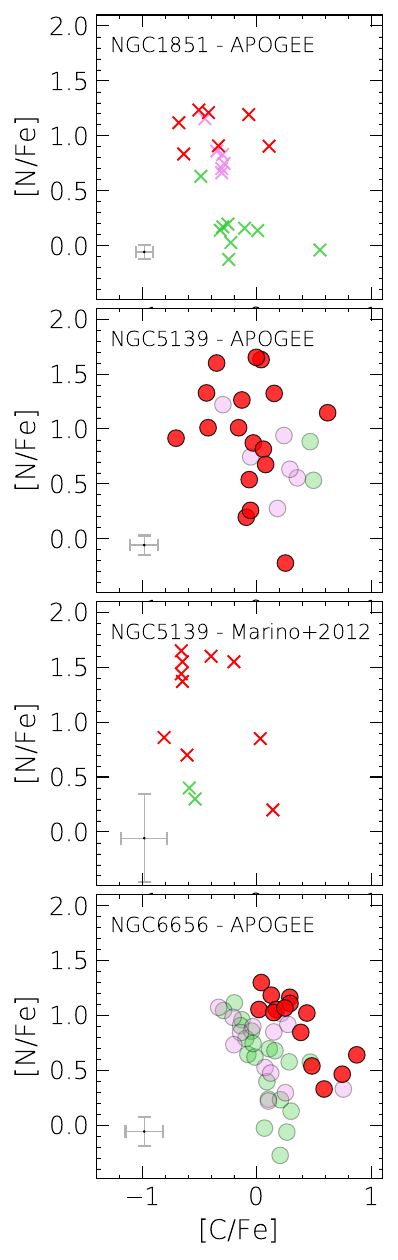}
\includegraphics[trim={0cm 0cm 0cm 0cm},clip,height=13.4cm]{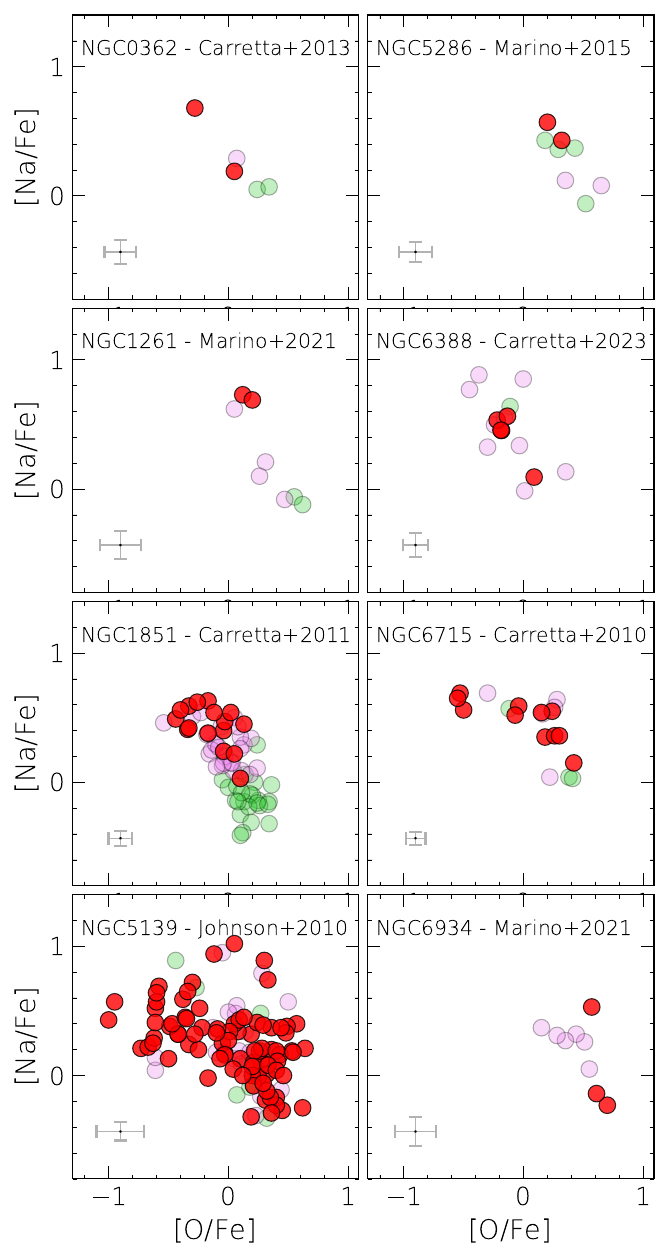}
\includegraphics[trim={0cm 0cm 0cm 0cm},clip,height=13.4cm]{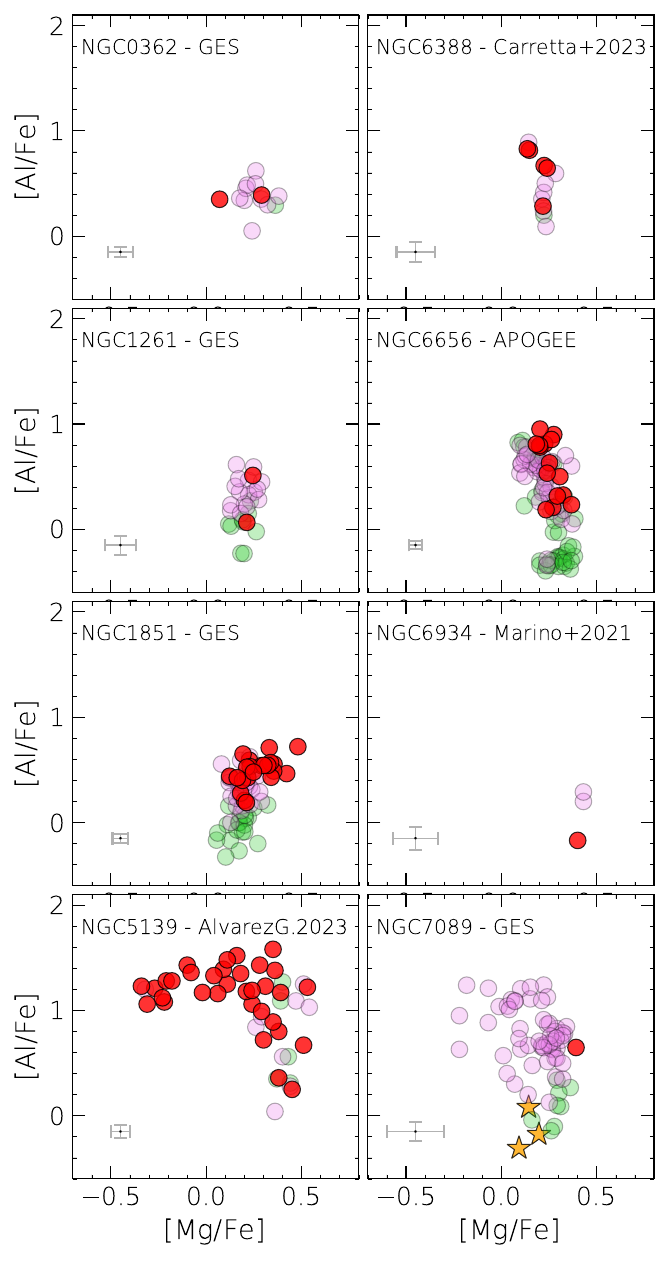}
    \caption{{\it{From top to bottom rows:}} [C/Fe] vs. [N/Fe], [O/Fe] vs. [Na/Fe], and [Mg/Fe] vs. [Al/Fe] diagrams of the anomalous GCs in our sample with available spectroscopy. 1P and 2P stars are represented by the dim green and violet dots, while anomalous stars are represented in red. The iron-rich population of NGC\,7089 (see text) is indicated with orange starred symbols.}
    \label{fig:ans_light}
\end{figure*}

Figure~\ref{fig:ans_light} presents the [C/Fe] vs. [N/Fe], [O/Fe] vs. [Na/Fe], and [Mg/Fe] vs. [Al/Fe] diagrams in the left, central, and right rows, respectively, for anomalous stars (red filled dots). For comparison, 1P and 2P stars are also displayed. Each plot specifies the dataset considered.

[C/Fe] and [N/Fe] measurements for stars brighter than the RGB bump are marked with crosses. For $\omega$Cen, we include two separate datasets (APOGEE and \citealt{marino2012}), as they capture different segments of the complex anomalous star population: while most APOGEE measurements involve stars near the 1P and 2P in the ChM, stars from \citealt{marino2012} populate regions with higher average $\Delta_{\rm F275W, F814W}$ values. Although the \citet{marino2012} measurements cover stars above the RGB bump, the stars shown in the figure span a relatively small $\sim$2 mag range, minimizing magnitude effects. In the [O/Fe] vs. [Na/Fe] and [Mg/Fe] vs. [Al/Fe] plots, employing two datasets is unnecessary, as the available measurements sufficiently span the entire anomalous population in the ChM.
Notably, in four GCs (NGC\,1851, $\omega$Cen, NGC\,6656, and NGC\,6715), anomalous stars exhibit chemical inhomogeneities similar to 1P-2P patterns but with more chemically extreme average values. In NGC\,6656, anomalous stars display higher average carbon than the canonical population, consistent with \citet{marino2011}. A similar trend is observed in $\omega$Cen using the \citet{marino2012} dataset, where some anomalous stars show high [C/Fe] values compared to the rest.
In the remaining six GCs, limited sample sizes hinder the identification of clear trends. However, anomalous stars in NGC\,0362, NGC\,1261, and NGC\,5286 exhibit light-element abundances comparable to or more extreme than 2P stars, aligning with previous spectroscopic results \citep{carretta2013b, marino2015, marino2021}. Conversely, anomalous stars in NGC\,6934 are depleted in Al and Na relative to 2P stars, though one star shows Na enhancement. For NGC\,7089, we include the metal-rich population identified by \citet{yong2014}, likely associated with a remnant host galaxy \citep[see][]{milone2015a}, which exhibits Mg and Al abundances similar to 1P stars; its sole anomalous star overlaps the 2P region. In NGC\,6388, the anomalous population shows no significant differences from canonical stars in the [Na/Fe] vs. [O/Fe] and [Mg/Fe] vs. [Al/Fe] diagrams.

\begin{figure*}
\includegraphics[trim={0cm 0cm 0cm 0cm},clip,height=8.3cm]{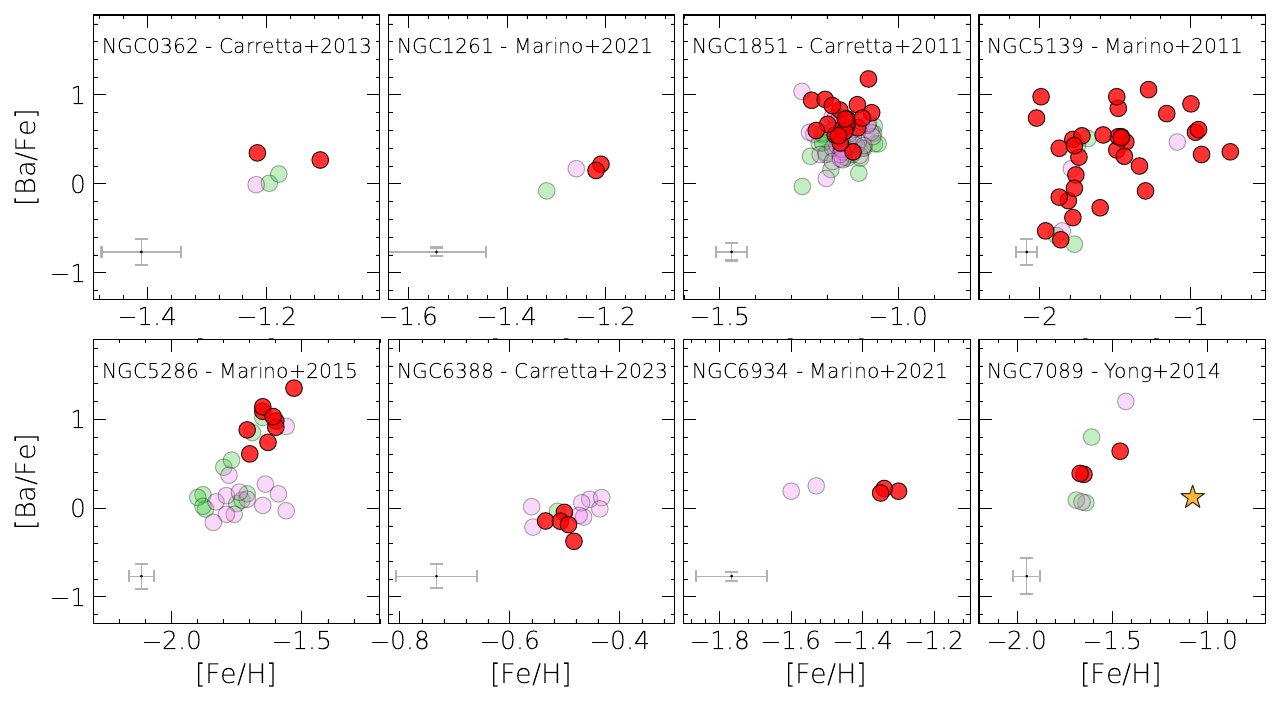}
    \caption{Same as Figure~\ref{fig:ans_bafe} but for the [Ba/Fe] vs. [Fe/H] relation.}
    \label{fig:ans_bafe}
\end{figure*}

Anomalous stars are linked with enrichment in iron, s-process elements, and total C$+$N$+$O compared to the rest of the clusters' stars \citep[e.g.,][]{yong2008, marino2009, carretta2011, marino2015, mckenzie2022, dondoglio2023}. Accordingly, we explore their [Fe/H], [Ba/Fe] (as an s-process element proxy), and [C+N+O/Fe] distributions. Figure~\ref{fig:ans_bafe} illustrates the [Ba/Fe] vs. [Fe/H] diagram for eight Type II GCs. In NGC\,1851 and NGC\,5286, anomalous stars cluster around higher average barium levels and, for NGC\,5286, higher iron levels, while still overlapping somewhat with canonical stars, confirming the spectroscopic investigations \citet{carretta2011, marino2015}. In $\omega$Cen, this effect is more pronounced, with anomalous stars showing a broad distribution (spanning over 1 dex) in both elements, far exceeding observational error. Iron-rich stars also tend to show greater barium abundance.
In NGC\,0362, NGC\,1261, NGC\,6934, and NGC\,7089, the limited sample size results in low statistics. In NGC\,0362, anomalous stars span a similar [Fe/H] range as canonical stars but show slightly higher barium levels, consistent with the enhancement identified by \citet{carretta2013a}. In NGC\,1261 and NGC\,6934, anomalous stars are iron-richer than canonical stars but have comparable [Ba/Fe], aligning with \citet{marino2021}. For NGC\,7089, anomalous stars exhibit similar barium and iron levels, consistent with \citet{yong2014}, who identified a group of Ba- and Fe-rich stars. Discrepancies likely arise from the small sample size in our ChM tagging. The single star associated with the remnant of NGC\,7089’s host galaxy, as estimated by \citet{yong2014}, is approximately 0.5 dex [Fe/H]-richer than the other stellar populations.

\begin{figure*}
\includegraphics[trim={0cm 0cm 0cm 0cm},clip,height=4.8cm]{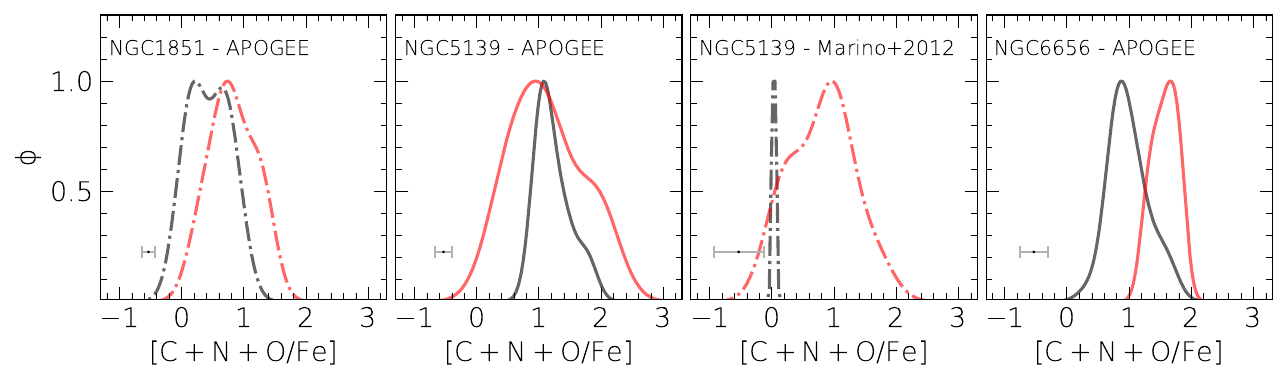}
    \caption{Kernel density distribution of the [C+N+O/Fe] values of NGC\,0362, NGC\,1851, $\omega$Cen, and NGC\,6656. The distribution of canonical and anomalous stars are colored in black and red, respectively. Median spectroscopic errors are represented in gray. For NGC\,0362, having only one anomalous star, we highlight its position with a vertical red line. The narrow canonical stars distribution in NGC\,5139 (\citealt{marino2012} dataset) is due to having only two stars with ChM tagging.}
    \label{fig:ans_cno}
\end{figure*}

Figure~\ref{fig:ans_cno} displays the kernel density distributions of [C$+$N$+$O/Fe] for canonical (black) and anomalous (red) stars in NGC\,1851, $\omega$Cen, and NGC\,6656. These distributions are derived using APOGEE data, except for $\omega$Cen, where abundances from \citet{marino2012} are also used. Dash-dotted lines indicate distributions for stars above the RGB bump.
In NGC\,1851, the distributions partially overlap, but anomalous stars generally show higher [C$+$N$+$O/Fe] values, suggesting CNO enhancement \citep[as derived by][]{yong2015, dondoglio2023}. APOGEE data reveal significant variability among anomalous populations within this massive GC, with some stars showing CNO enhancement relative to canonical stars and others displaying similar levels. Anomalous stars in the dataset from \citet{marino2012} shows a similar spread, but centered at higher values, highlighting greater enhancement. This difference likely reflects the inclusion of anomalous stars farther from the 1P and 2P bulk along the ChM in the Marino and collaborators' dataset, which tend to exhibit wider chemical inhomogeneities.
In NGC\,6656, anomalous stars are clearly CNO-enhanced, with their [C$+$N$+$O/Fe] distribution peaking about 1 dex higher than that of canonical stars.

Next, we assess the relationship between iron and barium enrichment in anomalous stars and the mass of their host cluster. Following the approach in Section~\ref{sec:4}, we measure the spreads $W_{\rm [Fe/H]}$ and $W_{\rm [Ba/Fe]}$. Results are shown as black dots, while clusters with fewer than ten stars (excluded following Section~\ref{sec:4} guidelines) are in cyan. Iron spread tends to increase in more massive clusters, with the highest values seen in $\omega$Cen. Among lower-mass clusters (log(M/M${\odot}$) < 5.6), three of four clusters have minimal iron spread. Notable exceptions include NGC\,6934, with significant $W{\rm [Fe/H]}$ despite being the least massive, and NGC\,6388, the third most massive Type II cluster, yet showing negligible iron spread \citep[see also][]{marino2021, carretta2022}.
Similarly, $W_{\rm [Ba/Fe]}$ correlates with mass, though NGC\,6388 remains an outlier with nil barium spread. 

Our analysis demonstrates that, relative to canonical stars, anomalous stars are generally enhanced in iron, barium (and likely other s-process elements), and total C$+$N$+$O, even tough with few exceptions.
This indicates that these stars formed from a gas polluted with the product of different polluters than the ones that concurred in originating the 2P stars, such as core-collapse supernovae of massive stars and intermediate-mass AGB stars \citep[see the discussion in][and reference therein]{dondoglio2023}.

\section{Conclusions}
\label{sec:9}

In this work, we identified 1P and 2P stars in 38 Galactic GCs using ChMs derived by \citet{milone2017} and \citet{jang2022} from {\it HST} and ground-based photometry, along with newly introduced ChMs for $\omega$Cen, NGC\,5286, and NGC\,6656. We also scanned $\sim$20 years of spectroscopic analyses to infer chemical abundances for photometrically tagged stars. This sample includes $\sim$3,200 stars across 38 GCs, focusing on the abundances of 14 elements (Li, C, N, O, Na, Mg, Al, Si, K, Ca, Ti, Fe, Ni, and Ba).

The observed carbon and oxygen depletion and the enrichment in nitrogen and sodium among 2P stars are consistent with a contribution from material that underwent CNO-cycling and p-capture processes at high temperatures (like the Ne-Na cycle) in the formation of this population.
Magnesium is modestly depleted in 2P stars (except for NGC\,2419), while aluminum displays the largest observed enhancements. The most metal-rich GCs in our sample tend to display small-to-nil [Mg/Fe] and [Al/Fe] variations, while their 1P-2P differences increase by decreasing metallicity. These findings strongly suggest that the forming material of 2P stars also underwent the Mg-Al cycle, which requires temperatures above $\sim $70 MK and hence operates more effectively at lower metallicities, where it depletes magnesium and produces aluminum. 
Similarly, significant potassium spread occurs below [Fe/H]$\sim -$1.5, increasing further at lower metallicities (Figure\ref{fig:width}), likely due to activation of the Ar-K chain.
Two GCs emerged as notable outliers in our analysis. In NGC\,6402, 2P stars show $\sim$0.9 dex [O/Fe] depletion relative to 1P stars, a much larger variation than in any other GC in our sample.
The peculiarity of NGC\,6402 is that, even if the [O/Fe] depletion among the most extreme 2P is roughly similar to NGC\,2808, all its 2P stars have very low oxygen amount, while NGC\,2808 host different 2P populations with intermediate composition.
Although other light elements (Na, Mg, and Al) exhibit some of the highest 1P-2P variations, they align more closely with the general GC trend. \citet{dantona2022} proposed that this cluster may be Type II, potentially explaining its unusual behavior, though the limited number of photometrically tagged stars with abundance measurements complicates drawing firm conclusions. In NGC\,2419, 2P stars show extreme magnesium depletion and potassium enhancement. \citet{mucciarelli2011} suggest that the cluster’s large distance from the Galactic center, and its limited interaction with the Milky Way, could facilitate retention of processed material. \citet{ventura2012} further propose that extreme Mg and K variations are more likely at low metallicities, though additional measurements, especially for O, Na, and Al (available only for a few stars without ChM tagging, \citealt{cohen2011, cohen2012}), are needed to confirm this model. These factors, combined with NGC\,2419's large mass (typically associated with greater light-element inhomogeneity), may explain its unique chemical profile.

The 1P-2P chemical patterns observed in this work align well with previous studies on multiple populations, based on both photometry and spectroscopy \citep[e.g.,][]{carretta2009, pancino2017, milone2018, marino2019, meszaros2020}. Our sample constitutes the largest census to date of chemical differences among multiple populations in GCs, in terms of clusters, stars, and elemental diversity.

We then find that the mass of the host cluster plays an important role in the phenomenon of multiple populations. Indeed, the combined spread of the six light elements that show widespread 1P-2P differences correlates with this quantity, a trend that becomes clearer when metallicity dependence is removed. Furthermore, the spread in these elements is strongly linked to helium variation, with GCs showing greater He enhancement also displaying broader spreads in elements linked to multiple populations. These findings provide spectroscopic evidence supporting earlier photometric results \citep[e.g.,][]{milone2017, milone2018, lagioia2024}.

For 22 GCs in our sample, we measured the [Fe/H] spread among 1P stars, $W^{\rm 1P}_{\rm [Fe/H]}$, finding that only three clusters are consistent with negligible spread at a 1-$\sigma$-level, suggesting a constant [Fe/H]. This is the largest sample study conducted on this phenomenon based on direct [Fe/H] measurements from spectroscopy.
Our results generally agree with photometric predictions by \citet{legnardi2022}, supporting the role of GC mass and metallicity in influencing this behavior. In eight GCs, we observed a correlation between iron abundance in 1P stars and their position on the ChM x-axis, where redder stars tend to be more metal-rich. In three clusters, a similar correlation with $\alpha$ elements amount supports the idea that chemical differences among 1P stars reflect broader metallicity differences beyond iron alone \citep[see][]{marino2023}. Notably, this is the first time these patterns have been detected in ChMs constructed with UBVI filters.

We then investigated the lithium variation in nine GCs, focusing on the composition of their most chemically extreme 2P stars, 2Pext. Even among these stars, the lithium depletion reaches a maximum of $\sim$0.3 dex in the clusters with the largest mass and He-richer 2Pext. Conversely, it is consistent with zero in the least massive one, typically with the smallest helium variations. These results suggest that lithium depletion is not a common feature of all clusters hosting multiple populations but rather is experienced only by the most chemically extreme 2P, which typically reside in the most massive GCs with large helium variations, and even in this latter category some of the 2Pext stars exhibit a lithium amount that overlaps the 1P distribution.
Our extended investigation on literature A(Li) abundances combined with the population tagging provided by the ChM gives a robust confirmation of the conclusions of previous studies \citep[e. g.,][]{dorazi2014, dorazi2015, aguilera2022, schiappacasse-ulloa2022}.
Our findings support a scenario where the lithium present among 2P stars would not entirely come from dilution with pristine material, thus requiring a process that produces this element in the polluted gas. AGB stars constitute an appealing candidate, as they can provide extra lithium through the \citet{cameron1971} process \citep[see the discussion in ][]{dantona2019}, but it remains to be seen to which extend other polluters, such as e.g., massive interacting binaries could also produce some lithium.

Finally, we analyzed the chemical composition of anomalous stars in ten Type II GCs, finding that light-element inhomogeneities are common in several clusters (e.g., NGC\,1851, $\omega$Cen, NGC\,6656, and NGC\,6715), though not universally. In most clusters, anomalous stars have chemical compositions closer to 2P than 1P stars,
with the possible exception of NGC\,6934.
Anomalous stars typically show larger barium (a proxy for s-process elements) and iron abundances, though these enhancements do not always co-occur. Some clusters display enrichment in only one of these elements, with NGC\,6388 showing neither. Remarkably, $\omega$Cen exhibits pronounced enhancement in both, with its most Fe-rich anomalous stars also being the Ba-richest.

In NGC\,1851, $\omega$Cen, and NGC\,6656, we combined photometric tagging with carbon, nitrogen, and oxygen abundances, finding that anomalous stars (in the case of $\omega$Cen only a fraction of them) generally show higher [C$+$N$+$O/Fe] than canonical stars. The presence of iron, barium (and likely other s-process elements), and C$+$N$+$O enhancement indicates that different polluters contributed to the formation of anomalous stars compared to 2P stars. As discussed in Section~\ref{sec:7}, candidate sources include core-collapse supernovae and $\sim$3.5–4 $M_{\odot}$ AGB stars. We find that more massive GCs tend to exhibit larger abundance spreads in iron and barium, with notable exceptions like NGC\,6934 and NGC\,6388. The latter, despite displaying Type II GC photometric features (e.g., a secondary CMD sequence with a fainter SGB and redder RGB), shows no iron or barium enhancement and spread. This may suggest a formation history distinct from other anomalous clusters \citep[see discussion in][]{carretta2022, carretta2023}, as NGC\,6388 is the only bulge GC in our Type II sample. Further exploration of potential C$+$N$+$O enrichment in its anomalous stars, not feasible with the current data, would provide crucial insights into the cluster’s nature.

\begin{figure}
\includegraphics[trim={0cm 0cm 0cm 0cm},clip,height=8.5cm]{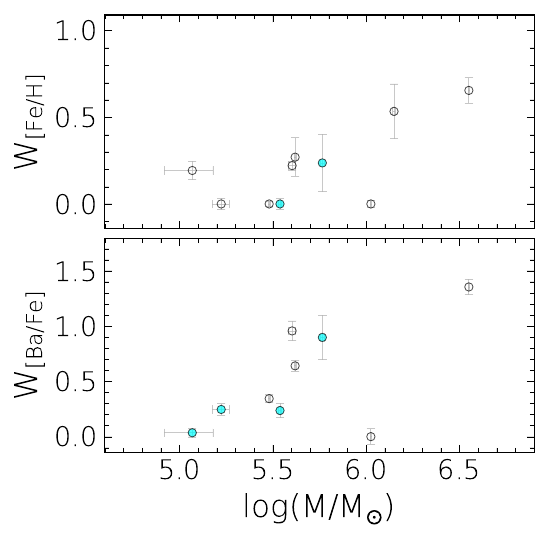}
    \caption{[Fe/H], [Ba/Fe], and [C+N+O/Fe] width of Type II GCs in our sample vs. their mass (in upper, central, and lower panel, respectively).
    Dots colored in cyan indicate the clusters for which the width has been measured with less than ten stars (see text for details).
    }
    \label{fig:ans_mass}
\end{figure}

\begin{acknowledgements}
This work has been funded by the European Union – NextGenerationEU RRF M4C2 1.1 (PRIN 2022 2022MMEB9W: “Understanding the formation of globular clusters with their multiple stellar generations”, CUP C53D23001200006).
T. Ziliotto acknowledges funding from the European Union’s Horizon 2020 research and innovation programme under the Marie Skłodowska-Curie Grant Agreement No. 101034319 and from the European Union – NextGenerationEU".
SJ acknowledges support from the NRF of Korea (2022R1A2C3002992, 2022R1A6A1A03053472).

\end{acknowledgements}

\bibliographystyle{aa}
\bibliography{aanda}

\begin{appendix}
\section{details on our dataset} \label{sec:ap1}

In this appendix, we provide additional details about our dataset. Figure~\ref{fig:chms_hst} presents all previously published ChMs based on {\it{HST}} photometry used in this work. The ChMs of NGC\,6402 and NGC\,2419 are sourced from \citet{dantona2022} and \citet{zennaro2019}, respectively, while the remaining data come from \citet{milone2017}. Similarly, Figure~\ref{fig:chms_grd} shows the ChMs published by \citet{jang2022}, which utilize UBVI photometry from ground-based facilities. In both figures, stars with available spectroscopy measurements from the literature are marked with green, violet, and red dots, corresponding to their belonging to the 1P, 2P, and anomalous populations, respectively. The stars associated with the remnant of the host galaxy of NGC\,7089 are indicated with orange starred symbols. While in the HST-based ChM their separation from the bulk of anomalous stars is evident, in the ChM from Jang and collaborators, based on a smaller color scale on the ChM x-axis, the separation is not clear-cut. To separate the two populations, we rely on their large [Fe/H] differences \citep{yong2014}, tagging with orange starred symbols the stars with [Fe/H] larger than -1.2 dex.

At the same signal-to-noise ratio, the separation between different populations on the ChM becomes less clear-cut when considering stars in the most metal-poor GCs. In particular, this effect is more prominent in the UBVI photometric system compared to others that have filters that more efficiently cover the crucial absorption bands to disentangle multiple populations, like the HST WFC3 or the JWL indices \citep[e. g.,][]{lee2017}. For example, as pointed out by \citet{lee2024}, the separation between populations in ChMs made with the UBVI bands tends to become less clear-cut, as the CN absorption band becomes weaker when considering M92 ([Fe/H]$\sim -$2.3). For GCs in this regime for which we exploited ground-based photometry, we verified that the $\Delta$X derived relying on the UBVI ChM is in agreement within uncertainties with $\Delta$X obtained through the more robust HST tagging, thus showing how any possible limitation among metal-poor GCs does not significantly affect our results.

Table~\ref{tab:data} summarizes the complete dataset used in this work. For each GC, we list the source studies and the number of stars with ChM tagging (from both {\it{HST}} and ground-based photometry) for each of the 14 species included in our analysis.

Figures~\ref{fig:ax_ab1},~\ref{fig:ax_ab2}, and~\ref{fig:ax_ab3}, illustrate the median abundances of 1P, 2P, and (when present) anomalous stars for each dataset covered in Section~\ref{sec:3.2}, selecting only those with at least four stars per population. In these plots, open circles represent the median abundance value for each population on the y-axis, while the x-axis positions are arbitrarily adjusted to display 1P, 2P, and anomalous stars from left to right. Grey lines connect median values from the same study. The APOGEE dataset measurements, which covers the largest number of common stars, element abundances, and clusters, are highlighted with open squares. Each point is accompanied by a colored rectangular band representing the interquartile range of the abundance distribution, according to the corresponding population.

With over 50 spectroscopic studies considered, systematic errors and disagreements in element abundances across studies are unavoidable. Indeed, we are dealing with different works, each using instruments with varying resolutions, distinct spectral features, and different data reduction techniques, thus making it inevitable to encounter systematic biases in abundance ratios measured across separate studies. To mitigate these effects, our analysis relies exclusively on relative metrics, such as $\Delta$ and W. Since systematic errors typically shift the entire abundance scale uniformly, our approach effectively mitigates their impact.
We emphasize that absolute abundance measurements are never combined in our analysis. Instead, we calculate $\Delta$ and W independently for each dataset and then average these quantities across the datasets. The only exception occurs in Section 5, where we combine [Fe/H] measurements from multiple datasets to compare them to the x-axis position on the ChM (Figure~\ref{fig:1G_spread}). However, since [Fe/H] is one of the most extensively measured elements in spectroscopy, we were able to leverage a large sample of stars measured in multiple studies whenever combining datasets. This allowed us to robustly estimate and account for any potential systematics.

\section{Spatial biases} \label{sec:ap2}

Several GCs exhibit radial gradients in their population fractions, with the proportion of 1P stars increasing towards the outskirts and 2P stars dominating in the central regions \citep[e.g.,][]{dondoglio2021, leitinger2023, metha2025}. Spectroscopic observations, however, often avoid the densely crowded central areas of GCs, focusing instead on stars located further out. This observational bias could lead to undersampling of 2P stars in the inner regions. If the fraction of 2P stars decreases towards the core, their overall contribution could be underestimated when measuring the Width, potentially introducing a spatial bias.

To evaluate whether this effect significantly impacts our results, we conducted simulations modeling a realistic sample of 50 stars, a typical number available for Width measurements. Two population distributions were considered: one with a 1P fraction of 0.3 and the other with 0.7, based on estimates reported in the literature. These simulations incorporated three scenarios with differences $\Delta$ between the median 1P and 2P abundance values of 0.2, 0.7, and 1.2 dex, which represent typical 1P-2P light-element variations observed in Section~\ref{sec:3}. Figure~\ref{fig:simulation} illustrates the simulated distributions for each $\Delta$, shown in the left, central, and right panels, respectively. To mimic real observations, we applied the Width calculation method described in Section~\ref{sec:4} and assumed typical spectroscopic measurement errors of 0.15 dex.

In the $\Delta =$ 0.2 case, we measured Width values of 0.24$\pm$0.06 and 0.20$\pm$0.07 for 1P fractions of 0.3 and 0.7, respectively. For $\Delta =$ 0.7, the results were 0.84$\pm$0.06 and 0.85$\pm$0.06, while for $\Delta =$ 1.2, the values were 1.26$\pm$0.07 and 1.34$\pm$0.09. Across all three scenarios, the differences introduced by radial population gradients were negligible. These findings demonstrate that, under realistic observational conditions, spatial biases due to population gradients do not significantly affect our Width measurements. This ensures that our results remain robust and unbiased by the radial distribution of 1P and 2P stars.

\renewcommand{\thefigure}{A.\arabic{figure}}
\renewcommand{\thetable}{A.\arabic{table}}
\setcounter{figure}{0}
\setcounter{table}{0}

\begin{figure*}
\includegraphics[trim={0cm 4.3cm 0cm 0cm},clip,width=15.7cm]{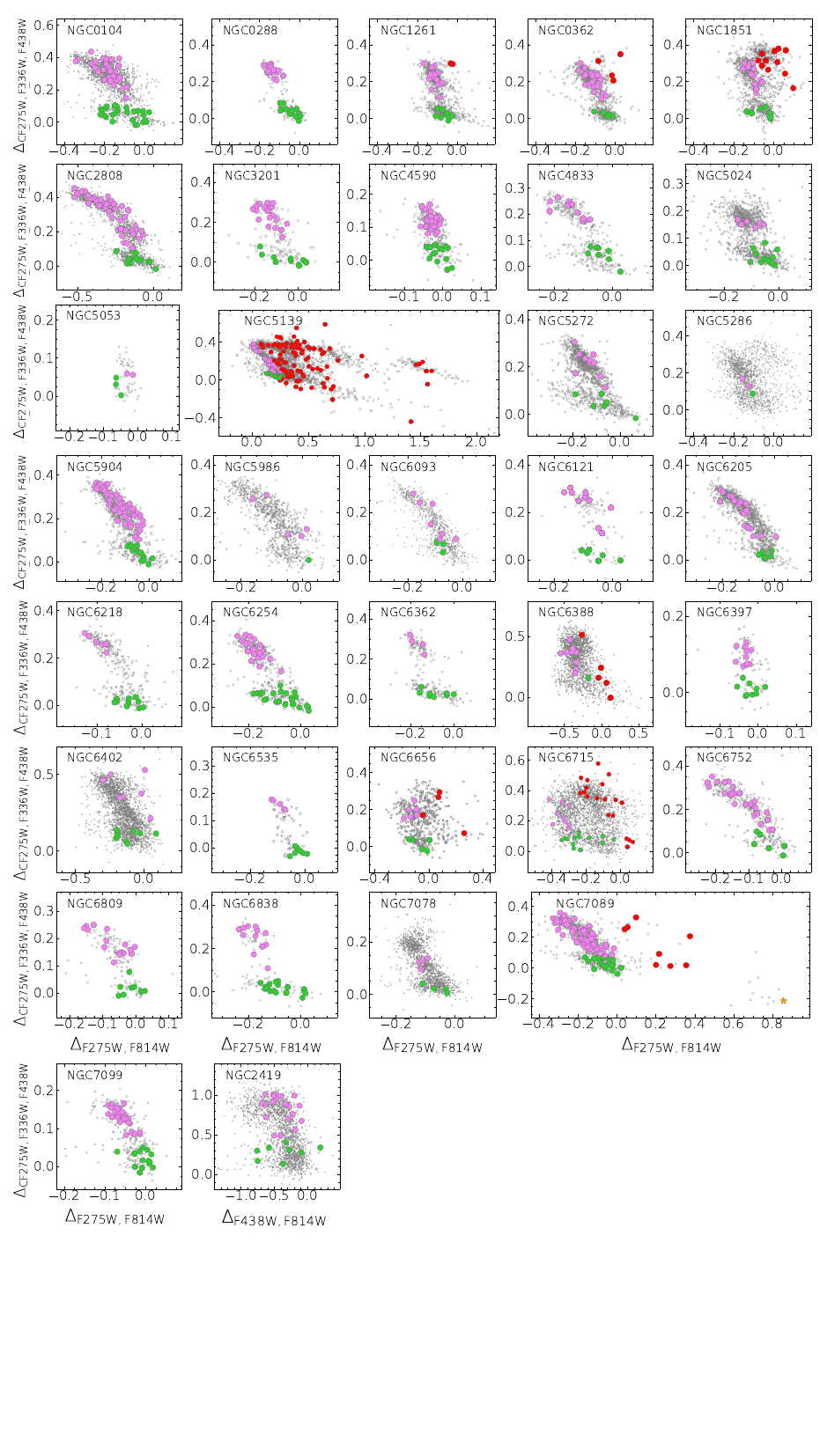}
    \caption{$\Delta_{\rm C F275W,F336W,F438W}$ vs. $\Delta_{\rm F275W,F814W}$ ChM from \citet{milone2017}. The ChM of NGC\,6402 comes from \citet{dantona2022}, while the $\Delta_{\rm C F275W,F336W,F438W}$ vs. $\Delta_{\rm F438W,F814W}$ was published by \citet{zennaro2019}. Stars with spectroscopy abundances are highlighted with the color-code introduced in Figure~\ref{fig:example}.}
    \label{fig:chms_hst}
\end{figure*}

\begin{figure*}
\includegraphics[trim={0cm 12cm 0cm 0cm},clip,width=16.8cm]{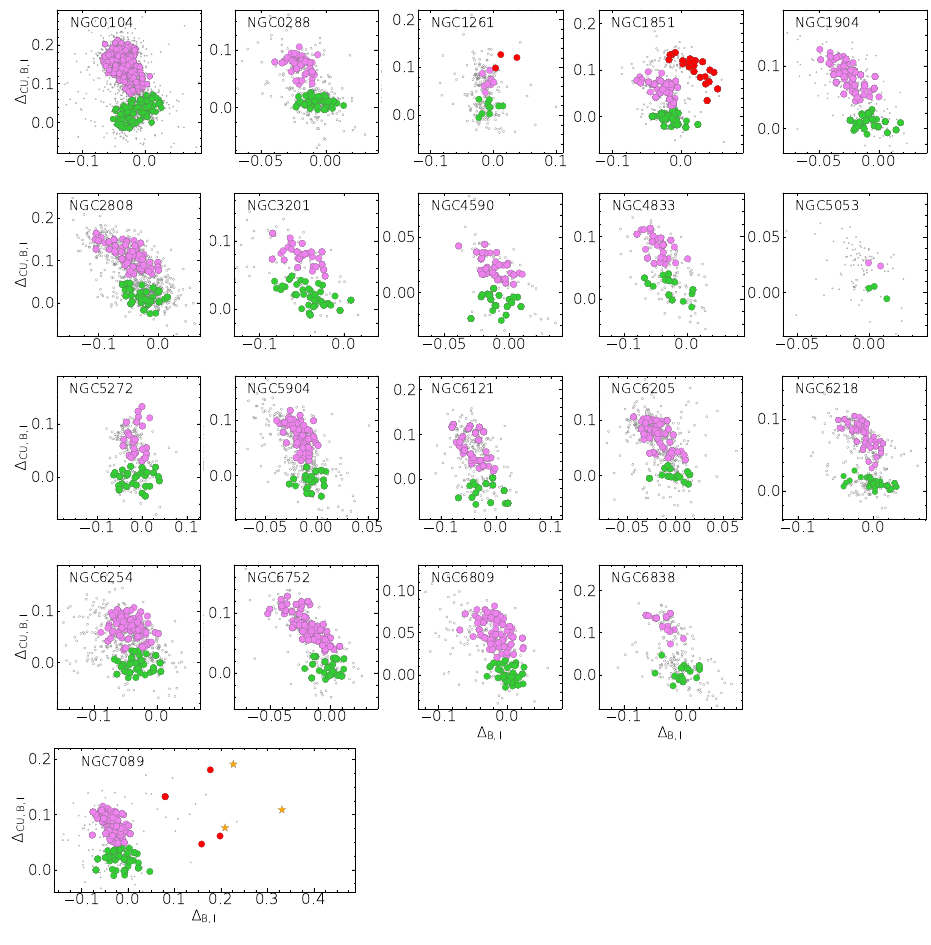}
    \caption{Same as Figure~\ref{fig:chms_hst} but for the  $\Delta_{\rm C U,B,I}$ vs. $\Delta_{\rm B,I}$ ChM from \citet{jang2022} of the GCs in our sample.}
    \label{fig:chms_grd}
\end{figure*}

\begin{figure*}
\includegraphics[width=18cm, clip, trim={0cm 0.15cm 0cm 0.25cm}]{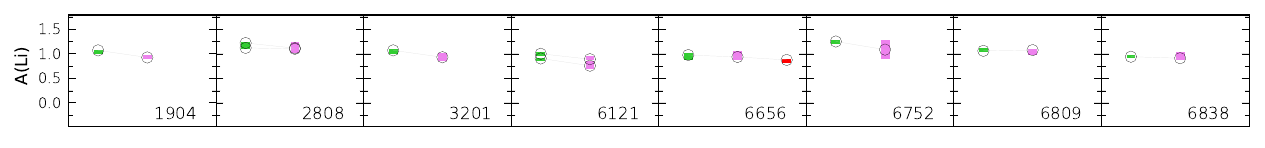}
\includegraphics[width=18cm, clip, trim={0cm 0.15cm 0cm 0.15cm}]{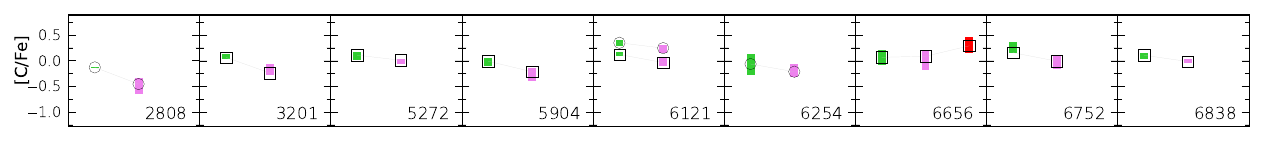}
\includegraphics[width=18cm, clip, trim={0cm 0.15cm 0cm 0.15cm}]{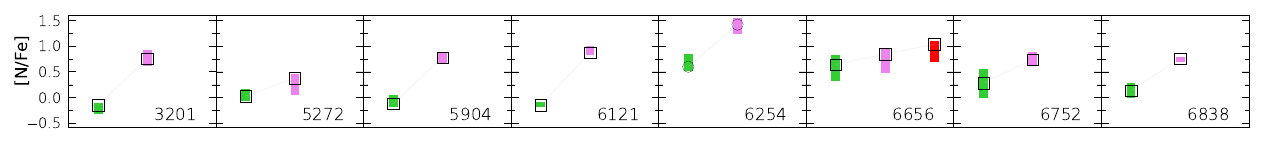}
\includegraphics[width=18cm, clip, trim={0cm 0.15cm 0cm 0.15cm}]{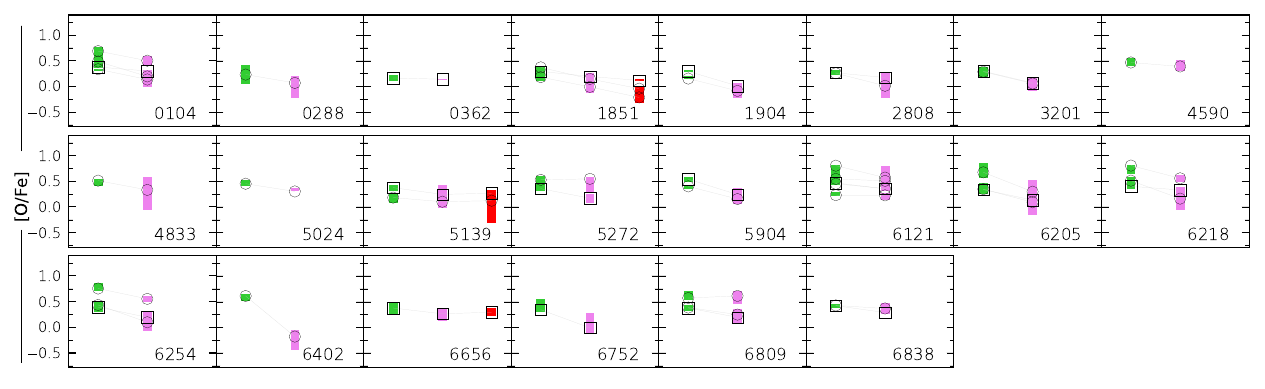}
\includegraphics[width=18cm, clip, trim={0cm 0.15cm 0cm 0.15cm}]{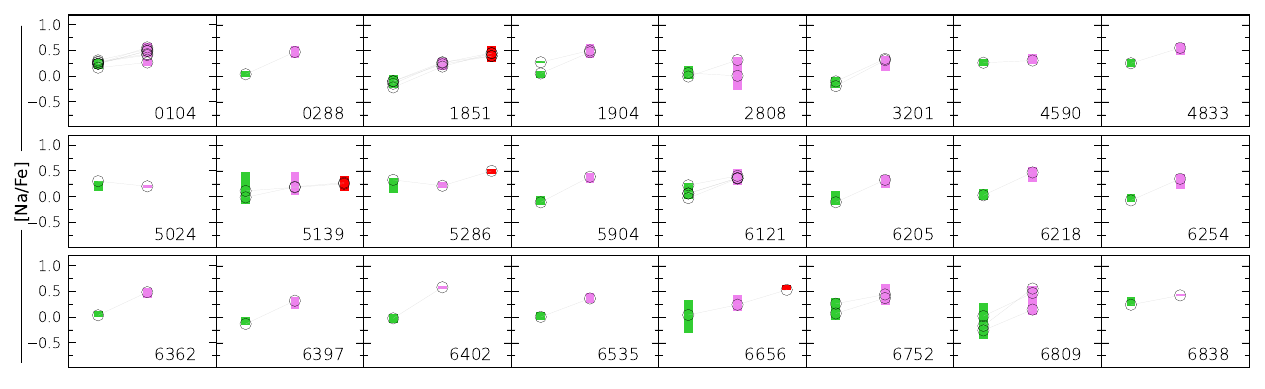}
\includegraphics[width=18cm, clip, trim={0cm 0.35cm 0cm 0.15cm}]{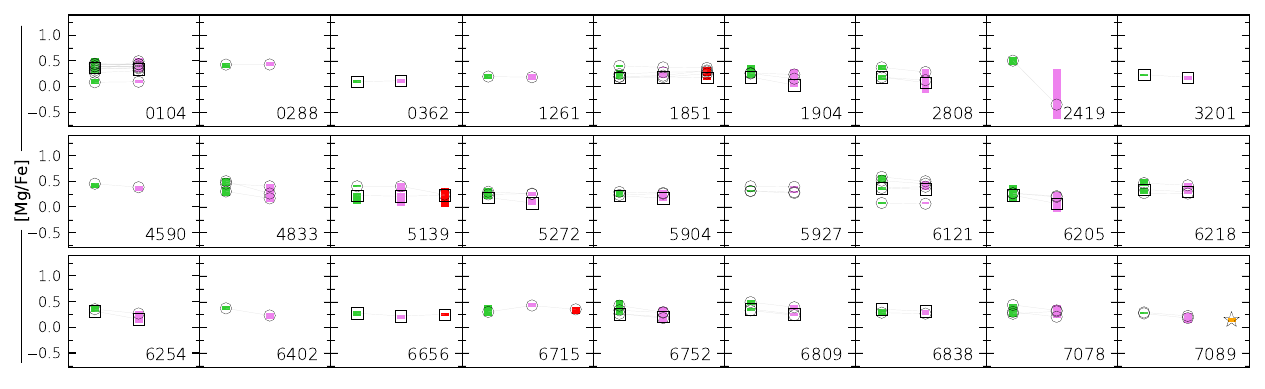}
    \caption{Median abundances (open dots) of lithium, carbon, nitrogen, oxygen, sodium, and amgnesium for the different populations identified in our sample of GCs with at least four stars with both photometric tagging and spectroscopic abundances. The open squares highlight APOGEE measurements. The vertical bands represent the interquartile range, and are color coded following the prescriptions introduced in Figure~\ref{fig:example}. Median values from the same spectroscopic dataset are connected by gray lines.}
    \label{fig:ax_ab1}
\end{figure*}

\begin{figure*}
\includegraphics[width=18cm, clip, trim={0cm 0.15cm 0cm 0.15cm}]{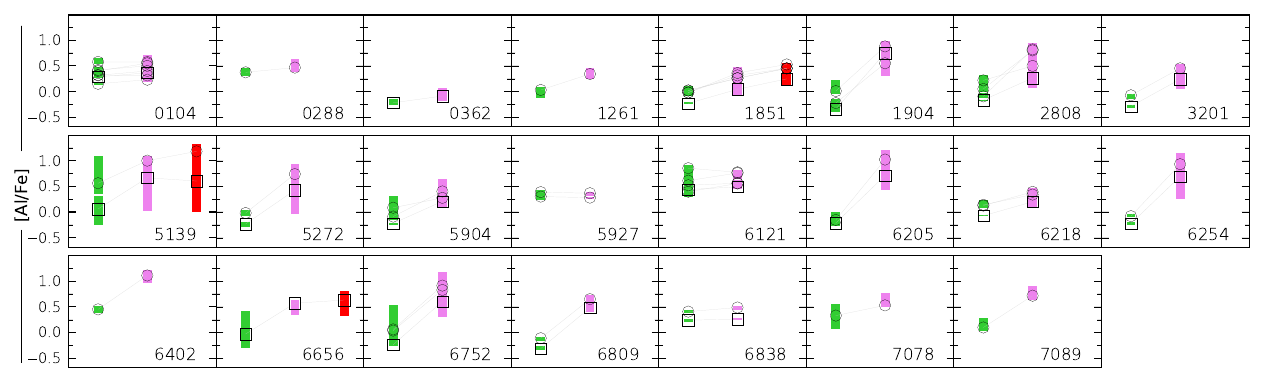}
\includegraphics[width=18cm, clip, trim={0cm 0.15cm 0cm 0.15cm}]{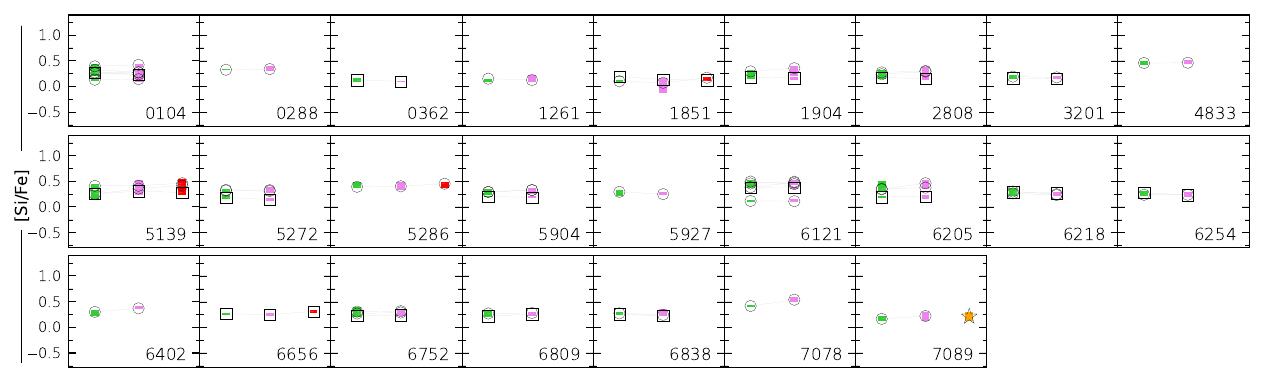}
\includegraphics[width=18cm, clip, trim={0cm 0.15cm 0cm 0.15cm}]{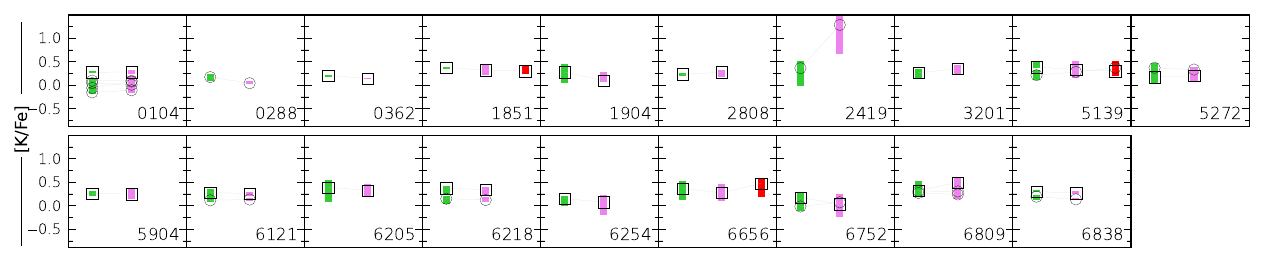}
\includegraphics[width=18cm, clip, trim={0cm 0.35cm 0cm 0.15cm}]{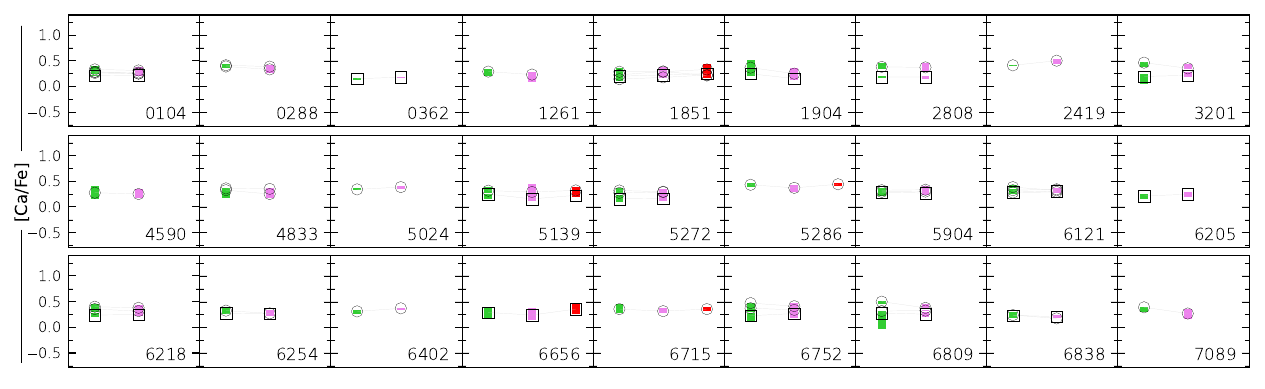}
    \caption{Same as Figure~\ref{fig:ax_ab1} but for aluminum, silicon, potassium, and calcium.}
    \label{fig:ax_ab2}
\end{figure*}

\begin{figure*}
\includegraphics[width=18cm, clip, trim={0cm 0.2cm 0cm 0.2cm}]{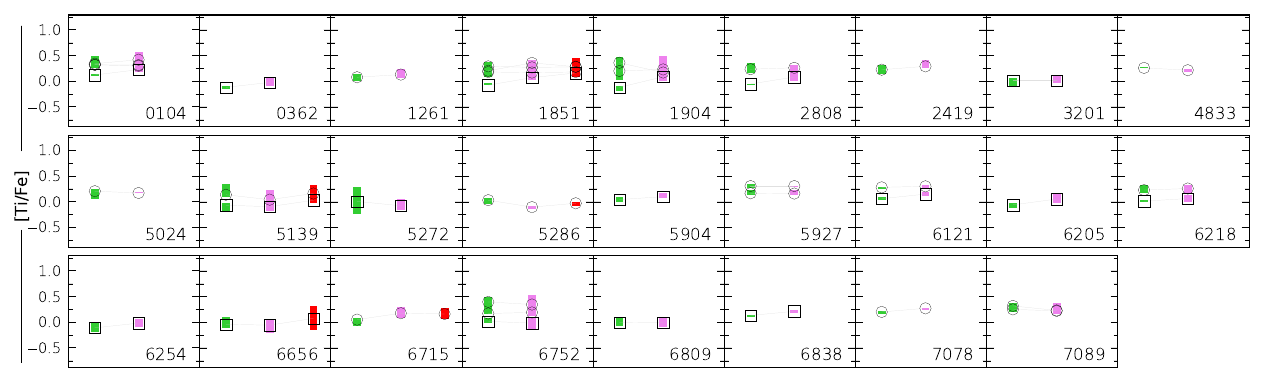}
\includegraphics[width=18cm, clip, trim={0cm 0.2cm 0cm 0.2cm}]{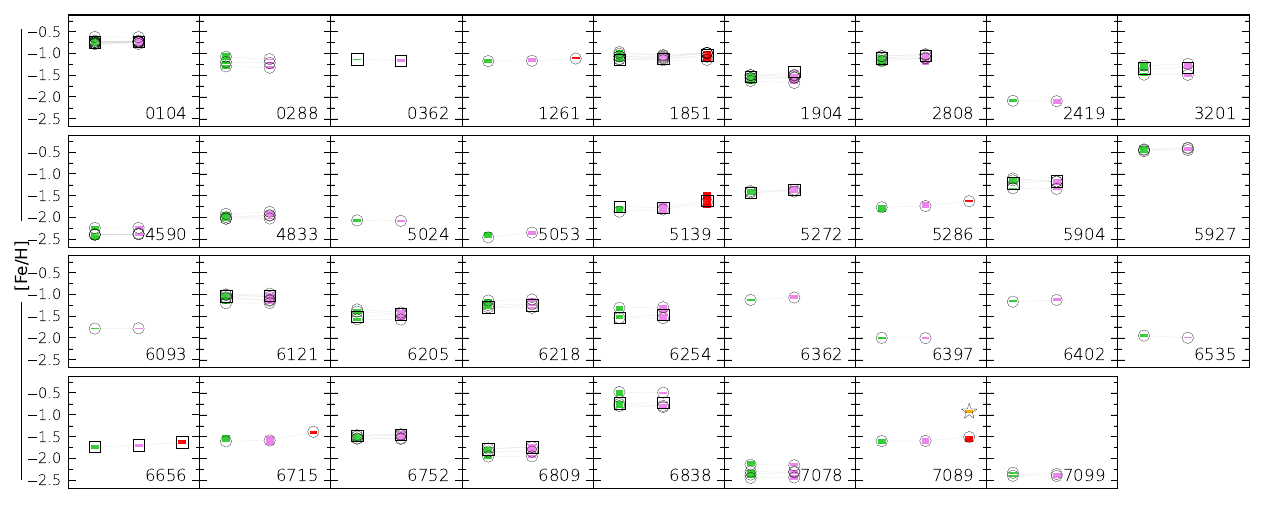}
\includegraphics[width=18cm, clip, trim={0cm 0.2cm 0cm 0.2cm}]{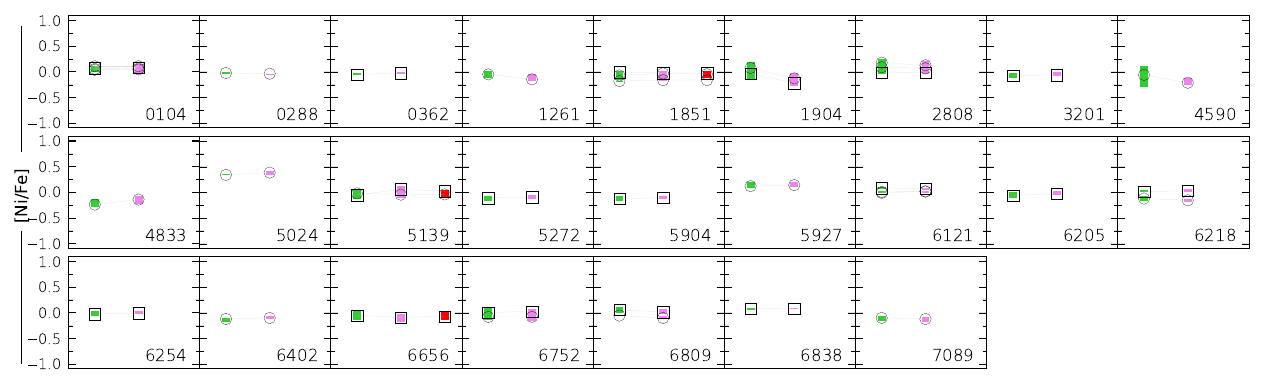}
\includegraphics[width=18cm, clip, trim={0cm 0.35cm 0cm 0.2cm}]{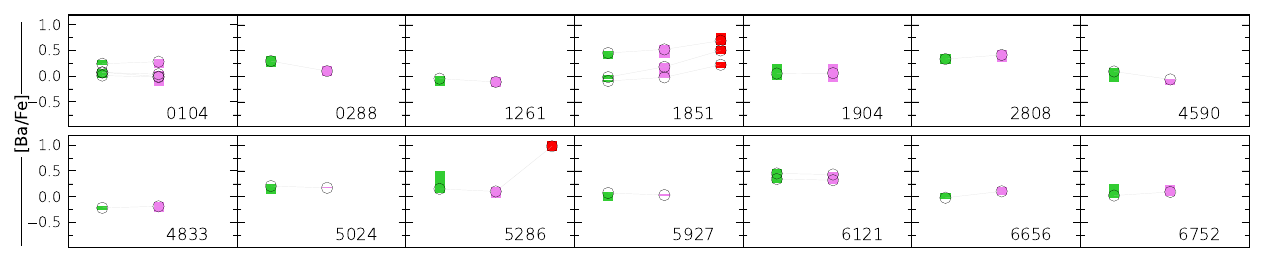}
    \caption{Same as Figure~\ref{fig:ax_ab1} but for titanium, iron, nickel, and barium.}
    \label{fig:ax_ab3}
\end{figure*}

\clearpage
{
\fontsize{6}{7}\selectfont
\onecolumn

\captionsetup{width=\linewidth}
\begin{longtable}{llrrrrrrrrrrrrrr}

\caption{
{\fontsize{9}{11}\selectfont
Summary of the spectroscopic datasets used in this work. For each dataset, we report the number of stars with available population tagging from the ChM. 
The '$-$' symbol indicates the lack of measurements for a given elemental ratio.} }
    \label{tab:data} \\
    \text{Cluster} & \text{Reference} & \text{A(Li)} & \text{[C/Fe]} & \text{[N/Fe]} & \text{[O/Fe]} & \text{[Na/Fe]} & \text{[Mg/Fe]} & \text{[Al/Fe]} & \text{[Si/Fe]} & \text{[K/Fe]} & \text{[Ca/Fe]} & \text{[Ti/Fe]} & \text{[Fe/H]} & \text{[Ni/Fe]} & \text{[Ba/Fe]} \\ 
    \\
    \hline
    \\
    \endfirsthead
    \multicolumn{3}{l}{\textit{\fontsize{9}{11}\selectfont Table~\ref{tab:data} $-$ continued}} \\
    \\
    \text{Cluster} & \text{Reference} & \text{A(Li)} & \text{[C/Fe]} & \text{[N/Fe]} & \text{[O/Fe]} & \text{[Na/Fe]} & \text{[Mg/Fe]} & \text{[Al/Fe]} & \text{[Si/Fe]} & \text{[K/Fe]} & \text{[Ca/Fe]} & \text{[Ti/Fe]} & \text{[Fe/H]} & \text{[Ni/Fe]} & \text{[Ba/Fe]} \\ 
    \\
    \hline
    \\
    \endhead
    \hline
    \multicolumn{3}{l}{\textit{\fontsize{9}{11}\selectfont Continued on next page}} \\
    \endfoot
    \hline
    \endlastfoot
NGC\,0104 & APOGEE                   &   $-$ &    98 &     98 &     98 &     $-$ &      98 &      98 &      98 &     98 &      98 &      98 &     98 &     98 &      $-$ \\[0pt]
          & \citet{carretta2009}     &   $-$ &   $-$ &    $-$ &     57 &      75 &     $-$ &     $-$ &     $-$ &    $-$ &     $-$ &     $-$ &     75 &    $-$ &      $-$ \\[0pt]
          & \citet{carretta2013a}    &   $-$ &   $-$ &     58 &    $-$ &     $-$ &      58 &      58 &      63 &    $-$ &     $-$ &     $-$ &    $-$ &    $-$ &      $-$ \\[0pt]
          & \citet{cernia2017}       &   $-$ &   $-$ &    $-$ &    $-$ &      21 &      18 &     $-$ &     $-$ &     21 &     $-$ &     $-$ &    $-$ &    $-$ &      $-$ \\[0pt]
          & \citet{dobrovolskas2021} &   $-$ &   $-$ &    $-$ &    $-$ &     159 &     $-$ &     $-$ &     $-$ &    $-$ &     $-$ &     $-$ &    159 &    $-$ &      151 \\[0pt]
          & GALAH                    &     6 &    66 &    $-$ &     77 &      87 &      87 &      86 &      85 &     87 &      87 &     $-$ &     62 &     80 &       86 \\[0pt]
          & GES                      &   $-$ &     9 &     10 &    $-$ &      25 &     101 &     123 &      92 &    $-$ &     119 &     113 &    126 &    122 &       44 \\[0pt]
          & \citet{kolomiecas2022}   &   $-$ &   $-$ &    $-$ &    $-$ &     144 &     $-$ &     $-$ &     $-$ &    $-$ &     $-$ &     $-$ &    144 &    $-$ &      $-$ \\[0pt]
          & \citet{kovalev2019}      &   $-$ &   $-$ &    $-$ &    $-$ &     $-$ &      46 &     $-$ &     $-$ &    $-$ &     $-$ &      46 &     46 &    $-$ &      $-$ \\[0pt]
          & \citet{marino2023}       &   $-$ &   $-$ &    $-$ &    $-$ &      26 &      26 &      26 &      26 &    $-$ &      26 &      26 &     26 &     26 &       26 \\[0pt]
          & \citet{meszaros2020}     &   $-$ &    60 &     61 &     49 &     $-$ &      63 &      50 &      63 &     58 &      63 &     $-$ &     65 &    $-$ &      $-$ \\[0pt]
          & \citet{mucciarelli2017}  &   $-$ &   $-$ &    $-$ &    $-$ &     $-$ &     $-$ &     $-$ &     $-$ &     63 &     $-$ &     $-$ &    $-$ &    $-$ &      $-$ \\[0pt]
          & \citet{pancino2017}      &   $-$ &   $-$ &    $-$ &     15 &      15 &      62 &      55 &     $-$ &    $-$ &     $-$ &     $-$ &     63 &    $-$ &      $-$ \\[0pt]
NGC\,0288 & APOGEE                   &   $-$ &     9 &      9 &      9 &     $-$ &       9 &       9 &       9 &      9 &       9 &       9 &      9 &      9 &      $-$ \\[0pt]
          & \citet{carretta2009}     &   $-$ &   $-$ &    $-$ &     38 &      67 &     $-$ &     $-$ &     $-$ &    $-$ &     $-$ &     $-$ &    $-$ &     69 &      $-$ \\[0pt]
          & GALAH                    &     2 &   $-$ &    $-$ &    $-$ &       9 &       9 &       2 &       8 &     10 &      10 &     $-$ &      8 &     10 &       10 \\[0pt]
          & \citet{kirby2016}        &     8 &   $-$ &    $-$ &    $-$ &     $-$ &     $-$ &     $-$ &     $-$ &    $-$ &     $-$ &     $-$ &    $-$ &     66 &      $-$ \\[0pt]
          & \citet{meszaros2020}     &   $-$ &     6 &    $-$ &      6 &     $-$ &      12 &      11 &      12 &      8 &      12 &     $-$ &     12 &    $-$ &      $-$ \\[0pt]
NGC\,0362 & APOGEE                   &   $-$ &     9 &      9 &      9 &     $-$ &       9 &       9 &       9 &      9 &       9 &       9 &      9 &      9 &      $-$ \\[0pt]
          & \citet{carretta2013b}    &   $-$ &   $-$ &    $-$ &      5 &       6 &       6 &     $-$ &       6 &    $-$ &       6 &       5 &      5 &    $-$ &        5 \\[0pt]
          & GES                      &   $-$ &   $-$ &    $-$ &    $-$ &     $-$ &      13 &      13 &     $-$ &    $-$ &      20 &     $-$ &     28 &     22 &      $-$ \\[0pt]
          & \citet{kirby2016}        &     8 &   $-$ &    $-$ &    $-$ &     $-$ &     $-$ &     $-$ &     $-$ &    $-$ &     $-$ &     $-$ &    $-$ &     66 &      $-$ \\[0pt]
          & \citet{meszaros2020}     &   $-$ &     6 &    $-$ &      6 &     $-$ &      12 &      11 &      12 &      8 &      12 &     $-$ &     12 &    $-$ &      $-$ \\[0pt]
          & \citet{pancino2017}      &   $-$ &   $-$ &    $-$ &    $-$ &     $-$ &       5 &     $-$ &     $-$ &    $-$ &     $-$ &     $-$ &      5 &    $-$ &      $-$ \\[0pt]
NGC\,1261 & GES                      &    12 &   $-$ &    $-$ &    $-$ &       7 &      41 &      23 &      12 &    $-$ &      30 &      24 &     45 &     28 &       28 \\[0pt]
          & \citet{marino2021}       &   $-$ &   $-$ &    $-$ &      8 &       9 &       4 &       4 &       4 &    $-$ &       4 &       4 &      9 &      3 &        4 \\[0pt]
NGC\,1851 & APOGEE                   &   $-$ &    26 &     26 &     26 &     $-$ &      26 &      26 &      26 &     26 &      26 &      23 &     26 &     26 &      $-$ \\[0pt]
          & \citet{carretta2011}     &   $-$ &   $-$ &    $-$ &     66 &      85 &      86 &       5 &     $-$ &    $-$ &     $-$ &     $-$ &    $-$ &     88 &       81 \\[0pt]
          & \citet{carretta2014a}    &   $-$ &   $-$ &     49 &    $-$ &     $-$ &     $-$ &     $-$ &     $-$ &    $-$ &     $-$ &     $-$ &    $-$ &    $-$ &      $-$ \\[0pt]
          & GES                      &     6 &   $-$ &    $-$ &    $-$ &      29 &      82 &      56 &      45 &    $-$ &      69 &      57 &     84 &     73 &       27 \\[0pt]
	       & GALAH                    &   $-$ &   $-$ &    $-$ &      4 &       4 &       4 &       5 &       4 &      4 &       4 &     $-$ &      4 &      4 &        4 \\[0pt]
          & \citet{kovalev2019}      &   $-$ &   $-$ &    $-$ &    $-$ &     $-$ &      60 &     $-$ &     $-$ &    $-$ &     $-$ &      60 &    $-$ &     60 &      $-$ \\[0pt]
          & \citet{meszaros2020}     &   $-$ &    14 &     10 &     14 &     $-$ &      14 &      17 &     $-$ &      6 &      16 &     $-$ &     12 &    $-$ &      $-$ \\[0pt]          
          & \citet{tatsu2022}        &   $-$ &    11 &     10 &     27 &      28 &      28 &      28 &      17 &    $-$ &      16 &     $-$ &    $-$ &     18 &       28 \\[0pt]
          & \citet{pancino2017}      &   $-$ &   $-$ &    $-$ &      6 &      27 &      14 &      29 &     $-$ &    $-$ &     $-$ &     $-$ &     65 &    $-$ &      $-$ \\[0pt]
NGC\,1904 & APOGEE                   &   $-$ &    10 &     10 &     10 &     $-$ &      10 &      10 &      10 &     10 &      10 &      10 &     10 &     10 &      $-$ \\[0pt]
          & \citet{carretta2009}     &   $-$ &   $-$ &    $-$ &     29 &      27 &     $-$ &     $-$ &     $-$ &    $-$ &     $-$ &     $-$ &    $-$ &     34 &      $-$ \\[0pt]
          & \citet{dorazi2015}       &    14 &   $-$ &    $-$ &    $-$ &     $-$ &     $-$ &      31 &     $-$ &    $-$ &     $-$ &     $-$ &    $-$ &    $-$ &      $-$ \\[0pt]
          & GES                      &     6 &   $-$ &    $-$ &    $-$ &      12 &      30 &      19 &      18 &    $-$ &      36 &      18 &     26 &     39 &       25 \\[0pt]
          & \citet{kirby2016}        &     2 &   $-$ &    $-$ &    $-$ &     $-$ &     $-$ &     $-$ &     $-$ &    $-$ &     $-$ &     $-$ &    $-$ &     33 &      $-$ \\[0pt]
          & \citet{kovalev2019}      &   $-$ &   $-$ &    $-$ &    $-$ &     $-$ &      22 &     $-$ &     $-$ &    $-$ &     $-$ &      23 &    $-$ &     23 &      $-$ \\[0pt]
          & \citet{pancino2017}      &   $-$ &   $-$ &    $-$ &    $-$ &       4 &       3 &       5 &     $-$ &    $-$ &     $-$ &     $-$ &     18 &    $-$ &      $-$ \\[0pt]
NGC\,2419 & \citet{mucciarelli2012}  &   $-$ &   $-$ &    $-$ &    $-$ &     $-$ &      26 &     $-$ &     $-$ &     19 &      30 &      27 &     30 &    $-$ &      $-$ \\[0pt]
NGC\,2808 & APOGEE                   &   $-$ &    36 &     36 &     36 &     $-$ &      36 &      36 &      36 &     36 &      36 &      36 &     36 &     36 &       36 \\[0pt]
          & \citet{carlos2023}       &   $-$ &    28 &     28 &     25 &     $-$ &     $-$ &      28 &     $-$ &    $-$ &     $-$ &     $-$ &     42 &     42 &       $-$ \\[0pt]
          & \citet{carretta2014c}    &   $-$ &   $-$ &    $-$ &    $-$ &     $-$ &      11 &      11 &     $-$ &    $-$ &     $-$ &     $-$ &    $-$ &    $-$ &      $-$ \\[0pt]
          & \citet{carretta2015b}    &   $-$ &   $-$ &    $-$ &     39 &      52 &      52 &     $-$ &      52 &    $-$ &     $-$ &     $-$ &    $-$ &    $-$ &       52 \\[0pt]
          & \citet{dorazi2015}       &    38 &   $-$ &    $-$ &    $-$ &     $-$ &     $-$ &      29 &     $-$ &    $-$ &     $-$ &     $-$ &    $-$ &    $-$ &      $-$ \\[0pt]
          & GES                      &    43 &     6 &    $-$ &    $-$ &      10 &      27 &      76 &      81 &    $-$ &      80 &      86 &     94 &     74 &       64 \\[0pt]
	       & GALAH                    &   $-$ &     5 &      5 &      6 &       6 &       6 &       5 &       6 &      6 &       6 &     $-$ &      6 &      6 &        6 \\[0pt]
          & \citet{kovalev2019}      &   $-$ &   $-$ &    $-$ &    $-$ &     $-$ &      13 &     $-$ &     $-$ &    $-$ &     $-$ &      13 &    $-$ &     13 &      $-$ \\[0pt]
          & \citet{pancino2017}      &   $-$ &   $-$ &    $-$ &    $-$ &       2 &     $-$ &      10 &     $-$ &    $-$ &     $-$ &     $-$ &     17 &    $-$ &      $-$ \\[0pt]
NGC\,3201 & \citet{aguilera2022}     &    23 &   $-$ &    $-$ &    $-$ &      31 &     $-$ &     $-$ &     $-$ &    $-$ &     $-$ &     $-$ &    $-$ &    $-$ &      $-$ \\[0pt]
          & APOGEE                   &   $-$ &    27 &     26 &     27 &      27 &      27 &      27 &      27 &     27 &      27 &      27 &     27 &     27 &      $-$ \\[0pt]
          & \citet{carretta2009}     &   $-$ &   $-$ &    $-$ &     44 &      60 &     $-$ &     $-$ &     $-$ &    $-$ &     $-$ &     $-$ &    $-$ &     61 &      $-$ \\[0pt]
          & \citet{meszaros2020}     &   $-$ &    10 &    $-$ &    $-$ &     $-$ &       4 &      28 &      29 &     14 &     $-$ &     $-$ &     29 &    $-$ &      $-$ \\[0pt]          
NGC\,4590 & APOGEE                   &   $-$ &   $-$ &    $-$ &     10 &     $-$ &      10 &     $-$ &      10 &      9 &       8 &     $-$ &     10 &      8 &      $-$ \\[0pt]
          & \citet{carretta2009}     &   $-$ &   $-$ &    $-$ &     38 &      58 &     $-$ &     $-$ &     $-$ &    $-$ &     $-$ &     $-$ &    $-$ &     73 &      $-$ \\[0pt]
          & \citet{duggan2018}       &   $-$ &   $-$ &    $-$ &    $-$ &     $-$ &     $-$ &     $-$ &     $-$ &    $-$ &     $-$ &     $-$ &    $-$ &     13 &       13 \\[0pt]
          & GES                      &   $-$ &   $-$ &    $-$ &    $-$ &     $-$ &      48 &     $-$ &     $-$ &    $-$ &      24 &     $-$ &     48 &     19 &      $-$ \\[0pt]
          & \citet{kirby2016}        &     5 &   $-$ &    $-$ &    $-$ &     $-$ &     $-$ &     $-$ &     $-$ &    $-$ &     $-$ &     $-$ &    $-$ &     46 &      $-$ \\[0pt]
          & \citet{meszaros2020}     &   $-$ &   $-$ &    $-$ &    $-$ &     $-$ &      10 &      11 &      11 &    $-$ &     $-$ &     $-$ &     11 &    $-$ &      $-$ \\[0pt]
NGC\,4833 & \citet{carretta2014b}    &   $-$ &   $-$ &    $-$ &     24 &      23 &      18 &     $-$ &      33 &    $-$ &      36 &     $-$ &     36 &     32 &       32 \\[0pt]
          & GES                      &   $-$ &   $-$ &    $-$ &    $-$ &       3 &      19 &     $-$ &       3 &    $-$ &      29 &       3 &     33 &     19 &        3 \\[0pt]
          & \citet{kovalev2019}      &   $-$ &   $-$ &    $-$ &    $-$ &     $-$ &      27 &     $-$ &     $-$ &    $-$ &     $-$ &      27 &     27 &    $-$ &      $-$ \\[0pt]
          & \citet{pancino2017}      &   $-$ &   $-$ &    $-$ &    $-$ &       2 &       4 &       4 &     $-$ &    $-$ &     $-$ &     $-$ &     12 &    $-$ &      $-$ \\[0pt]          
NGC\,5024 & \citet{boberg2016}       &   $-$ &   $-$ &    $-$ &     18 &       9 &     $-$ &     $-$ &     $-$ &    $-$ &      19 &      17 &     19 &     19 &       17 \\[0pt]
          & \citet{gerber2021}       &   $-$ &    10 &     10 &    $-$ &     $-$ &     $-$ &     $-$ &     $-$ &    $-$ &     $-$ &     $-$ &      4 &    $-$ &      $-$ \\[0pt]
NGC\,5053 & APOGEE                   &   $-$ &     3 &      4 &      4 &     $-$ &       4 &     $-$ &       4 &    $-$ &     $-$ &       4 &      4 &    $-$ &      $-$ \\[0pt]
          & \citet{kirby2016}        &   $-$ &   $-$ &    $-$ &    $-$ &     $-$ &     $-$ &     $-$ &     $-$ &    $-$ &     $-$ &     $-$ &     23 &    $-$ &      $-$ \\[0pt]
          & \citet{meszaros2020}     &   $-$ &   $-$ &    $-$ &    $-$ &     $-$ &       4 &     $-$ &       4 &    $-$ &     $-$ &     $-$ &      4 &    $-$ &      $-$ \\[0pt]
NGC\,5139 & \citet{alvarez2022}      &   $-$ &   $-$ &    $-$ &    $-$ &      40 &      38 &     $-$ &     $-$ &     50 &     $-$ &     $-$ &     50 &    $-$ &      $-$ \\[0pt]
          & \citet{alvarez2024}      &   $-$ &   $-$ &    $-$ &    $-$ &     $-$ &     $-$ &      50 &      50 &    $-$ &     $-$ &     $-$ &    $-$ &    $-$ &      $-$ \\[0pt]
          & APOGEE                   &   $-$ &    21 &     18 &     21 &     $-$ &      21 &      21 &      21 &     21 &      21 &      21 &     21 &     21 &      $-$ \\[0pt]
          & \citet{johnson2010}      &   $-$ &   $-$ &    $-$ &     71 &      71 &     $-$ &     $-$ &      74 &    $-$ &      74 &      73 &     74 &     72 &      $-$ \\[0pt]
          & \citet{marino2011}       &   $-$ &   $-$ &    $-$ &     35 &      37 &     $-$ &     $-$ &      $-$ &    $-$ &     $-$ &     $-$ &     40 &   $-$ &       40 \\[0pt]
          & \citet{marino2012}       &   $-$ &    16 &     16 &     $-$ &    $-$ &     $-$ &     $-$ &     $-$ &    $-$ &     $-$ &     $-$ &    $-$ &    $-$ &      $-$ \\[0pt]
          & \citet{mucciarelli2018}  &    41 &   $-$ &    $-$ &    $-$ &      40 &     $-$ &     $-$ &     $-$ &    $-$ &     $-$ &     $-$ &     41 &    $-$ &      $-$ \\[0pt]          
NGC\,5272 & APOGEE                   &   $-$ &    62 &     59 &     62 &     $-$ &      62 &      62 &      62 &     61 &      62 &      46 &     62 &     61 &      $-$ \\[0pt]
          & \citet{meszaros2020}     &   $-$ &    32 &     26 &     15 &     $-$ &      62 &      59 &      61 &     21 &      57 &     $-$ &     60 &    $-$ &      $-$ \\[0pt]
          & \citet{masseron2019}     &   $-$ &     5 &      9 &      5 &     $-$ &      14 &     $-$ &      13 &      4 &      13 &     $-$ &     13 &    $-$ &      $-$ \\[0pt]
NGC\,5286 & \citet{marino2015}       &   $-$ &   $-$ &    $-$ &     12 &      20 &     $-$ &     $-$ &      31 &    $-$ &      33 &      32 &     33 &      3 &       33 \\[0pt]
NGC\,5904 & APOGEE                   &   $-$ &    80 &     79 &     80 &     $-$ &      80 &      80 &      80 &     79 &      80 &      80 &     80 &     80 &      $-$ \\[0pt]
          & \citet{carretta2009}     &   $-$ &   $-$ &    $-$ &     62 &      81 &     $-$ &     $-$ &     $-$ &    $-$ &     $-$ &     $-$ &     81 &    $-$ &      $-$ \\[0pt]
	       & GALAH                    &   $-$ &   $-$ &    $-$ &    $-$ &       4 &       4 &     $-$ &     $-$ &      4 &     $-$ &     $-$ &    $-$ &      4 &        4 \\[0pt]
          & \citet{meszaros2020}     &   $-$ &    10 &      6 &      3 &     $-$ &      86 &      82 &      83 &     13 &      67 &     $-$ &     85 &    $-$ &      $-$ \\[0pt]
          & \citet{masseron2019}     &   $-$ &     6 &      5 &      3 &     $-$ &      27 &      27 &      27 &      6 &      25 &     $-$ &     26 &    $-$ &      $-$ \\[0pt]
NGC\,5927 & GES                      &   $-$ &   $-$ &    $-$ &    $-$ &     $-$ &      25 &      25 &      19 &    $-$ &       7 &      21 &     25 &     15 &       24 \\[0pt]
          & \citet{kovalev2019}      &   $-$ &   $-$ &    $-$ &    $-$ &     $-$ &      20 &     $-$ &     $-$ &    $-$ &     $-$ &      20 &     20 &     20 &      $-$ \\[0pt]
          & \citet{pancino2017}      &   $-$ &   $-$ &    $-$ &    $-$ &       6 &      13 &      16 &     $-$ &    $-$ &     $-$ &     $-$ &    $-$ &     32 &      $-$ \\[0pt]
NGC\,5986 & \citet{johnson2017}      &   $-$ &   $-$ &    $-$ &      6 &       6 &       6 &       6 &       6 &    $-$ &       6 &     $-$ &      6 &    $-$ &      $-$ \\[0pt]
NGC\,6093 & \citet{carretta2015a}    &   $-$ &   $-$ &    $-$ &    $-$ &     $-$ &     $-$ &     $-$ &       7 &    $-$ &      10 &     $-$ &     11 &    $-$ &      $-$ \\[0pt]
NGC\,6121 & APOGEE                   &   $-$ &    42 &     41 &     42 &     $-$ &      42 &      42 &      42 &     42 &      42 &      42 &     42 &     42 &      $-$ \\[0pt]
          & \citet{carretta2009}     &   $-$ &   $-$ &    $-$ &     23 &      29 &     $-$ &     $-$ &     $-$ &    $-$ &     $-$ &     $-$ &     29 &    $-$ &      $-$ \\[0pt]
          & \citet{carretta2013a}    &   $-$ &   $-$ &     25 &    $-$ &     $-$ &      23 &      23 &      28 &    $-$ &     $-$ &     $-$ &    $-$ &    $-$ &      $-$ \\[0pt]
          & \citet{dorazi2010}       &    12 &   $-$ &    $-$ &    $-$ &     $-$ &     $-$ &     $-$ &     $-$ &    $-$ &     $-$ &     $-$ &    $-$ &    $-$ &      $-$ \\[0pt]
	       & GALAH                    &    21 &    24 &     21 &     32 &      35 &      35 &      34 &      35 &     35 &      35 &     $-$ &     34 &     35 &       35 \\[0pt]
          & \citet{marino2008}       &   $-$ &   $-$ &    $-$ &     25 &      31 &      31 &      26 &      31 &    $-$ &      31 &      31 &     31 &     26 &       30 \\[0pt]
          & \citet{meszaros2020}     &   $-$ &    14 &     19 &     16 &     $-$ &      29 &      29 &      30 &     23 &      27 &     $-$ &     29 &    $-$ &      $-$ \\[0pt]
          & \citet{mucciarelli2011}  &    17 &   $-$ &    $-$ &    $-$ &     $-$ &     $-$ &     $-$ &     $-$ &    $-$ &     $-$ &     $-$ &     19 &    $-$ &      $-$ \\[0pt]          
NGC\,6205 & APOGEE                   &   $-$ &    49 &     49 &     48 &     $-$ &      49 &      49 &      49 &     48 &      48 &      48 &     49 &     49 &      $-$ \\[0pt]
          & \citet{johnson2012}      &   $-$ &   $-$ &    $-$ &     56 &      57 &     $-$ &     $-$ &     $-$ &    $-$ &     $-$ &     $-$ &    $-$ &     48 &      $-$ \\[0pt]
          & \citet{meszaros2020}     &   $-$ &    14 &     14 &     12 &     $-$ &      58 &      39 &      50 &    $-$ &       9 &     $-$ &     37 &    $-$ &      $-$ \\[0pt]
          & \citet{masseron2019}     &   $-$ &     6 &     12 &      9 &     $-$ &      19 &      12 &      16 &    $-$ &       7 &       7 &     13 &    $-$ &      $-$ \\[0pt]
NGC\,6218 & APOGEE                   &   $-$ &    29 &     29 &     29 &     $-$ &      29 &      29 &      29 &     28 &      29 &      29 &     29 &     29 &      $-$ \\[0pt]
          & \citet{carretta2007b}    &   $-$ &   $-$ &    $-$ &     49 &      45 &     $-$ &     $-$ &     $-$ &    $-$ &     $-$ &     $-$ &    $-$ &     58 &      $-$ \\[0pt]
          & GES                      &   $-$ &   $-$ &    $-$ &    $-$ &       7 &      62 &      55 &      54 &    $-$ &      55 &      52 &     62 &     54 &        8 \\[0pt]
          & \citet{meszaros2020}     &   $-$ &    17 &     13 &     13 &     $-$ &      43 &      36 &      46 &     15 &      25 &     $-$ &     46 &    $-$ &      $-$ \\[0pt]
NGC\,6254 & APOGEE                   &   $-$ &    42 &     38 &     42 &     $-$ &      42 &      42 &      42 &     37 &      42 &      42 &     42 &     41 &      $-$ \\[0pt]
          & \citet{carretta2009}     &   $-$ &   $-$ &    $-$ &     61 &      72 &     $-$ &     $-$ &     $-$ &    $-$ &     $-$ &     $-$ &    $-$ &     90 &      $-$ \\[0pt]
          & \citet{gerber2018}       &   $-$ &    87 &     89 &    $-$ &     $-$ &     $-$ &     $-$ &     $-$ &    $-$ &     $-$ &     $-$ &    $-$ &    $-$ &      $-$ \\[0pt]
          & \citet{meszaros2020}     &   $-$ &    10 &     20 &     18 &     $-$ &      37 &      40 &      38 &      9 &      33 &     $-$ &     40 &    $-$ &      $-$ \\[0pt]
NGC\,6362 & \citet{mucciarelli2016}  &   $-$ &   $-$ &    $-$ &    $-$ &      12 &     $-$ &     $-$ &     $-$ &    $-$ &     $-$ &     $-$ &     12 &    $-$ &      $-$ \\[0pt]
NGC\,6388 & \citet{carretta2022}     &   $-$ &   $-$ &    $-$ &    $-$ &     $-$ &     $-$ &     $-$ &     $-$ &    $-$ &     $-$ &     $-$ &     10 &    $-$ &      $-$ \\[0pt]
          & \citet{carretta2023}     &   $-$ &   $-$ &    $-$ &     10 &      10 &      10 &       8 &      10 &    $-$ &      10 &      10 &    $-$ &     10 &        9 \\[0pt]
NGC\,6397 & \citet{carretta2009}     &   $-$ &   $-$ &    $-$ &      4 &      13 &     $-$ &     $-$ &     $-$ &    $-$ &     $-$ &     $-$ &     20 &    $-$ &      $-$ \\[0pt]
	       & GALAH                    &     7 &     3 &    $-$ &    $-$ &       7 &       6 &     $-$ &       6 &      8 &       8 &     $-$ &      8 &      8 &        8 \\[0pt]
NGC\,6402 & \citet{johnson2019}      &   $-$ &   $-$ &    $-$ &     15 &      15 &      14 &      16 &      16 &    $-$ &      16 &     $-$ &     15 &     15 &      $-$ \\[0pt]
NGC\,6535 & \citet{bragaglia2017}    &   $-$ &   $-$ &    $-$ &     11 &      10 &      10 &     $-$ &     $-$ &    $-$ &       2 &       2 &     12 &      2 &      $-$ \\[0pt]
NGC\,6656 & \citet{aguilera2022}     &    28 &   $-$ &    $-$ &    $-$ &      28 &     $-$ &     $-$ &     $-$ &    $-$ &     $-$ &     $-$ &    $-$ &    $-$ &      $-$ \\[0pt]
          & APOGEE                   &   $-$ &    46 &     45 &     46 &     $-$ &      46 &      46 &      46 &     43 &      44 &      24 &     46 &     44 &      $-$ \\[0pt]
NGC\,6715 & \citet{carretta2010}     &   $-$ &   $-$ &    $-$ &     19 &      19 &      18 &     $-$ &      19 &    $-$ &      19 &      18 &     19 &     19 &      $-$ \\[0pt]
          & \citet{mucciarelli2017}  &   $-$ &   $-$ &    $-$ &    $-$ &     $-$ &      27 &     $-$ &     $-$ &    $-$ &      26 &      24 &     27 &    $-$ &      $-$ \\[0pt]
NGC\,6752 & APOGEE                   &   $-$ &    43 &     40 &     42 &     $-$ &      43 &      43 &      43 &     42 &      43 &      42 &     43 &     43 &      $-$ \\[0pt]
          & \citet{carretta2007a}    &   $-$ &   $-$ &    $-$ &     43 &      60 &     $-$ &     $-$ &     $-$ &    $-$ &     $-$ &     $-$ &    $-$ &     66 &      $-$ \\[0pt]
          & GES                      &    21 &   $-$ &    $-$ &    $-$ &      20 &      63 &      49 &      36 &    $-$ &      70 &      40 &     79 &     64 &       31 \\[0pt]
	       & GALAH                    &     8 &   $-$ &    $-$ &    $-$ &       8 &       7 &     $-$ &       7 &      8 &       8 &     $-$ &      7 &      8 &        8 \\[0pt]
          & \citet{kovalev2019}      &   $-$ &   $-$ &    $-$ &    $-$ &     $-$ &      51 &     $-$ &     $-$ &    $-$ &     $-$ &      50 &     51 &    $-$ &      $-$ \\[0pt]
          & \citet{meszaros2020}     &   $-$ &     9 &     22 &      3 &     $-$ &      52 &      50 &      53 &      3 &      32 &     $-$ &     51 &    $-$ &      $-$ \\[0pt]
          & \citet{mucciarelli2017}  &   $-$ &   $-$ &    $-$ &    $-$ &     $-$ &     $-$ &     $-$ &     $-$ &     61 &     $-$ &     $-$ &    $-$ &    $-$ &      $-$ \\[0pt]
          & \citet{pancino2017}      &   $-$ &   $-$ &    $-$ &    $-$ &       8 &     $-$ &       7 &     $-$ &    $-$ &     $-$ &     $-$ &      9 &    $-$ &      $-$ \\[0pt]     
NGC\,6809 & \citet{aguilera2022}     &    24 &   $-$ &    $-$ &    $-$ &      27 &     $-$ &     $-$ &     $-$ &    $-$ &     $-$ &     $-$ &    $-$ &    $-$ &      $-$ \\[0pt]
          & APOGEE                   &   $-$ &    24 &     32 &     33 &     $-$ &      33 &      33 &      33 &     32 &      28 &      33 &     13 &     33 &      $-$ \\[0pt]
          & \citet{carretta2009}     &   $-$ &   $-$ &    $-$ &     66 &      76 &     $-$ &     $-$ &     $-$ &    $-$ &     $-$ &     $-$ &    $-$ &    100 &      $-$ \\[0pt]
	       & GALAH                    &   $-$ &   $-$ &    $-$ &    $-$ &      13 &      10 &     $-$ &       8 &     13 &      13 &     $-$ &     13 &     13 &      $-$ \\[0pt]
          & \citet{mucciarelli2017}  &   $-$ &   $-$ &    $-$ &    $-$ &     $-$ &     $-$ &     $-$ &     $-$ &     96 &     $-$ &     $-$ &    $-$ &    $-$ &      $-$ \\[0pt]
          & \citet{meszaros2020}     &   $-$ &     8 &      2 &     10 &     $-$ &      30 &      29 &       7 &    $-$ &      10 &     $-$ &     36 &    $-$ &      $-$ \\[0pt]
NGC\,6838 & \citet{aguilera2022}     &     9 &   $-$ &    $-$ &    $-$ &     $-$ &     $-$ &     $-$ &     $-$ &    $-$ &     $-$ &     $-$ &    $-$ &    $-$ &      $-$ \\[0pt]
          & APOGEE                   &   $-$ &    20 &     20 &     20 &     $-$ &      20 &      20 &      20 &     20 &      20 &      20 &     20 &     20 &       20 \\[0pt]
          & \citet{carretta2009}     &   $-$ &   $-$ &    $-$ &     23 &      36 &     $-$ &     $-$ &     $-$ &    $-$ &     $-$ &     $-$ &    $-$ &     36 &      $-$ \\[0pt]
          & \citet{kirby2016}        &     3 &   $-$ &    $-$ &    $-$ &     $-$ &     $-$ &     $-$ &     $-$ &    $-$ &     $-$ &     $-$ &    $-$ &     14 &      $-$ \\[0pt]
          & \citet{meszaros2020}     &   $-$ &    11 &      7 &      6 &     $-$ &      11 &      10 &      11 &     10 &      11 &     $-$ &     11 &    $-$ &      $-$ \\[0pt]
          & \citet{masseron2019}     &   $-$ &     2 &      2 &      2 &     $-$ &       2 &       2 &       2 &      2 &     $-$ &     $-$ &      2 &    $-$ &      $-$ \\[0pt]
NGC\,6934 & \citet{marino2021}       &   $-$ &   $-$ &    $-$ &     10 &     $-$ &     $-$ &     $-$ &     $-$ &    $-$ &     $-$ &     $-$ &     11 &    $-$ &      $-$ \\[0pt]
NGC\,7078 & GES                      &   $-$ &   $-$ &    $-$ &    $-$ &     $-$ &      15 &     $-$ &       3 &    $-$ &       7 &       3 &     16 &      5 &      $-$ \\[0pt]
          & \citet{carretta2009}     &   $-$ &   $-$ &    $-$ &     13 &      13 &     $-$ &     $-$ &     $-$ &    $-$ &     $-$ &     $-$ &    $-$ &     20 &      $-$ \\[0pt]
          & \citet{kovalev2019}      &   $-$ &   $-$ &    $-$ &    $-$ &     $-$ &      10 &     $-$ &     $-$ &    $-$ &     $-$ &      10 &     10 &    $-$ &      $-$ \\[0pt]
          & \citet{meszaros2020}     &   $-$ &   $-$ &    $-$ &      5 &     $-$ &       8 &       8 &       8 &    $-$ &     $-$ &     $-$ &      8 &    $-$ &      $-$ \\[0pt]
          & \citet{masseron2019}     &   $-$ &     3 &    $-$ &      7 &     $-$ &      10 &      11 &      11 &      2 &       6 &     $-$ &     10 &    $-$ &      $-$ \\[0pt]
NGC\,7089 & APOGEE                   &   $-$ &    13 &     13 &     13 &     $-$ &      13 &      13 &      13 &     13 &      13 &      13 &     13 &     13 &      $-$ \\[0pt]
          & GES                      &   $-$ &   $-$ &      3 &    $-$ &       6 &      76 &      71 &      25 &    $-$ &      69 &      39 &     83 &     73 &        6 \\[0pt]
          & \citet{kirby2016}        &     9 &   $-$ &    $-$ &    $-$ &     $-$ &     $-$ &     $-$ &     $-$ &    $-$ &     $-$ &     $-$ &    $-$ &    199 &      $-$ \\[0pt]
          & \citet{kovalev2019}      &   $-$ &   $-$ &    $-$ &    $-$ &     $-$ &      63 &     $-$ &     $-$ &    $-$ &     $-$ &      63 &     63 &    $-$ &      $-$ \\[0pt]
          & \citet{yong2014}         &   $-$ &   $-$ &    $-$ &      4 &      10 &       9 &       6 &      10 &    $-$ &     $-$ &     $-$ &     10 &    $-$ &      $-$ \\[0pt]
NGC\,7099 & \citet{carretta2009}     &   $-$ &   $-$ &    $-$ &      3 &       5 &     $-$ &     $-$ &     $-$ &    $-$ &     $-$ &     $-$ &    $-$ &     14 &      $-$ \\[0pt]
          & \citet{kirby2016}        &   $-$ &   $-$ &    $-$ &    $-$ &     $-$ &     $-$ &     $-$ &     $-$ &    $-$ &     $-$ &     $-$ &    $-$ &     45 &      $-$ \\[0pt]
          & \citet{pancino2017}      &   $-$ &   $-$ &    $-$ &    $-$ &       2 &     $-$ &      10 &     $-$ &    $-$ &     $-$ &     $-$ &     17 &    $-$ &      $-$ \\[0pt]
\end{longtable}
}
\twocolumn

\begin{table*}
    \centering
\tiny
\setlength{\tabcolsep}{5.5pt}
\renewcommand{\arraystretch}{1.6}

\caption{
Average abundance differences between 1P and 2P stars. The rows indicated with 'A' are referred to the differences between 1P and anomalous stars. Entries with the '$-$' symbol refers to elements with no measurements. Measurements with the '$^{\rm *}$' notation are based on carbon and nitrogen abundances above the RGB bump.}

\scalebox{0.8}{

\begin{tabular}{c|cccccccccccccc}
     {NGC} &   $\Delta$A(Li) &   $\Delta$[C/Fe] &   $\Delta$[N/Fe] &   $\Delta$[O/Fe] &   $\Delta$[Na/Fe] &   $\Delta$[Mg/Fe] &   $\Delta$[Al/Fe] &   $\Delta$[Si/Fe] &   $\Delta$[K/Fe] &   $\Delta$[Ca/Fe] &   $\Delta$[Ti/Fe] &   $\Delta$[Fe/H] &   $\Delta$[Ni/Fe] &   $\Delta$[Ba/Fe] \\
\hline
\\
0104 & $-$ & -0.17±0.02$^{\rm *}$ & 0.41±0.03$^{\rm *}$ & -0.10±0.01 & 0.21±0.01 & -0.01±0.01 & 0.08±0.01 & -0.03±0.01 & 0.01±0.01 & -0.00±0.01 & 0.04±0.01 & 0.02±0.00 & 0.01±0.00 & -0.05±0.02 \\
0288 & $-$ & $-$ & $-$ & -0.16±0.14 & 0.43±0.05 & 0.00±0.07 & 0.09±0.17 & 0.01±0.03 & -0.13±0.10 & -0.03±0.03 & $-$ & -0.02±0.01 & -0.01±0.04 & -0.19±0.11 \\     
0362 & $-$ & $-$ & $-$ & -0.02±0.04 & $-$ & 0.02±0.06 & 0.13±0.19 & -0.02±0.04 & -0.07±0.04 & 0.03±0.02 & 0.10±0.09 & -0.03±0.04 & 0.03±0.03 & $-$ \\
\hspace{4mm}A & $-$ & $-$ & $-$ & $-$ & $-$ & $-$ & $-$ & $-$ & $-$ & $-$ & $-$ & $-$ & $-$ & $-$ \\
1261 & $-$ & $-$ & $-$ & $-$ & $-$ & -0.01±0.04 & 0.31±0.10 & -0.03±0.07 & $-$ & -0.06±0.07 & 0.05±0.15 & 0.01±0.03 & -0.10±0.05 & -0.06±0.18 \\
\hspace{4mm}A & $-$ & $-$ & $-$ & $-$ & $-$ & $-$ & $-$ & $-$ & $-$ & $-$ & $-$ & 0.06±0.05 & $-$ & $-$ \\
1851 & $-$ & -0.10±0.09$^{\rm *}$ & 0.47±0.10$^{\rm *}$ & -0.15±0.04 & 0.39±0.04 & -0.02±0.01 & 0.32±0.05 & -0.05±0.03 & -0.06±0.09 & 0.03±0.02 & 0.02±0.02 & -0.01±0.01 & 0.01±0.02 & 0.09±0.04 \\
\hspace{4mm}A & $-$ & -0.26±0.10$^{\rm *}$ & 0.73±0.11$^{\rm *}$ & -0.27±0.04 & 0.58±0.05 & -0.01±0.01 & 0.50±0.05 & -0.06±0.02 & -0.08±0.10 & 0.06±0.02 & 0.02±0.02 & 0.04±0.01 & 0.01±0.02 & 0.33±0.04 \\
1904 & -0.15±0.05 & -0.50±0.19$^{\rm *}$ & 0.53±0.12$^{\rm *}$ & -0.27±0.09 & 0.38±0.07 & -0.11±0.04 & 0.80±0.06 & -0.01±0.03 & -0.18±0.20 & -0.10±0.07 & 0.05±0.04 & 0.03±0.01 & -0.18±0.10 & 0.01±0.13 \\
2419 & $-$ & $-$ & $-$ & $-$ & $-$ & -0.86±0.19 & $-$ & $-$ & 0.92±0.35 & 0.09±0.05 & 0.06±0.08 & -0.01±0.06 & $-$ & $-$ \\      
2808 & -0.05±0.03 & -0.32±0.07 & 80±0.08$^{\rm *}$ & -0.13±0.04 & 0.31±0.08 & -0.09±0.03 & 0.59±0.06 & 0.00±0.01 & 0.04±0.04 & -0.00±0.02 & 0.12±0.02 & 0.02±0.01 & -0.00±0.02 & 0.08±0.08 \\
3201 & -0.14±0.07 & -0.30±0.10 & 0.91±0.13 & -0.23±0.06 & 0.47±0.04 & -0.05±0.02 & 0.53±0.06 & -0.01±0.01 & 0.08±0.08 & -0.04±0.06 & 0.00±0.09 & 0.00±0.01 & 0.01±0.05 & $-$ \\        
4590 & $-$ & $-$ & $-$ & -0.08±0.11 & 0.04±0.06 & -0.07±0.03 & $-$ & $-$ & $-$ & -0.03±0.08 & $-$ & 0.01±0.01 & -0.15±0.13 & -0.16±0.11 \\                                
4833 & $-$ & $-$ & $-$ & -0.18±0.21 & 0.30±0.13 & -0.11±0.05 & $-$ & 0.01±0.03 & $-$ & -0.01±0.01 & -0.05±0.03 & 0.01±0.01 & 0.10±0.15 & 0.03±0.07 \\                           
5024 & $-$ & $-$ & $-$ & -0.15±0.08 & -0.10±0.10 & $-$ & $-$ & $-$ & $-$ & 0.04±0.05 & -0.03±0.05 & -0.01±0.04 & 0.04±0.05 & -0.03±0.05 \\                                      
5053 & $-$ & $-$ & $-$ & $-$ & $-$ & $-$ & $-$ & $-$ & $-$ & $-$ & $-$ & 0.11±0.10 & $-$ & $-$ \\                                                                                 
5139 & $-$ & $-$ & $-$ & -0.11±0.10 & 0.13±0.15 & -0.00±0.09 & 0.53±0.21 & 0.05±0.03 & 0.05±0.12 & -0.03±0.05 & -0.07±0.08 & -0.00±0.08 & 0.04±0.05 & $-$ \\                    
\hspace{4mm}A & $-$ & $-$ & $-$ & 0.21±0.11 & -0.12±0.05 & 0.58±0.17 & 0.05±0.02 & 0.12±0.10 & 0.00±0.04 & 0.06±0.06 & 0.21±0.07 & 0.02±0.04 & 0.50±0.40 \\
5272 & $-$ & -0.10±0.13 & 0.35±0.28 & -0.16±0.05 & $-$ & -0.09±0.02 & 0.69±0.06 & -0.02±0.02 & -0.01±0.07 & -0.01±0.03 & -0.08±0.14 & 0.06±0.02 & 0.02±0.04 & $-$ \\
5286 & $-$ & $-$ & $-$ & $-$ & -0.12±0.12 & $-$ & $-$ & 0.01±0.06 & $-$ & -0.06±0.04 & -0.14±0.06 & 0.03±0.05 & $-$ & -0.05±0.18 \\                                 
\hspace{4mm}A & $-$ & $-$ & $-$ & $-$ & 0.18±0.12 & $-$ & $-$ & 0.06±0.06 & $-$ & 0.01±0.04 & -0.06±0.05 & 0.14±0.04 & $-$ & 0.83±0.18 \\
5904 & $-$ & -0.20±0.07 & 0.90±0.09 & -0.28±0.04 & 0.50±0.05 & -0.04±0.01 & 0.45±0.04 & -0.01±0.01 & -0.01±0.05 & -0.01±0.02 & 0.06±0.03 & -0.01±0.01 & 0.02±0.03 & $-$ \\              
5927 & $-$ & $-$ & $-$ & $-$ & $-$ & -0.03±0.01 & -0.02±0.03 & -0.04±0.06 & $-$ & $-$ & -0.00±0.03 & 0.03±0.02 & 0.02±0.05 & -0.04±0.07 \\                                              
5986 & $-$ & $-$ & $-$ & $-$ & $-$ & $-$ & $-$ & $-$ & $-$ & $-$ & $-$ & $-$ & $-$ & $-$ \\                                                                                             
6093 & $-$ & $-$ & $-$ & $-$ & $-$ & $-$ & $-$ & $-$ & $-$ & $-$ & $-$ & 0.01±0.02 & $-$ & $-$ \\                                                                                       
6121 & -0.12±0.05 & -0.14±0.04 & 1.02±0.11 & -0.11±0.02 & 0.30±0.03 & -0.03±0.01 & 0.05±0.03 & 0.00±0.01 & -0.01±0.04 & 0.01±0.01 & 0.04±0.01 & 0.00±0.01 & 0.01±0.01 & -0.02±0.05 \\   
6205 & $-$ & -0.35±0.09$^{\rm *}$ & 0.58±0.09$^{\rm *}$ & -0.23±0.05 & 0.43±0.08 & -0.13±0.04 & 1.03±0.10 & 0.02±0.02 & -0.06±0.15 & 0.03±0.06 & 0.11±0.05 & 0.03±0.02 & 0.04±0.05 & $-$ \\                              
6218 & $-$ & -0.04±0.10$^{\rm *}$ & 0.62±0.19$^{\rm *}$ & -0.13±0.04 & 0.44±0.08 & -0.03±0.02 & 0.25±0.03 & -0.02±0.01 & -0.03±0.09 & -0.01±0.03 & 0.04±0.04 & 0.01±0.01 & -0.01±0.03 & $-$ \\                           
6254 & $-$ & -0.15±0.21 & 0.82±0.33 & -0.21±0.04 & 0.42±0.07 & -0.14±0.02 & 0.96±0.07 & -0.04±0.01 & -0.08±0.17 & -0.03±0.04 & 0.09±0.06 & 0.02±0.01 & 0.02±0.03 & $-$ \\               
6362 & $-$ & $-$ & $-$ & $-$ & 0.45±0.10 & $-$ & $-$ & $-$ & $-$ & $-$ & $-$ & 0.05±0.05 & $-$ & $-$ \\                                                                                 
6366 & $-$ & $-$ & $-$ & $-$ & $-$ & $-$ & $-$ & $-$ & $-$ & $-$ & $-$ & $-$ & $-$ & $-$ \\                                                                                             
6397 & $-$ & $-$ & $-$ & $-$ & 0.45±0.17 & $-$ & $-$ & $-$ & $-$ & $-$ & $-$ & -0.01±0.02 & $-$ & $-$ \\                                                                                
6402 & $-$ & $-$ & $-$ & -0.79±0.25 & 0.60±0.13 & -0.14±0.05 & 0.66±0.16 & 0.07±0.07 & $-$ & 0.06±0.03 & $-$ & 0.04±0.04 & 0.02±0.05 & $-$ \\                                           
6535 & $-$ & $-$ & $-$ & $-$ & 0.36±0.16 & $-$ & $-$ & $-$ & $-$ & $-$ & $-$ & -0.05±0.05 & $-$ & $-$ \\                                                                                
6656 & -0.04±0.07 & 0.02±0.16 & 0.20±0.20 & -0.11±0.05 & 0.31±0.11 & -0.03±0.02 & 0.60±0.10 & -0.01±0.01 & -0.01±0.04 & -0.02±0.02 & -0.01±0.10 & 0.04±0.02 & -0.03±0.02 & 0.12±0.09 \\ 
\hspace{4mm}A & -0.10±0.07 & 0.22±0.11 & 0.39±0.16 & -0.08±0.07 & 0.49±0.13 & -0.02±0.02 & 0.66±0.13 & 0.05±0.02 & 0.11±0.14 & 0.07±0.11 & 0.12±0.14 & 0.10±0.03 & -0.02±0.06 & $-$ \\
6715 & $-$ & $-$ & $-$ & $-$ & $-$ & 0.12±0.13 & $-$ & $-$ & $-$ & -0.04±0.08 & 0.12±0.13 & 0.02±0.23 & $-$ & $-$ \\
\hspace{4mm}A & $-$ & $-$ & $-$ & $-$ & $-$ & 0.05±0.13 & $-$ & $-$ & $-$ & 0.00±0.07 & 0.11±0.10 & 0.22±0.19 & $-$ & $-$ \\
6752 & -0.17±0.11 & -0.16±0.10 & 0.45±0.22 & -0.35±0.11 & 0.24±0.10 & -0.07±0.02 & 0.83±0.10 & 0.00±0.02 & 0.08±0.05 & -0.05±0.03 & 0.02±0.02 & 0.01±0.01 & 0.01±0.04 & 0.07±0.11 \\
6809 & 0.01±0.07 & -0.13±0.25$^{\rm *}$ & 0.52±0.18$^{\rm *}$ & -0.15±0.05 & 0.46±0.07 & -0.10±0.02 & 0.78±0.06 & 0.04±0.02 & -0.02±0.02 & 0.00±0.01 & 0.01±0.09 & 0.02±0.01 & -0.05±0.02 & $-$ \\                   
6838 & -0.03±0.09 & -0.12±0.10 & 0.62±0.21 & -0.10±0.03 & 0.18±0.04 & -0.04±0.01 & 0.03±0.03 & -0.03±0.01 & -0.04±0.03 & -0.02±0.02 & 0.08±0.02 & -0.02±0.01 & 0.00±0.01 & $-$ \\   
6934 & $-$ & $-$ & $-$ & $-$ & $-$ & $-$ & $-$ & $-$ & $-$ & $-$ & $-$ & $-$ & $-$ & $-$ \\                                                                                   
7078 & $-$ & $-$ & $-$ & $-$ & $-$ & -0.03±0.08 & 0.20±0.34 & 0.12±0.11 & $-$ & $-$ & 0.07±0.05 & 0.04±0.02 & $-$ & $-$ \\                                                    
7089 & $-$ & $-$ & $-$ & $-$ & $-$ & -0.09±0.03 & 0.62±0.10 & 0.06±0.07 & $-$ & -0.12±0.05 & -0.04±0.03 & 0.02±0.02 & -0.02±0.04 & $-$ \\                                     
\hspace{4mm}A & $-$ & $-$ & $-$ & $-$ & $-$ & $-$ & $-$ & $-$ & $-$ & $-$ & $-$ & 0.09±0.05 & $-$ & $-$ \\
7099 & $-$ & $-$ & $-$ & $-$ & $-$ & $-$ & $-$ & $-$ & $-$ & $-$ & $-$ & -0.01±0.02 & $-$ & $-$ \\
\hline
\\
\end{tabular}
}

\label{tab:deltas}
\end{table*}

\begin{table*}
    \centering
\tiny
\setlength{\tabcolsep}{5.5pt}
\renewcommand{\arraystretch}{1.6}

\caption{
Elemental spreads $W_{\rm [X/Fe]}$ and $W_{\rm C,N,O,Na,Mg,Al}$ derived in Section~\ref{sec:4} and reported in Figure~\ref{fig:width} and~\ref{fig:total}. Entries with the '$-$' symbol refers to elements with no measurements, while $^{\rm *}$ is referred to C and N measurements above the RGB bump.}

\scalebox{0.8}{

\begin{tabular}{c|ccccccccccccc}
     {NGC} &   $W_{\rm [C/Fe]}$ &   $W_{\rm [N/Fe]}$ &   $W_{\rm [O/Fe]}$ &   $W_{\rm [Na/Fe]}$ &   $W_{\rm [Mg/Fe]}$ &   $W_{\rm [Al/Fe]}$ &   $W_{\rm [Si/Fe]}$ &   $W_{\rm [K/Fe]}$ &   $W_{\rm [Ca/Fe]}$ &   $W_{\rm [Ti/Fe]}$ &   $W_{\rm [Fe/H]}$ &   $W_{\rm [Ni/Fe]}$ &   $W_{\rm C,N,O,Na,Mg,Al}$ \\
\hline
\\
0104 & 0.48±0.12           & 1.03±0.34           & 0.35±0.03 & 0.35±0.03 & 0.04±0.01 & 0.28±0.03 & 0.13±0.02 & 0.13±0.02 & 0.11±0.02 & 0.27±0.02 & 0.08±0.02 & 0.09±0.01 & 2.11±0.56 \\
0288 &                 $-$ &                 $-$ & 0.41±0.07 & 0.61±0.06 & 0.04±0.03 & 0.42±0.06 & 0.06±0.02 & 0.18±0.06 & 0.19±0.06 & 0.13±0.06 & 0.02±0.02 & 0.08±0.03 &       $-$ \\
0362 &                 $-$ &                 $-$ & 0.10±0.04 &       $-$ & 0.11±0.02 & 0.49±0.11 & 0.07±0.02 & 0.16±0.04 & 0.07±0.03 & 0.19±0.05 & 0.13±0.03 & 0.08±0.05 &       $-$ \\
1261 &                 $-$ &                 $-$ &       $-$ &       $-$ & 0.11±0.03 & 0.47±0.11 & 0.18±0.04 &       $-$ & 0.00±0.07 & 0.35±0.15 & 0.12±0.03 & 0.00±0.03 &       $-$ \\
1851 & 0.46±0.10$^{\rm *}$ & 0.76±0.09$^{\rm *}$ & 0.50±0.06 & 0.69±0.07 & 0.05±0.02 & 0.49±0.06 & 0.20±0.04 & 0.23±0.04 & 0.20±0.04 & 0.21±0.04 & 0.04±0.02 & 0.18±0.03 &       $-$ \\
1904 & 0.61±0.08$^{\rm *}$ & 0.96±0.09$^{\rm *}$ & 0.30±0.06 & 0.66±0.08 & 0.22±0.04 & 0.80±0.06 & 0.33±0.04 & 0.32±0.02 & 0.32±0.03 & 0.25±0.04 & 0.41±0.06 & 0.34±0.05 &       $-$ \\
2419 &                 $-$ &                 $-$ &       $-$ &       $-$ & 1.22±0.14 &       $-$ &       $-$ & 1.37±0.21 & 0.16±0.06 & 0.24±0.06 & 0.00±0.05 &       $-$ &       $-$ \\
2808 & 0.49±0.07           & 0.99±0.13           & 0.81±0.12 & 0.53±0.06 & 0.42±0.07 & 0.98±0.05 & 0.23±0.03 & 0.18±0.03 & 0.20±0.02 & 0.30±0.04 & 0.16±0.02 & 0.01±0.02 & 3.43±0.50 \\
3201 & 0.26±0.07           & 1.19±0.14           & 0.43±0.07 & 0.57±0.06 & 0.17±0.02 & 0.92±0.10 & 0.04±0.02 & 0.85±0.10 & 0.34±0.07 & 0.35±0.10 & 0.14±0.02 & 0.21±0.06 & 2.33±0.47 \\
4590 &                 $-$ &                 $-$ & 0.34±0.11 & 0.50±0.08 & 0.10±0.04 &       $-$ &       $-$ &       $-$ & 0.07±0.07 &       $-$ & 0.08±0.03 & 0.34±0.06 &       $-$ \\
4833 &                 $-$ &                 $-$ & 0.34±0.13 & 0.55±0.05 & 0.27±0.06 &       $-$ & 0.00±0.01 &       $-$ & 0.00±0.02 & 0.00±0.03 & 0.02±0.05 & 0.46±0.15 &       $-$ \\
5024 &                 $-$ &                 $-$ & 0.32±0.08 & 0.25±0.06 &       $-$ &       $-$ &       $-$ &       $-$ & 0.00±0.06 & 0.15±0.06 & 0.00±0.06 & 0.00±0.06 &       $-$ \\
5053 &                 $-$ &                 $-$ &       $-$ &       $-$ &       $-$ &       $-$ &       $-$ &       $-$ &       $-$ &       $-$ & 0.00±0.04 &       $-$ &       $-$ \\
5139 & 0.85±0.05           & 1.27±0.12           & 0.78±0.04 & 0.67±0.05 & 0.52±0.06 & 1.21±0.05 & 0.24±0.02 & 0.58±0.06 & 0.27±0.03 & 0.40±0.03 & 0.62±0.07 & 0.21±0.02 & 4.13±0.36 \\
5272 & 0.40±0.21           & 0.87±0.17           & 0.39±0.03 &       $-$ & 0.18±0.02 & 0.95±0.05 & 0.06±0.03 & 0.53±0.10 & 0.21±0.03 & 0.79±0.14 & 0.26±0.03 & 0.19±0.02 &       $-$ \\
5286 &                 $-$ &                 $-$ & 0.36±0.07 & 0.52±0.10 &       $-$ &       $-$ & 0.19±0.03 &       $-$ & 0.00±0.04 & 0.06±0.04 & 0.22±0.02 &       $-$ &       $-$ \\
5904 & 0.48±0.05           & 1.12±0.13           & 0.52±0.06 & 0.62±0.04 & 0.21±0.02 & 0.84±0.05 & 0.04±0.02 & 0.28±0.03 & 0.20±0.02 & 0.21±0.02 & 0.13±0.01 & 0.16±0.01 & 2.84±0.35 \\
5927 &                 $-$ &                 $-$ &       $-$ &       $-$ & 0.00±0.02 & 0.06±0.02 & 0.26±0.04 &       $-$ &       $-$ & 0.00±0.04 & 0.03±0.02 & 0.00±0.02 &       $-$ \\
5986 &                 $-$ &                 $-$ &       $-$ &       $-$ &       $-$ &       $-$ &       $-$ &       $-$ &       $-$ &       $-$ &       $-$ &       $-$ &       $-$ \\
6093 &                 $-$ &                 $-$ &       $-$ &       $-$ &       $-$ &       $-$ &       $-$ &       $-$ &       $-$ &       $-$ & 0.00±0.01 &       $-$ &       $-$ \\
6121 & 0.29±0.06           & 0.82±0.08           & 0.34±0.04 & 0.42±0.03 & 0.07±0.02 & 0.30±0.05 & 0.06±0.02 & 0.22±0.03 & 0.12±0.02 & 0.13±0.03 & 0.11±0.02 & 0.05±0.01 & 1.41±0.28 \\
6205 & 0.56±0.05           & 0.82±0.31           & 0.76±0.17 & 0.59±0.06 & 0.38±0.16 & 1.33±0.31 & 0.13±0.16 & 0.60±0.08 & 0.22±0.04 & 0.36±0.05 & 0.15±0.03 & 0.25±0.05 & 3.25±1.10 \\
6218 & 0.34±0.05           & 0.76±0.22           & 0.53±0.07 & 0.67±0.07 & 0.06±0.03 & 0.35±0.04 & 0.01±0.02 & 0.32±0.09 & 0.21±0.04 & 0.07±0.05 & 0.05±0.03 & 0.04±0.03 & 1.69±0.48 \\
6254 & 0.59±0.16           & 1.03±0.42           & 0.35±0.07 & 0.69±0.07 & 0.15±0.03 & 1.22±0.07 & 0.04±0.02 & 0.79±0.11 & 0.17±0.05 & 0.27±0.06 & 0.14±0.02 & 0.11±0.03 & 2.82±0.82 \\
6362 &                 $-$ &                 $-$ &       $-$ & 0.52±0.07 &       $-$ &       $-$ &       $-$ &       $-$ &       $-$ &       $-$ & 0.00±0.03 &       $-$ &       $-$ \\
6388 &                 $-$ &                 $-$ &       $-$ &       $-$ &       $-$ &       $-$ &       $-$ &       $-$ &       $-$ &       $-$ &       $-$ &       $-$ &       $-$ \\
6397 &                 $-$ &                 $-$ &       $-$ & 0.50±0.12 &       $-$ &       $-$ &       $-$ &       $-$ &       $-$ &       $-$ & 0.00±0.01 &       $-$ &       $-$ \\
6402 &                 $-$ &                 $-$ & 1.23±0.20 & 0.71±0.08 & 0.17±0.05 & 0.82±0.15 & 0.11±0.08 &       $-$ & 0.00±0.02 &       $-$ & 0.00±0.03 & 0.00±0.03 &       $-$ \\
6535 &                 $-$ &                 $-$ &       $-$ & 0.32±0.12 & 0.00±0.02 &       $-$ &       $-$ &       $-$ &       $-$ &       $-$ & 0.09±0.03 &       $-$ &       $-$ \\
6656 & 0.63±0.07           & 0.91±0.08           & 0.47±0.02 & 0.87±0.08 & 0.23±0.03 & 1.10±0.03 & 0.15±0.02 & 0.56±0.08 & 0.40±0.04 & 0.69±0.08 & 0.29±0.17 & 0.25±0.03 & 2.87±0.31 \\
6715 &                 $-$ &                 $-$ &       $-$ & 0.76±0.10 & 0.68±0.14 &       $-$ & 0.25±0.05 &       $-$ & 0.31±0.08 & 0.42±0.10 & 1.10±0.32 & 0.13±0.03 &       $-$ \\
6752 & 0.48±0.09           & 0.84±0.18           & 0.63±0.07 & 0.61±0.07 & 0.15±0.03 & 1.05±0.06 & 0.32±0.02 & 0.29±0.06 & 0.19±0.04 & 0.64±0.07 & 0.15±0.04 & 0.37±0.06 & 2.56±0.49 \\
6809 & 1.09±0.10$^{\rm *}$ & 0.68±0.15$^{\rm *}$ & 0.23±0.07 & 0.68±0.07 & 0.14±0.03 & 1.06±0.09 & 0.10±0.02 & 0.21±0.03 & 0.29±0.10 & 0.33±0.06 & 0.07±0.02 & 0.15±0.03 &       $-$ \\
6838 & 0.23±0.04           & 0.87±0.10           & 0.10±0.04 & 0.24±0.05 & 0.12±0.04 & 0.16±0.03 & 0.04±0.01 & 0.19±0.05 & 0.16±0.05 & 0.16±0.02 & 0.06±0.03 & 0.04±0.01 & 1.25±0.29 \\
6934 &                 $-$ &                 $-$ &       $-$ &       $-$ &       $-$ &       $-$ &       $-$ &       $-$ &       $-$ &       $-$ &       $-$ &       $-$ &       $-$ \\
7078 &                 $-$ &                 $-$ &       $-$ & 0.44±0.10 & 0.34±0.13 & 0.86±0.24 & 0.12±0.18 &       $-$ &       $-$ & 0.00±0.04 & 0.07±0.06 &       $-$ &       $-$ \\
7089 &                 $-$ &                 $-$ &       $-$ &       $-$ & 0.10±0.04 & 0.80±0.10 & 0.14±0.02 &       $-$ & 0.00±0.03 & 0.09±0.03 & 0.04±0.03 & 0.00±0.02 &       $-$ \\
7099 &                 $-$ &                 $-$ &       $-$ &       $-$ &       $-$ &       $-$ &       $-$ &       $-$ &       $-$ &       $-$ & 0.03±0.03 &       $-$ &       $-$ \\

\hline

\\
\end{tabular}
}

\label{tab:deltas}
\end{table*}

\renewcommand{\thefigure}{B.\arabic{figure}}
\renewcommand{\thetable}{B.\arabic{table}}
\setcounter{figure}{0}
\setcounter{table}{0}

\begin{figure*}
\includegraphics[clip,width=15cm]{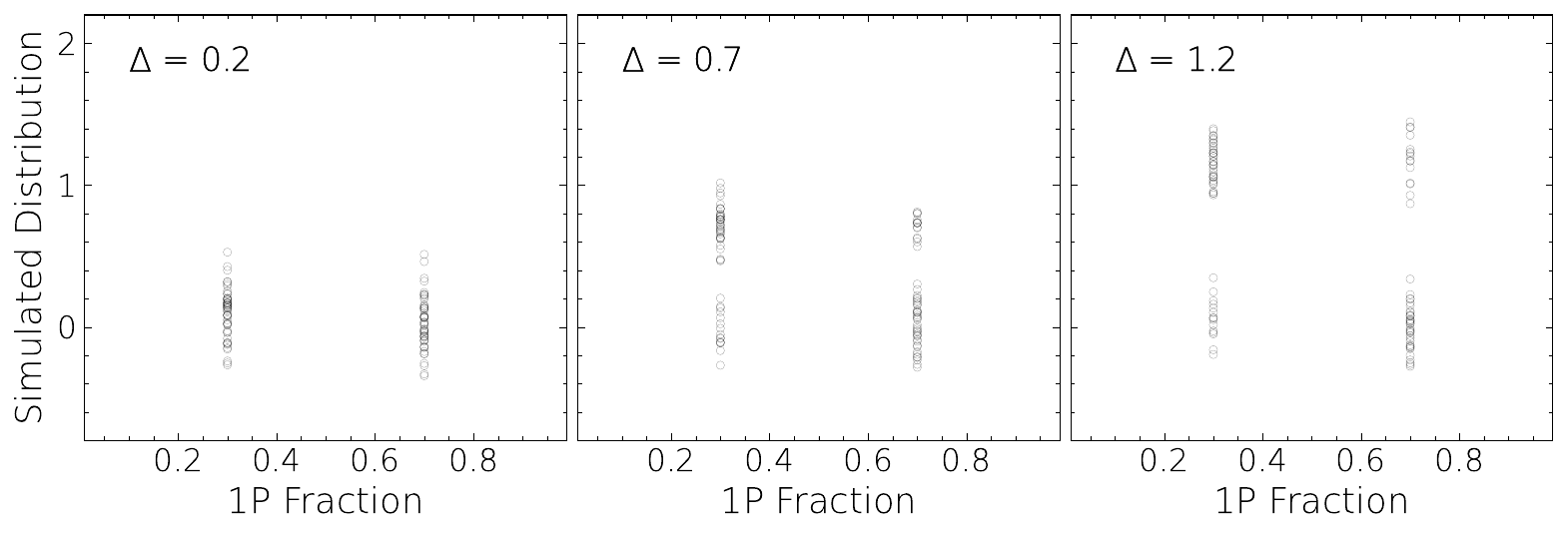}
    \caption{Simulated distributions of spectroscopic abundances when the 1P fraction is 0.3 and 0.7. Left, central, and right panels represents simulation for $\Delta =$ 0.2, 0.7, and 1.2, repsectively (see text for details).}
    \label{fig:simulation}
\end{figure*}

\end{appendix}

\end{document}